\documentclass[prd,preprint,eqsecnum,amsfonts,amssymb,amsmath,floatfix,nofootinbib]{revtex4}

\usepackage{bm}
\usepackage{slashed}
\usepackage{graphicx}
\usepackage{MnSymbol}
\usepackage{braket}
\usepackage{appendix}
\usepackage{float}
\usepackage{mathtools}
\usepackage{enumitem}

\makeatletter
\newcommand{\raisemath}[1]{\mathpalette{\raisem@th{#1}}}
\newcommand{\raisem@th}[3]{\raisebox{#1}{$#2#3$}}
\makeatother

\DeclareMathOperator{\sgn}{sgn}

\begin{document}

\title{\textbf{The cosmological constant, dark matter, \\ and other unsolved problems from a fresh perspective}}

\author{Roland E. Allen}

\affiliation{Department of Physics and Astronomy \\ Texas A\&M University,  College Station, Texas 77843, USA \\ 
allen@tamu.edu, orcid 0000-0001-5951-4717}

\date{\today}

\begin{abstract}
Quantum theory, general relativity, the standard model of particle physics, and the $\Lambda$CDM model of cosmology have all been spectacularly successful within their respective regimes of applicability, but many central problems remain unsolved. Here we propose a fresh perspective on these problems, based on a description which has elements in common with other efforts toward a fundamental theory, since it also is based on higher dimensions (with an internal space), a form of supersymmetry, important topological structures, and the implication of a multiverse: Our particular universe is topologically stable because it contains the product of two vortex-like (or instanton-like) configurations of a primordial condensate -- one in 4-dimensional external spacetime, with the big bang at its origin; and the second in a 10-dimensional (naturally compactified) internal space, which automatically yields an SO(10) grand-unified gauge theory. A third 3-dimensional (also compactified) internal space yields family replication. The arguments leading up to and following this basic picture imply a radically modified view of many unsolved problems and related issues, including the absence of an enormous cosmological constant, the nature of dark matter, the origin of supersymmetry, the origin of Lorentz invariance, the origin of gravitational and gauge interactions, the gravitational metric and its signature (which distinguishes time from space and characterizes spacetime as 4-dimensional),  the fundamental action for fermionic and bosonic fields, the regularization of quantum gravity, the origin of quantum fields, the origin of spacetime coordinates, the entropy of black holes, and the probability interpretation of quantum mechanics. 
Among other near-term predictions, calculations for the dark matter WIMP based on these ideas show that it should be observable at the high-luminosity LHC,  nominally at the 5$\sigma$ level with 500 fb$^{-1}$ of integrated luminosity, and it may already have been observed by Fermi-LAT and AMS-02.
 \end{abstract}


\maketitle


\section{\label{sec:sec1}Introduction}

The great achievements of modern physics and astronomy have actually \textit{increased} the number of perceived mysteries as we strive for a fundamental understanding of Nature. Some of these mysteries were already appreciated about a century ago, by Einstein and others, and some have arisen from recent discoveries. Many excellent reviews have been given of the current situation~\cite{pdg}, but it may be worthwhile to begin with a brief summary.

Most recently, the particle discovered by the ATLAS and CMS collaborations at the LHC is now known to be a Higgs boson~\cite{pdg,ATLAS,CMS}. A naive conclusion is that the standard model of particle physics is now complete. But the more profound interpretation is that the discovery of a scalar boson immediately points to physics beyond the standard model, since otherwise radiative corrections should push the mass of this particle up to an absurdly large value. The most natural candidate for such new physics is supersymmetry~\cite{baer-2025,Constantin-2025,Baer-Tata,Dreiner,Nath,Mohapatra,Drees,Binetruy,Djouadi,Kane-susy,Haber-Kane,Arkani-Hamed-Giudice,Ellis-Olive,Baer-Barger-2016,Baer-Barger-Tata,Roszkowski-2018,Baer-Barger-2020,Tata-2020,3}, for which there is already indirect experimental evidence: The coupling constants of the 3 nongravitational forces are found to converge to a common value, as they are run up to high energy in a grand unified theory, only if the calculation includes superpartners. So, instead of acting as an endpoint for physics, and a mere capstone of the standard model, the observation of a Higgs boson may open the door to a plethora of new particles and effects. But supersymmetry has not yet been observed, despite widespread initial optimism, and even at the theoretical level a generally accepted mechanism for supersymmetry breaking has not been found.

Another major advance has been the discovery and exploration of neutrino masses~\cite{nu-2021}, which appear to open the door to a more fundamental understanding of forces and matter via grand unification~\cite{Nath,Mohapatra,SO(10)-1975,Nanopoulos-1979,Nanopoulos-1984,barger,Pernow,Ozer,cheng,ross}. There are two possibilities for a neutrino mass, either of which is inconsistent with the requirements of the standard model: For a Dirac mass an extra field has to be added for each generation of fermions, and for a Majorana mass lepton number conservation has to be violated, whereas either or both types of mass are natural with grand unification, and in addition a seesaw mechanism can explain the small observed values of neutrino masses. Still, it is not yet demonstrated whether neutrinos have Majorana masses or Dirac masses or both, and the unification of coupling constants at high energy appears to require supersymmetry, which again has an uncertain status currently.

There are many other gaps in fundamental understanding. For example, the discovery and exploration of cosmic acceleration~\cite{riess,perlmutter} has suggested the need for truly revolutionary new physics. The cause of this acceleration has increasingly been found to resemble a cosmological constant $\Lambda$, and has therefore been a strong reminder of the original cosmological constant problem~\cite{weinberg-1989}: Because of the various contributions to the vacuum energy, conventional general relativity predicts that $\Lambda$ should be vastly larger than permitted by observation, unless there is a somewhat magical cancellation of positive and negative contributions. This problem is so difficult that a number of outstanding theorists, including Steven Weinberg~\cite{weinberg-1987,weinberg-2005}, have felt that there is no convincing solution other than an anthropic argument.

A parallel astronomical mystery is the origin of dark matter~\cite{hooper}. Many dark matter candidates for which there was initial optimism have now been ruled out by the remarkable achievements of the experimentalists, in obtaining ever-increasing sensitivity and sophistication.

Yet another and even older theoretical problem is the difficulty of reconciling general relativity with quantum mechanics: A new fundamental theory must somehow regularize quantum gravity, in addition to reducing the value of $\Lambda$ by many orders of magnitude.

The next level of theoretical understanding is not likely to be a ``theory of everything'', since ``everything'' surely transcends our current observational capabilities and imagination.  But the following are a few among the many problems that a fundamental theory might hope to explain: the absence of an enormous cosmological constant, the nature of dark matter, the origin of supersymmetry, the origin of Lorentz invariance, the origin of gravitational and gauge interactions, the gravitational metric and its signature (which distinguishes time from space and characterizes spacetime as 4-dimensional),  the fundamental action for fermionic and bosonic fields, the regularization of quantum gravity, the origin of quantum fields, the origin of spacetime coordinates, the entropy of black holes, and the probability interpretation of quantum mechanics. 

The present theory addresses all of the issues listed above and leads to new predictions, some of which can be tested in near-term experiments, 
including the dark matter candidate discussed below and in our previous papers~\cite{DM2021a,Tallman}.

\section{\label{sec:sec1a}Overview}

DeWitt has provided an elegant survey of contemporary fundamental physics~\cite{dewitt}, which is based on path-integral quantization over classical
trajectories in the combined space of coordinates and fields:
``A classical dynamical system is described globally by a 
\textit{trajectory} or \textit{history}. A history is a section of a fibre
bundle $E$ having the manifold $M$ of spacetime as its base space. 
The
typical fibre is known as configuration space and will be denoted by $C$....
Denote by $\Phi $ the set, or space, of all possible
field histories, both those that do and those that do not satisfy the
dynamical equations .... The nature and dynamical properties of a classical
dynamical system are completely determined by specifying an \textit{action
functional} $S$ for it.''

This is the basic picture used in most versions of fundamental physics
that are investigated by sizable communities of physicists. Notice that
coordinates and fields have essentially the same status. This is consistent
with the way they are defined in the present theory, starting near the
beginning of the next section. In standard field theory, the spacetime
coordinates $x^{\mu }$ correspond to $M$ and the fields to $C$. In
conventional string (or $p$-brane) theory, $M$ corresponds to a $2-$dimensional
worldsheet (or  $\left( p+1\right) -$dimensional worldvolume) and $C$ to 
bosonic and fermionic coordinates. Note also that coordinates (in nonrelativistic 
quantum mechanics) and fields (in quantum field theory) 
are treated essentially the same way in both canonical
and path-integral quantization.

The arguments below involve many unfamiliar definitions and redefinitions of fields, with 
physically accessible fields emerging from more primitive 
 ``hidden'' degrees of freedom. But this kind of treatment is already necessary and familiar in standard physics -- 
 for example, when the fundamental gauge fields are redefined after Higgs condensation, 
 and when many kinds of elementary excitations are defined in condensed matter physics. 
 More broadly, the deeper hidden origin
of observed phenomena is a common theme in science. 

The physical fields that have emerged by the end of this paper are interpreted as
those chosen by Nature to yield a stable vacuum $\left\vert 0\right\rangle $ satisfying
$a\left\vert 0\right\rangle =0$, where $a$ is a typical destruction operator
for one of these fields. For
example, the excitations $a^{\dag}\left\vert 0\right\rangle $ must have positive 
energy.

The principal ideas and results of this paper are as follows.

(1) In standard physics, all the events of the world are simply a progression through 
the states of fields on a spacetime manifold. In the present picture, they are a progression 
through the states of a single fundamental system. This second unified picture leads back to (and beyond) the 
first via the arguments below.

Time will be defined as a parameter describing a trajectory through the space of states. 
There are, of course, precedents for defining time internally, within a stationary state, using some internal parameter. 
For example, in minisuperspace models based on the Wheeler-DeWitt equation~\cite{DeWitt-1967,Hartle-Hawking}, this parameter is the cosmic scale factor, and time is defined by the expansion of the universe. In the present picture, \textit{all} coordinates and fields are defined internally, with the progression of events in Nature regarded as progression through a space of states. In the most primitive space, each point is defined by a microstate 
$\ket{m}$. In the emergent space corresponding to DeWitt's $\Phi $ (see above), each point is defined by a macrostate, as described below.

For a concrete characterization of both microstates and macrostates, 
we adopt the picture that Nature is composed of discrete distinguishable constituents that are called 
``dits'' because each can exist in any of $d$ states labeled by $i=1,2,...,d$. 
The number of dits in the $i$th state 
determines the size of the field or coordinate labeled by $i$.

The fact that fields and coordinates have the same basic status in standard
physics, as noted above, is then explained by the fact that they have the
same fundamental origin.

This appears to be the simplest imaginable picture that can still lead back to standard physics. The fundamental objects are dits (a generalization of Wheeler's ``it from bit''~\cite{it}), and these are the ``atoms'' of the fundamental system. Coordinates and fields are then emergent quantities, associated with the possible paths through states of the fundamental system. It is somewhat miraculous that such a simple picture can lead back to standard physics and its natural generalizations.

(2) The action is defined to be essentially the negative of the entropy for an observable macrostate (which corresponds to a particular set of values of coordinates and fields). Action conventionally has the units of $\hbar $, and entropy the units
of the Boltzmann constant $k$, but here we use natural units, with 
$\hbar =k=c=1$.

(3) Much later in the paper, the resulting path integrals with Euclidean form are 
transformed to equivalent Lorentzian path integrals, with the action left
unchanged. 

The standard formulation of quantum field theory has then been regained. One
can subsequently transform from path-integral quantization to canonical
quantization in the usual way (since the action has a standard form).

Since spacetime points are discrete (although very closely spaced) in the present description, path integrals are well-defined. All aspects of the theory are in fact clearly mathematically consistent, since both coordinates and fields are initially defined on a lattice specified by integers, and the continuum is only an approximation. In contrast, both the meaning of path integrals and overall mathematical consistency are issues in field theory and in extensions such as string theory.

(4) The assumption that all states of the fundamental system are realized
leads inevitably to a multiverse picture. There are still many who reject the
possibility of a multiverse, but one should recall that most people at the
time of Galileo would have rejected the possibility of  hundreds of billions
of galaxies, or a single galaxy, or even a heliocentric Solar System, and that 
the history of physics shows a steady progression toward more expansive 
views of Nature.

Among the vast number of states of the fundamental system, there are some trajectories through these states which can
legitimately be assigned to universes, in the sense that the states or configurations can be
coherently connected with high probability. This will be the case if there is a 
path through the space of these configurations along which a local 
minimum is stably maintained in the action.

In the present picture our own universe is stabilized \textit{topologically} 
-- specifically, it has an extremely stable geometry determined by 
two topological defects in a 17-dimensional primordial condensate:  
one in $4$-dimensional external
spacetime, and the other in an internal space with 10 dimensions.
These are ``vortex-like'' (or ``instanton-like''), in the sense that they involve 
 a rotation of the primordial condensate around a central point. 
 The stability of our universe is then analogous to the stability 
 of a vortex in a 2-dimensional fluid. (In the remaining 3-dimensional space the order parameter is isotropic, with the geometry of a tiny 3-dimensional ball, curled up to have finite action.)
 
This stability has an additional  feature, in that there is a natural mechanism for compactification of the internal space, as mentioned at the end of Appendix \ref{sec:appB}. (The term ``compactification'' as used here means that the internal space is restricted to a finite and very small volume.) Since the relevant functions in the internal space are exponentially decaying solutions to the fundamental higher-order differential equation, they will be confined to an extremely small region (with a length scale $\sim \mathrm{\textit{few}} \times a_0$, where $a_0$ is the extremely small fundamental length defined below). There is then a natural explanation of the difference between external spacetime and the internal space: The condensate and basis functions in each internal space decay exponentially (with distance from a central point), but the external condensate begins in one of the nondecaying solutions, with normal and unlimited expansion of the universe afterward. 
 
 Despite early attempts by Candelas and Weinberg~\cite{weinberg-com}, and others, a convincing mechanism for compactification in Kaluza-Klein, supergravity, and string theories is an unsolved problem, and the qualitatively different nature of the internal space is ordinarily just postulated.

There is an additional nuance, in that a universe can
be made stable through an effect which is exhibited in the behavior  of the
quartic self-coupling of the Higgs: Within the
standard model, the unrenormalized value of this coupling appears to be very
nearly equal to zero~\cite{elias-miro,alekhin,degrassi,buttazzo}, but at low energies it is made appreciably finite (and positive) by
radiative corrections, so that a stable Higgs condensate forms. This
suggests that more generally there will be configurations, within the
complete path integral of all possibilities, where radiative corrections will similarly yield a
nonzero quartic self-coupling for a primordial condensate which allows it to form. 
In order to search for such a possible solution (i.e. a persistent local minimum in the
action), we adopt the artifice of adding an imaginary random potential
to the action, proportional to a parameter $b$ which is ultimately taken
to go to zero: $b\rightarrow 0+$.

Known physics is regained in a very simple picture based on a
primordial condensate with an SU(2)$\times$U(1) order parameter in 4-dimensional external
spacetime, a Spin(10)$\times$U(1)
order parameter in the 10-dimensional internal space, and an isotropic order parameter in the 3-dimensional internal space.
Each of the first two factors in the overall order parameter has a vortex-like
topological singularity at the origin. The external topological defect is
interpreted as the big bang, and the internal topological defect gives rise 
to a Spin(10) grand-unified gauge group. The 3-dimensional space yields 3 ($\ell=1$) families of fields from the $32=16 + \overline{16}$ representation of Spin(10), with Yukawa couplings to one ($\ell=0$) Higgs family.

The gravitational vierbein and the gauge fields of nongravitational forces are interpreted as ``superfluid velocities'',
with arbitrary curvatures permitted by a background of ``rapidly fluctuating'' topological
defects (in the complete path integral) that are analogous to vortices and vortex rings (or extended and closed flux tubes) -- in roughly the same way that, as shown by Feynman and Onsager, vortices permit rotation of a superfluid. Background quantum fluctuations including topological defects \textemdash \, constituting ``quantum foam'' \textemdash \, have been previously proposed in various contexts including quantum gravity~\cite{foam,foam2}.

(5) In the detailed treatment, fermion fields and scalar-boson
fields are found to automatically couple in the correct way to both the
gravitational field and the gauge fields of the other forces. 

(6) At the same time, local Lorentz invariance automatically emerges (rather
than being postulated), and external spacetime is automatically  $(3+1)-$dimensional.

(7) The present theory unavoidably predicts SO(10) grand unification if the first internal space is 10-dimensional.

I.e., gauge symmetry results from rotational symmetry in the internal space, and this explains why forces can be described by a fundamental gauge symmetry which is so similar to rotational symmetry.

The second (3-dimensional) internal space explains why there are 3 families of fermions with only one Higgs family (containing the two Higgs doublets of supersymmetry).

(8) The present theory also unavoidably predicts an initial mathematical supersymmetry (susy), involving standard fermion fields but unphysical boson fields which must be transformed to obtain a stable vacuum with Lorentz invariance. In the simplest possibility, the final kinetic action (after this transformation)  is the same as in conventional supersymmetry (SUSY). It is then natural to follow the usual SUSY models by postulating a conventional superpotential, in which the postulated interactions have the same symmetry as the derived kinetic terms. 
I.e., in the simplest scenario the present theory is consistent with conventional broken SUSY~\cite{baer-2025,Baer-Tata,Dreiner,Nath,Mohapatra,Drees,Binetruy,Djouadi,Kane-susy,Haber-Kane,Arkani-Hamed-Giudice,Ellis-Olive,Baer-Barger-2016,Baer-Barger-Tata,Roszkowski-2018,Baer-Barger-2020,Tata-2020,3}.

There is no generally accepted solution for the breaking of conventional SUSY, but here the primitive susy is automatically broken for scalar bosons when the transformed scalar-boson fields acquire mass-squared terms from radiative corrections (as in e.g. Eq. (1.4) of \cite{Drees}, which still holds when the bare masses are zero, or in the emergence of a negative mass-squared term with renormalization in conventional SUSY~\cite{Baer-Tata}), while their fermion superpartners remain massless. (The reverse is true for gauge bosons and gauginos.) The two hierarchy problems for low-mass scalar bosons are resolved if their masses (i) result only from relatively small radiative corrections and (ii) are protected by supersymmetry. (The bosonic condensations at high energies that result in SO(10)$ \rightarrow \,$SU(3)$\times$SU(2)$\times$U(1) are beyond the scope of the present paper.) Other aspects of supersymmetry breaking require further radiative corrections, or additional fields as in conventional SUSY.

There will still be a lightest supersymmetric particle~\cite{Baer-Tata,Nath,Mohapatra,Kane-susy} which (as a subdominant component) can stably coexist with the dark matter candidate of the present theory, as can axion-like particles~\cite{PQ,weinberg-axion,wilczek-axion,Sikivie,Tanner} (and possibly others), because these various species are stable for different reasons.

(9) The usual cosmological constant (regarded as one of the deepest
problems in standard physics) automatically vanishes for two independent
reasons:

(i) For fermion fields and scalar-boson fields, there is initially no factor of $e=\sqrt{-g}$ 
in the integrals giving their action. As shown below, this implies that their conventional contribution to a cosmological constant vanishes. The standard familiar form of the action can be regained,
\begin{equation}
 S_{matter}  = \int d^{4}x\, e \, \overline{\mathcal{L}}_{matter}  \; ,
\end{equation}
but the stress-energy tensor for a fermion or scalar-boson field $\chi$ must then be calculated from 
 \begin{align}
T^{\mu \nu }_{\chi} =  2 \, \chi^{\dag} \, \frac{\delta A_{\chi} } {\delta g_{\mu \nu} }  \, \chi 
\label{nocos}
\end{align}
and not the conventional
\begin{align}
T^{\mu \nu }_{\chi} = 2 \, e^{-1} \frac{\delta \left( e \chi^{\dag} A_{\chi} \chi  \right)}{\delta g_{\mu \nu}} 
\end{align}
where the notation is defined below.

(ii) When the gauge-field action is quantized, the operators must be
normal-ordered, in accordance with the interpretation of the origin of this
action in the present theory: It arises from the response of vacuum fields
to the curvature of the external gauge fields, and it must therefore vanish
when there are no external fields. The vacuum stress-energy  tensor for the
gauge fields then also vanishes.

As shown below, standard physics is regained in each case:

(i) Classical matter (which follows the on-shell classical
equations of motion) acts as a source for Einstein gravity in the same way
as in standard physics, and all matter and fields move in the same way.

(ii) The results are consistent with experiment and observation, even though
there is no vacuum zero-point energy for the gauge fields.

Many people (including some who are otherwise expert in this area) will
naively object that the Casimir effect, as verified experimentally,
demonstrates that the electromagnetic field does have a zero-point energy in
the vacuum. 

This belief is common but incorrect~\cite{Jaffe}. The experimentally-observed Casimir effect demonstrates only that the static
electromagnetic field energy is \textit{changed} by the modification of
boundary conditions~\cite{casimir1,casimir2}. In the simplest model, two metal plates are inserted
and the force between them calculated. There are two ways to do the
calculation: The first is indeed to assume zero-point vibrations of the
electromagnetic field, whose energy is modified when the boundary conditions
are modified. The second approach is instead to consider the processes
involving virtual photons which mediate the interaction of the plates, with
no reference to zero-point vibrations and no need for a vacuum energy. The
first method is more popular because it is easier (and has been assigned as a homework problem by the present author). But the two
methods give the same basic answer, as they should. (The second method regards the force as 
mediated by virtual photons, and the first obtains the force from the derivative of the 
energy with respect to a displacement.) The second method is more
difficult, but is consistent with the way other virtual processes are
calculated in e.g. quantum electrodynamics. Of course, the second method
also implies a change in the static electromagnetic field energy 
(interpreted as a van der Waals interaction), but this 
\textit{change} does not imply an initially nonzero vacuum zero-point energy. 

In summary, the observed Casimir effect is perfectly consistent with the
present theory, in which there is no vacuum zero-point energy due to the
electromagnetic field or other gauge fields.

(10) Although the usual cosmological constant vanishes, there will
still be a weaker response of the vacuum to the curvature imposed by external
fields. Within the present picture, the Maxwell-Yang-Mills action for gauge fields and the Einstein-Hilbert 
action for the gravitational field must arise from the modification of vacuum states (and the terms in their action) when they adjust to the curvature of these fields, just as the free energy of a metal associated with Landau diamagnetism arises from the modification of the electronic ground state when it qualitatively changes to accommodate the gauge curvature of a magnetic field.

(11) Quantum gravity would be regularized by an energy and momentum cutoff $a_0^{-1}$, where $a_0$ is an extremely small minimum length defined below, if it were not already regularized by the finiteness of the assumed vacuum response, which gives the Einstein-Hilbert action only to lowest order in a perturbative treatment. The higher-order corrections in the same treatment can explain inflation~\cite{Guth,Linde,Kolb,Peebles} via the original proposal: Starobinsky inflation~\cite{Starobinsky,Vilenkin}.

(12) The Bekenstein-Hawking entropy of a black hole is obtained in the present picture when a well-known result of Gibbons and Hawking~\cite{Gibbons-Hawking} is used. The new feature is that the entropy ultimately results from counting the microstates of the collection of dits in a specified field configuration \textemdash \, in this case, a specific black hole.

(13) Topological singularities, including those associated with black holes and the big bang, are admissible in the present theory, for the same reason as in a standard superfluid: The density of the primordial condensate underlying standard physics goes to zero at each point where there is there is a gravitational or gauge-field singularity. (These singularities are present in the continuum approximation, on a length scale substantially larger than $a_0$.)

(14) The present theory also leads to the prediction of new particles, including new scalar bosons, new fermions, and a new dark matter WIMP~\cite{DM2021a,Tallman} \textemdash \, which is consistent with experiment and observation because it has no relevant couplings other than its second-order gauge couplings to $W$ and $Z$ bosons (plus couplings to higher-mass gauginos when supersymmetry is included). It should be observable in the high-luminosity LHC, and it may already have been detected via the gamma rays observed by Fermi-LAT and antiprotons observed by AMS-02. This particle is unique among viable dark matter candidates in that both its most relevant couplings and its mass are quantitatively determined (to a good approximation), making clean experimental tests possible in the near future.

(15) The probability interpretation of quantum mechanics is a central feature which has never been convincingly explained~\cite{born,many}. Here we do not attempt to derive this interpretation for a general measurement, which would require a treatment far beyond the scope of the present paper. We only note the suggestive form of the relation below,
\begin{align}
\phi _{k}^{2}\left( x\right) &=\rho _{k}\left( x\right) = \mathrm{density\ of\ microstates} \\
& \propto \mathrm{Boltzmann\ probability\ density\ for\ field\ strength} \; \phi _{k} \; \mathrm{at\ postion} \; x \; ,
\end{align}
which resembles the simplest and most familiar version of the Born rule,
\begin{align}
| \psi (\vec{r} ) |^2 = \mathrm{probability\ density\ for\ observing\ particle\ with\ wavefunction} \; \psi (\vec{r} ) \; \mathrm{at\ position} \; \vec{r} \; .
\end{align}
There is thus an underlying connection from the beginning between probabilities and the intensities of quantum fields. (It should be emphasized that conventional quantum field theory and other aspects of quantum mechanics are fully recovered in the present theory. For example, there are no ``hidden variables'' \textemdash \, only unobservable microstates that yield completely conventional quantum states, treated in either the path-integral or canonical formulation of quantum theory, as described below.) To proceed further, even for a single example, would require specifying detectors composed of the physical bosonic and fermionic fields defined later in the paper, obtaining the probabilities of outcomes for an ensemble of observers which are themselves treated as quantum systems, and then arguing that the probabilities in an ensemble are equivalent to the probabilities obtained by a single observer after multiple measurements. Since this problem is peripheral to the main goals of this paper, it will not be considered further in the remaining sections, which contain the detailed arguments behind the other claims above.

\section{\label{sec:sec2x}Statistical origin of the initial action}

For a theory to be viable, it must be mathematically (and philosophically) consistent, its premises must lead to testable predictions, and these predictions must be consistent with experiment and observation. The theory presented here satisfies these requirements, but it starts with an extremely unfamiliar point of view: There are initially no laws, and instead all possibilities are realized with equal weight. The observed laws of nature are emergent phenomena, which result from statistical counting and the geography (i.e. specific features) of our particular universe in $D$ dimensions. Standard physics (plus natural extensions such as grand unification and supersymmetry) finally emerges as an effective field theory at relatively low energies.

Our starting point is a single fundamental system which consists of
identical (but distinguishable) irreducible objects, which we will call
``dits''. Each dit can exist in any of $d$ states, with the number
of dits in the $i$th state represented by $n_{i}$. An unobservable
microstate of the fundamental system is specified by the number of dits and
the state of each dit. An observable macrostate is specified by only the
occupancies $n_{i}$ of the states. 

As discussed below, $D$ of the states are
used to define $D$ spacetime coordinates $x^{M}$, and $N_{F}$ of the states
are used to define fields $\phi _{k}$. A specific macrostate, which means a specific set $\{ n_{i} \}$ of occupancies, will be called a ``dot'', since it is equivalent to a specific state of all fields at a specific point in the $D$ dimensional coordinate space. 

I.e., a dot is a point in the full combined space of fields and coordinates (as in the above picture of DeWitt). 

Let us begin with the coordinates: 
\begin{equation}
x^M=\Delta n_M  \, a_0  \
\quad , \quad  \Delta n_M = n_M - \overline{n}_M   
\quad , \quad M=0,1,...,D-1 
\label{eq2.1}
\end{equation}
so that $\overline{n}_M$ specifies a shift in the origin of 
coordinates. It is convenient to include a (very small) fundamental length 
$a_{0}$ in this definition, so that we can later express the coordinates in
conventional units. The number of dits in state $M$ determines the size of this coordinate (just as the number of grains in the lower bulb of an hourglass measures the time).

In parallel with the treatment of fields in Appendix \ref{sec:sec2}, we eventually (for convenience) take the limit 
$\overline{n}_M  \rightarrow \infty$, with $\Delta n_M$ finite, and there will then be no lower bound 
to negative coordinates. I.e., $\Delta n_M$ can have any integer value. (A central feature of the present theory is that both 
coordinates and physical fields are defined by relatively 
small perturbations $\Delta n_i = n_i - \overline{n}_i $, analogous to waves 
on a deep ocean.)

Now define a set of initial fields $\phi _{k}$ by 
\begin{equation}
\phi _{k}^{2}\left( x\right) =\rho _{k}\left( x\right) 
\quad , \quad k=1,2,..., N_{F}  
\label{eq2.2}
\end{equation}
where 
\begin{equation}
\rho _{k}\left( x\right) =n_{k}\left( x\right) /a_{0}^{D}  \label{eq2.3}
\end{equation}
and $x$ represents all the coordinates. 
(To avoid awkward notation, we write $n_k$ for $n_{i=D+k}$.)
The most primitive fields $\phi _{k}$ are then real, and defined only up to a phase factor $\pm 1$.

In Appendices \ref{sec:sec2} and \ref{sec:sec3} we obtain the entropy $S$ for a given configuration of the 
$\phi _{k}$ at all points in spacetime, which then leads to the action for a 
given path (i.e.  specific classical field configuration) in the quantum path integral:
\begin{equation}
S_{E}=\int d^{D}x\,\left\{ \frac{1}{2m_{0}}\left[ \frac{\partial \Psi ^{\dagger }
}{\partial x^{M}}\frac{\partial \Psi }{\partial x^{M}}+\frac{a_{0}^{2}}{16}
\frac{\partial ^{2}\Psi ^{\dagger }}{\partial \left( x^{M}\right) ^{2}}
\frac{\partial ^{2}\Psi }{\partial \left( x^{M}\right) ^{2}}\right] -\mu_{0} \,
\Psi^{\dagger }\Psi +\frac{1}{2}b\left( \Psi ^{\dagger }\Psi \right) ^{2}
\right\}  \label{eq3.14x}
\end{equation}
with a sum over the repeated index $M$ in the first two terms.
Here $\Psi $ contains primordial bosonic and fermionic fields $\Psi _{b}$ and $\Psi _{f}$ 
\begin{equation}
\Psi =\left( 
\begin{array}{c}
\Psi _{b} \\ 
\Psi _{f}
\end{array}
\right) 
 \label{eq3.7x}
\end{equation}
which, after the transformations below, will become the more familiar spin $0$ and spin $1/2$ fields of standard physics. 
For any functional $F$ of these fields we have
\begin{equation}
\left\langle F\right\rangle =\frac{\int \mathcal{D}\,\Psi ^{\dagger }\,
\mathcal{D}\,\Psi \,F\,e^{-S_{E}} }
{\int \mathcal{D}\,\Psi ^{\dagger }\,\mathcal{D}\,\Psi \; e^{-S_{E}}}
\; . \label{eq3.17x}
\end{equation}
Ordinarily we can let $a_{0}\rightarrow 0$ in (\ref{eq3.14x}), so that
\begin{equation}
S_{E}=\int d^{D}x\,\left[ \frac{1}{2m_{0}}\partial _{M}\Psi ^{\dagger }\partial
_{M}\Psi -\mu_{0} \,\Psi ^{\dagger }\Psi +\frac{1}{2}b\left( \Psi ^{\dagger
}\Psi \right) ^{2}\right] \; . \label{eq3.18x}
\end{equation}
However, the higher-derivative term in (\ref{eq3.14x}) is relevant in the
internal space defined below, and a finite $a_{0}$ also automatically provides an ultimate ultraviolet cutoff. 

\section{\label{sec:sec4}Origin of fermion action and (3+1) dimensional spacetime}

The present theory is based on
(1) statistical counting (which ultimately produces the results of (\ref{eq3.14x})-(\ref{eq3.18x}) via the arguments of Appendices \ref{sec:sec2} and \ref{sec:sec3}) and (2) the
geography (or specific features) of our universe, to which we now turn. 

The most central assumption is that 
\begin{eqnarray}
\Psi _{b} = \Psi _{0}^{\prime } +\Psi _{b}^{\prime } 
\label{eq4.2} 
\end{eqnarray}
where $\Psi _{0}^{\prime } $ contains the order parameter $\Psi _{0}$ for a primordial bosonic 
condensate which is present from the beginning of the universe, and 
$\Psi _{b}^{\prime }$ contains all the other bosonic fields. I.e., in $\Psi _{0}^{\prime }$ only one set of components is nonzero and equal to $\Psi _{0}$, and this set of components is zero in $\Psi _{b}^{\prime } $.
The treatment of Appendix \ref{sec:appA} implies that 
\begin{eqnarray}
\Psi _{0} ^{\prime \, \dagger } \Psi _{b}^{\prime } = 0
\end{eqnarray}
everywhere. (The fields in other representations do not overlap the representation containing  $\Psi _{0}$. Fields in the same representation are orthogonal according to (\ref{eq12.6}) and the comments above (\ref{eq12.2}) and
(\ref{eq12.4}).) 

In the simplest description, this primordial condensate forms from a singlet bosonic field in a $16$ representation of SO(10) \textemdash \, or more precisely Spin(10) \textemdash \,  with an extremely large vacuum expectation value because (\ref{eq3.14x}) contains the term $-\mu_0 \Psi^{\dag}\Psi $. As will be seen below, it is distinct from the other such singlet fields (which are partners to the three right-handed neutrino fields of SO(10)) in that it does not belong to a family, and it is not a normal scalar-boson field. Instead it rotates as a spinor field in both 4-dimensional external spacetime and the 10-dimensional internal space (although it is spherically symmetric in the 3-dimensional space). This exceptional behavior of the primordial condensate, which forms near or above the Planck scale, is compatible with the Lorentz invariance of all other fields. Its condensation, to create our universe, breaks the initial 17-dimensional \textit{spatial} symmetry by forming a separate order parameter in each of the 4-, 10-, and 3-dimensional spaces. I.e., $\Psi _{0}$ breaks down to a product of a 2-component external order parameter $\Psi ^0_{ext}$, which is a function of 4 external coordinates $x^{\mu}$, and an internal order parameter $\Psi ^0_{int} \times \Psi ^{\prime \, 0}_{int}$, which is primarily a function of the internal coordinates $x^m$. The initial gauge symmetry \textemdash \, resulting from the rotational symmetry of the internal space, as described in Section \ref{sec:sec5} \textemdash \, is Spin(10). 

Below we will first limit attention to the early stage of our universe before the usual breaking of SO(10)~\cite{Nath,Nanopoulos-1984}, but after the formation of this condensate with $G_{ext} \times G_{int} \times G'_{int}$ symmetry, where $G_{ext} =$ SU(2)$\times$U(1) and the internal groups are $G_{int} =$ Spin(10)$\times$U(1) and $G'_{int} =$ SO(3)$\times$U(1). The full field in (\ref{eq3.7x}) is taken to have the form
\begin{align}
\Psi = \Psi_{ext} \times \Psi_{int} \times \Psi'_{int}
\end{align}
where the field $\Psi_{int}$ in the 10-dimensional internal space contains the full vectorial $10= 5 + \overline{5}$ and spinorial $32=16+\overline{16}$ representations, as well as possibly others,
\begin{align}
\Psi_{int} = \left( 
\begin{array}{c}
\Psi _{10} \\
\Psi _{32} \\
\Psi _{other}
\end{array} 
\right) =  
\left( 
\begin{array}{c}
\Psi _{5} \\
\Psi _{\overline{5}} \\
\Psi _{16 } \\
\Psi _{\overline{16}} \\
\Psi _{other}
\end{array} 
\right) 
\; ,
\label{repx}
\end{align}
and the field $\Psi'_{int}$ in the 3-dimensional internal space contains the scalar ($\ell=0$) and vector ($\ell=1$) representations with 1 and 3 components respectively. The rotational eigenstates in $\Psi_{int}$ will turn out to correspond to gauge eigenstates. The rotational eigenstates in $\Psi'_{int}$ correspond to family number for fermions and for bosons (sfermions and higgsons~\cite{Tallman}) in the $\ell=1$ states, and for bosons and fermions (Higgs bosons and higgsinos) in the $\ell=0$ state, with the primordial condensate belonging to the 16 and $\ell=0$ representations. The correspondence between field/particle content and these 10-dimensional and 3-dimensional representations will be detailed below when it is appropriate.

Like other physical systems \textemdash \, including galaxies, black holes, solar systems, stars, and planets \textemdash \, our universe is assumed to be  born with nonzero angular momentum \textemdash \, but primarily with rotation in the space of fields rather than coordinates. Again, it is born from the condensation of a bosonic field to form an order parameter with SU(2)$\times$Spin(10)$\times$SO(3) rotational symmetry in $(4+10+3)$-dimensional spacetime. The SU(2)$\times$U(1) order parameter in external 4-dimensional spacetime is a noncompactified solution to the full 4th-order equation of motion resulting from (\ref{eq3.14x}), which can absorb angular momentum on long distance scales ($\gg a_0$). External spacetime has 3 spatial dimensions associated with the SU(2) rotations involving the 3 Pauli matrices, and one time dimension associated with the U(1) rotations involving the 2$\times$2 identity matrix. This structure can exist because a vortex-like (or instanton-like) solution can exist in 4 dimensions. As seen below, it produces Lorentz invariance at observable energies, with the correct coupling of matter to the gravitational vierbein (or metric tensor). The Spin(10)$\times$U(1) order parameter in the internal 10-dimensional space is a solution to the 4th-order equation of motion which is compactified to have finite action at each point in external spacetime. It can absorb angular momentum on short distance scales ($\sim a_0$). This structure can exist because a vortex-like (or instanton-like) solution can exist in 10 dimensions. As seen below, it produces a Spin(10) gauge theory \textemdash \, the currently favorite candidate for grand unification of nongravitational forces \textemdash \, with the correct coupling of matter fields to the gauge potentials. Finally, the remaining SO(3)$\times$U(1) order parameter in the internal 3-dimensional space cannot absorb angular momentum in the same way as the other spaces (since it has an $\ell=0$ order parameter), but it is again a solution to the 4th-order equation of motion which is compactified (as a spherical drop of condensate) to have finite action. It accounts for 3 families of standard-model fermions with Yukawa couplings to a single family of Higgs bosons. 

This picture then leads to the prediction of other new fields and particles, some of  which should be observable in the near future.

{\allowdisplaybreaks
The action can now be written as 
\begin{eqnarray}
S_{E} &=&S_{cond}+S_{b}+S_{f}+S_{int}  \label{eq4.6} \\
S_{cond} &=&\int d^{D}x\,\left[ \frac{1}{2m_{0}}\partial _{M}
\Psi _{0}^{\dagger }\partial _{M}\Psi _{0} 
-\mu_{0} \Psi _{0}^{\dagger }\Psi _{0} +\frac{1}{2}
b\left( \Psi _{0}^{\dagger }\Psi _{0}\right) ^{2}\right]  
\label{eq4.7} \\
S_{b} &=&\int d^{D}x\,\left[ \frac{1}{2m_{0}}\partial _{M}\Psi _{b}^{\prime
\,\dagger }\partial _{M}\Psi _{b}^{\prime }\,\,+\left( V_{0}-\mu_{0} \right)
\,\Psi _{b}^{\prime \,\dagger }\Psi _{b}^{\prime }\,+\frac{1}{2}b\left( \Psi
_{b}^{\prime \,\dagger }\Psi _{b}^{\prime }\right) ^{2}\right]  
\label{eq4.8} \\
S_{f} &=&\int d^{D}x\,\left[ \frac{1}{2m_{0}}\partial _{M}\Psi _{f}^{
\dagger }\partial _{M}\Psi _{f}\,\,+\left( V_{0}-\mu_{0} \right)
\,\Psi _{f}^{\dagger }\Psi _{f}\,+\frac{1}{2}b\left( \Psi
_{f}^{\dagger }\Psi _{f}\right) ^{2}\right]  
\label{eq4.9} \\
S_{int} &=&\int d^{D}x\, b 
\left( \Psi _{f}^{\dagger }\Psi _{f} \right) 
\left( \Psi _{b}^{\prime \,\dagger }\Psi _{b}^{\prime } \right)
\label{eq4.9a} \\
V_{0} &=& b \, \Psi _{0}^{\dagger }\Psi _{0} \; . \label{eq4.10}
\end{eqnarray} }
In the remainder of this paper, (\ref{eq4.9a}) and the last terms in 
(\ref{eq4.8}) and (\ref{eq4.9}) will usually be neglected; we are 
then considering the theory prior to formation of further
condensates beyond the primordial $\Psi _{0}$. 

For a static condensate we could write 
$\Psi _{0}=n_{0}^{1/2}\eta _{0}$, 
where $\eta _{0}$ is constant, $\eta _{0}^{\dagger }\eta _{0}=1 $, 
and $n_{0}=\Psi _{0}^{\dagger }\Psi _{0}$ is the condensate
density. This picture is too simplistic, however, since the order parameter
can exhibit rotations that are analogous to the rotations in 
the complex plane of the order parameter $\psi _{s}=e^{i\theta
_{s}}n_{s}^{1/2}$ for an ordinary superfluid: 
\begin{equation}
\Psi _{0}\left( x\right) =U_{0}\left( x\right) \,n_{0}\left( x\right)
^{1/2}\eta _{0}\quad ,\quad U_{0}^{\dagger }U_{0}=1 \; .  \label{eq4.11a}
\end{equation}
After an integration by parts in (\ref{eq4.7}) (with boundary terms usually
neglected in the present paper), extremalization of the action gives 
the classical equation of motion for the order parameter:
\begin{equation}
-\frac{1}{2m_{0}}\partial _{M}\partial _{M}\Psi _{0}+\left( V_{0}-\mu_{0} 
\right) \Psi _{0}=0  \; .\label{eq4.11b}
\end{equation}

Now an important nuance, which requires some references to the gauge potentials $A_{\mu}^i$, vierbein $e_{\alpha}^{\mu}$, metric tensor $g_{\mu \nu}$, and covariant derivative $\nabla _{\mu }$ defined below: Because the primordial condensate density is extremely large, (\ref{eq4.11b}) is assumed to always hold, except exactly at the points where there are topological singularities (just as for the order parameter in an ordinary superfluid). More precisely, at normal energies this constraint is imposed in the configurations that are included in the path integral over $\Psi _0$ (with the higher-derivative terms of (\ref{eq3.14x}) restored when appropriate).
However, consistent with this constraint, its ``phase'' and ``superfluid velocities'' are allowed to vary within the path integral. 
This means that $A_{\mu}^i$ and $g_{\mu \nu}$ are quantized. I.e., the path integral over the original field $\Psi_0$ is replaced by path integrals over $A_{\mu}^i$ and $e_{\alpha}^{\mu}$ (with the usual gauge fixing). In the present treatment, we additionally assume, as usual, the minimal case of a torsion-free universe, so that
\begin{equation}
\nabla _{\mu }e^{\alpha}_{\nu} = 0 
\label{torsion-free}
\end{equation} 
always holds (with $e_{\mu} ^{\alpha }e^{\mu} _{\beta } = \delta_{\beta}^{\alpha}$). I.e., at normal energies the path integral over $e_{\alpha}^{\mu}$ (or $g_{\mu \nu}$) is restricted to torsion-free spacetime geometries.

Again, as mentioned above, in specifying the geography of our universe, it will be assumed that
$\Psi _{0}$ can be written as the product of a $2$-component external 
order parameter $\Psi ^0_{ext}$, which is a function of $4$ external coordinates 
$x^{\mu }$, and an internal order parameter $\overline{\Psi} ^{\,0}_{int}=\Psi ^0_{int} \times \Psi ^{\prime \,0}_{int}$, which is 
primarily a function of the internal coordinates $x^{m}$, but which also 
varies with $x^{\mu }$:
\begin{eqnarray}
\Psi _{0} &=&\Psi ^0_{ext}\left( x^{\mu }\right) \times \overline{\Psi} ^{\,0}_{int}\left(
x^{m},x^{\mu }\right)  \label{eq4.12}  \\
\Psi ^0_{ext}\left( x^{\mu }\right) &=&U_{ext}\left( x^{\mu }\right)
\,n_{ext}\left( x^{\mu }\right) ^{1/2}\eta _{ext}\quad ,\quad \mu =0,1,2,3 
\label{eq4.13}  \\
\overline{\Psi} ^{\,0}_{int}\left( x^{m},x^{\mu }\right) &=&\overline{U}_{int}\left( x^{m},x^{\mu
    }\right) \,\overline{n}_{int}\left( x^{m},x^{\mu }\right) ^{1/2}\overline{\eta} _{int}\quad ,\quad
m=4,...,D-1  \label{eq4.14}
\end{eqnarray}
where again $\eta _{ext}$ and $\overline{\eta} _{int}$ are constant, and 
$\eta _{ext}^{\dagger }\eta _{ext}=\overline{\eta} _{int}^{\dagger }\overline{\eta} _{int}=1 $.
Here, according
to a standard notation, $x^{\mu }$ actually represents the set of $x^{\mu }$,
and $x^{m}$ the set of $x^{m}$.

Let us define external and internal ``superfluid velocities'' by 
\begin{eqnarray}
m_{0}v_{\mu } = -iU_{ext}^{-1} \partial  _{\mu } U_{ext} \quad , \quad 
m_{0}\overline{v}_{m}    = -i\overline{U}_{int}^{-1}\partial _{m}   \overline{U}_{int} \;  . \label{eq4.17}
\end{eqnarray}
The fact that $U_{ext}^{\dagger } U_{ext} = 1$ implies that 
$\left( \partial _{\mu } U_{ext}^{\dagger } \right) U_{ext}
= - U_{ext}^{\dagger } \left( \partial_{\mu }U_{ext} \right) $ with 
$U_{ext}^{\dagger }=U_{ext}^{-1} $, or $m_{0}v_{\mu } = 
i \left( \partial _{\mu }U_{ext}^{\dagger } \right) U_{ext} $, 
so that 
\begin{equation}
v_{\mu }^{\dagger }=v_{\mu } \; . \label{eq4.19}
\end{equation}

For simplicity, let us first consider the case 
\begin{equation}
\partial _{\mu }\overline{U}_{int}=0  \label{eq4.20}
\end{equation}
for which there are separate external and internal equations of
motion: 
\begin{equation}
\left( -\frac{1}{2m_{0}}\partial_{\mu }\partial_{\mu }-\mu _{ext}\right) \Psi^0_{ext}=0  \quad , \quad
    \left( -\frac{1}{2m_{0}}\partial _{m}\partial _{m}-\mu _{int}+V_{0}\right) \overline{\Psi}^0_{int}=0  \label{eq4.21}
\end{equation}
with 
\begin{equation}
\mu _{int}=\mu_{0} -\mu _{ext} \; . 
\label{eq4.21a}
\end{equation}
The quantities $\mu _{int}$ and $V_{0}$ have a 
relatively slow parametric dependence on $x^{\mu }$.

When (\ref{eq4.13}) and (\ref{eq4.17}) are used in 
(\ref{eq4.21}), we obtain 
\begin{equation}
\eta _{ext}^{\dagger }\,n_{ext}^{1/2}\left[ \left( \frac{1}{2}
m_{0}v_{\mu }v_{\mu }-\frac{1}{2m_{0}}\partial _{\mu }\partial _{\mu }-\mu
_{ext}\right) -i\left( \frac{1}{2}\partial _{\mu }v_{\mu }+v_{\mu }\partial
_{\mu }\right) \right] n_{ext}^{1/2}\eta _{ext}=0  \label{eq4.22}
\end{equation}
and its Hermitian conjugate 
\begin{equation}
\eta _{ext}^{\dagger }\,n_{ext}^{1/2}\left[ \left( \frac{1}{2}
m_{0}v_{\mu }v_{\mu }-\frac{1}{2m_{0}}\partial _{\mu }\partial _{\mu }-\mu
_{ext}\right) +i\left( \frac{1}{2}\partial _{\mu }v_{\mu }+v_{\mu }\partial
_{\mu }\right) \right] n_{ext}^{1/2}\eta _{ext}=0 \; . \label{eq4.23}
\end{equation}
Subtraction gives the equation of continuity 
\begin{equation}
\partial_{\mu }\,j_{\mu }^{ext}=0~\quad ,\quad j_{\mu }^{ext}=n_{ext}\,\eta
_{ext}^{\dagger }\,v_{\mu }\eta _{ext}  \label{eq4.24}
\end{equation}
and addition gives the Bernoulli equation 
\begin{equation}
\frac{1}{2}m_{0}\bar{v}_{ext}^{2}+P_{ext}=\mu _{ext}  \label{eq4.25}
\end{equation}
where 
\begin{eqnarray}
\bar{v}_{ext}^{2}=\eta _{ext}^{\dagger }\,v_{\mu }v_{\mu }\,\eta _{ext}
\quad , \quad 
P_{ext}=-\frac{1}{2m_{0}}n_{ext}^{-1/2}\partial _{\mu }\partial _{\mu
}n_{ext}^{1/2}\; . \label{eq4.27}
\end{eqnarray}

Since the order parameter $\Psi ^0_{ext}$ in external spacetime has $2$
components, its (Hermitian) ``superfluid velocity'' 
$v_{\mu }$ can be written in terms of the identity matrix $\sigma ^{0}$ and
Pauli matrices $\sigma ^{a}$ : 
\begin{equation}
v_{\mu }=v_{\alpha \mu }\sigma ^{\alpha } \quad , \quad \alpha = 0,1,2,3 \; . \label{eq4.28}
\end{equation}
Let us now transform to a coordinate system in which 
\begin{equation}
v_{0 k}=v_{0}^{k}=v_{a 0}=v_{a}^{0}=0  \quad , \quad k = 1,2,3 \quad \mathrm{and} \quad a=1,2,3  
\label{eq4.29}
\end{equation}
(with the volume element held constant) so that (\ref{eq4.25}) becomes
\begin{eqnarray}
{\frac{1}{2}}m_{0} v_{\alpha }^{\mu  } v_{\alpha \mu }+P_{ext}=\mu _{ext}
\; .
\label{eq4.30}
\end{eqnarray}

\textit{To avoid notational complexity we will still use $x^{\mu}$ to label the new coordinates. At this point a physically meaningful metric tensor has not yet been introduced, and the notation in (\ref{eq4.30}) merely indicates that  $v_{\alpha }^{\mu  } v_{\alpha \mu }$ is to be kept constant under coordinate transformations. Later in the development, where Lorentz invariance holds (at energies far below $m_0$), the usual conventions are used for raising and lowering indices.}

The transformation to (\ref{eq4.29})  is trivial in, e.g., a (simplistic) cosmological model in which the big bang is at the origin of the new coordinates, with the U(1) phase of $\Psi _{0} $ varying only with respect to the radial coordinate $x^{0}$, and the ``SU(2) phase'' involving the Pauli matrices varying within successive $3$-spheres with coordinates $x^{k}$, confirming that $v_{a k}$ has a vortex-like (or instanton-like) configuration. 

More generally, the time coordinate $x^{0}$ is distinguished from the spatial coordinates $x^{k}$ in (\ref{eq4.29}) because it is the direction of U(1) rather than SU(2) rotations of the order parameter.

As $v_{\alpha }^{\mu  } v_{\alpha \mu } $ varies, $\mu _{ext}$ varies in 
response, with $\mu _{int} $ determined by (\ref{eq4.21a}).

Now expand $\Psi _{f}$ in terms of a complete set of basis
functions $\widetilde{\psi }_{int}^{r}$ in the internal space:
\begin{equation}
\Psi _{f}\left( x^{\mu },x^{m}\right) =\widetilde{\psi }
_{f}^{r}\left( x^{\mu }\right) \widetilde{\psi }_{int}^{r}
\left( x^{m}\right)  \label{eq4.31}
\end{equation}
with 
\begin{eqnarray}
\left( -\frac{1}{2m_{0}}\partial _{m}\partial _{m}-\mu _{int}+V_{0}\right) 
\widetilde{\psi }_{int}^{r}\left( x^{m}\right) &=&\varepsilon _{r}
\widetilde{\psi }_{int}^{r}\left( x^{m}\right) \label{eq4.32}  \\
\int d^{D-4}x\,\widetilde{\psi }_{int}^{r\dag }\left( x^{m}\right) 
\widetilde{\psi }_{int}^{r^{\prime }}\left( x^{m}\right) &=&\delta
_{rr^{\prime }}  \label{eq4.33}
\end{eqnarray}
and with the usual summation over repeated indices in (\ref{eq4.31}) (and $D-4 = 10 +3$). For reasons that will
become fully apparent below, but which are already suggested by the form of
the order parameter, each $\widetilde{\psi }_{f}^{r}\left( x^{\mu }\right) $
has two components. As usual, only the zero ($\varepsilon_{r}=0$) 
modes will be kept. (To simplify the presentation, the higher-derivative 
terms are not explicitly shown in 
the present section; they will be restored when appropriate.) When 
(\ref{eq4.31})-(\ref{eq4.33}) are then used in (\ref{eq4.8}) (with the
last term neglected), the result is 
\begin{equation}
S_{f}=\int d^{4}x\,\widetilde{\psi }_{f}^{\dagger }
\left( -\frac{1}{2m_{0}}\partial ^{\mu }\partial _{\mu }-\mu _{ext}\right) 
\widetilde{\psi }_{f}  \label{eq4.34}
\end{equation}
where $\widetilde{\psi }_{f}$ is the vector with components $\widetilde{\psi 
}_{f}^{r}$.

Let $\widetilde{\psi }_{f}$ be written in the form 
\begin{equation}
\widetilde{\psi }_{f}\left( x^{\mu }\right) =U_{ext}\left( x^{\mu }\right)
\psi _{f}\left( x^{\mu }\right)  \label{eq4.35}
\end{equation}
or equivalently
\begin{eqnarray}
\widetilde{\psi }_{f}^{r}\left( x^{\mu }\right) =U_{ext}\left( x^{\mu
}\right) \psi _{f}^{r}\left( x^{\mu }\right) \; . \label{eq4.36}
\end{eqnarray}
Here $\psi _{f}$ has a simple interpretation: It is the field seen by an
observer in the frame of reference that is moving with the condensate. In
the present theory, a (very high density) condensate $\Psi _{0}$ is present from the beginning of
the universe, and the other bosonic and fermionic fields 
are subsequently born into it. It is therefore natural to define the
fields $\psi _{f}^{r}$ in the condensate's frame. 

Equation (\ref{eq4.35}) is, in fact, exactly analogous to 
rewriting the wavefunction of 
a particle in an ordinary superfluid moving with velocity $v_{s}$: 
$\widetilde{\psi }_{par}\left( x\right) = \exp \left( imv_{s}x\right) \psi
_{par}\left( x\right) $. Here $\psi _{par}$ is the wavefunction in the 
superfluid's frame of reference. 

The frame of the rotating order parameter may be interpreted as the frame of an observer composed of the physical fields defined below (with (\ref{eq4.35}) or(\ref{eq4.36}) very roughly analogous to shifting to the frame of reference of a human observer as the Solar System revolves around the galaxy, or the Earth revolves within the Solar System, or a point on the Earth follows its daily rotation, although these are coordinate rather than field rotations).

When (\ref{eq4.35}) is substituted into (\ref{eq4.34}), the result is 
\begin{equation}
S_{f}=\int d^{4}x~\psi _{f}^{\dagger }\left[
\left( \frac{1}{2}m_{0}v^{\mu }v_{\mu }-\frac{1}{2m_{0}}\partial ^{\mu }\partial _{\mu }-\mu _{ext}\right) -i\left( \frac{1}{2}\partial _{\mu }v^{\mu}+v^{\mu }\partial _{\mu }\right) \right] \psi _{f}\; . \label{eq4.37}
\end{equation}
We assume that, in ordinary circumstances, the extremely dense primordial condensate can be treated as an ``incompressible fluid'' with the condensate density $n_{ext}$ constant, so that $P_{ext}=0$ (according to (\ref{eq4.27})), and with
\begin{equation}
\partial _{\mu }v^{\mu} = 0 
\label{incompressible}
\end{equation}
as is consistent with (\ref{eq4.24}). (Two subtleties: (i)~These statements refer to the quantities as experienced by the other fields, and used in their independent path integrals, which are averages over the ``rapidly fluctuating'' topologies of the external order parameter mentioned above and discussed later in the paper. (ii)~(\ref{incompressible}) can be regarded as an initial Lorenz gauge condition on the gravitational vierbein defined below, consistent with standard general relativity, which can later be relaxed in the final results.) Then (\ref{eq4.29}), (\ref{eq4.30}), and (\ref{incompressible}) lead to the simplification 
\begin{equation}
S_{f}=-\int d^{4}x~\psi _{f}^{\dagger }\left( 
\frac{1}{2m_{0}}\partial ^{\mu }\partial _{\mu }+iv_{\alpha }^{\mu  }\sigma
^{\alpha }\partial _{\mu }\right) \psi _{f} \; . \label{eq4.38}
\end{equation}
In most of the remainder of the paper it will be assumed that the first term in parentheses is negligible compared to the second for 
states $\psi $ with energies $\sim $ 1 TeV or less (as would be the case if we had, e.g., $m_{0}=a_{0}^{-1} \gtrsim 10^{15}$
TeV and $v_{\alpha }^{\mu }\sim 1$ for $\mu=\alpha$), so that (\ref{eq4.38}) reduces to just
\begin{eqnarray}
S_{f} &=& \int d^{4}x \, \psi _{f}^{\dagger }ie_{\alpha}^{\mu }\overline{\sigma} ^{\alpha }\partial _{\mu }\psi _{f}  \label{eq4.39} \\
e_{\alpha }^{0 } &=& - v_{\alpha }^{0 } \quad , \quad e_{\alpha }^{k } = v_{\alpha }^{k } \label{eq4.40x} \quad , \quad 
e_{\mu} ^{\alpha }e^{\mu} _{\beta } = \delta_{\beta}^{\alpha} \label{eq4.39z} \\
\overline{\sigma }^{\, 0} &=& \sigma ^{0} \quad , \quad  \overline{\sigma }^{k}=-\sigma^{k} \; . 
\label{eq4.40}
\end{eqnarray}
With this choice all fields are initially left-handed, as they are for fermions in conventional 
SU(5) and SO(10) grand-unified theories~\cite{cheng,Nanopoulos-1984}. 

To permit local Lorentz transformations as well as general coordinate transformations, let us rewrite (\ref{eq4.39}) as
\begin{eqnarray}
S_{f} = \int d^{4}x \,  \overline{\mathcal {L}}_f \quad , \quad  \overline{\mathcal {L}}_f  = \psi _{f}^{\dagger }ie_{\alpha}^{\mu }\overline{\sigma} ^{\alpha }\nabla _{\mu }^{L}\psi _{f}  
\label{invariant} 
\end{eqnarray}
where $\nabla _{\mu }^{L}$ gives the standard (curved spacetime) covariant derivative 
for a left-handed Weyl field. The full covariant derivative for a Dirac field, with left-handed and right-handed  Weyl components, is given by~\cite{Parker,GSW}
\begin{eqnarray}
\nabla_{\mu }^D&=&\partial _{\mu }+ i \omega _{\mu }^{\alpha \beta } \Sigma_{\alpha \beta } \; .
\end{eqnarray}
(The notation and conventions in this context usually follow those most common in the gravitational and string theory 
communities~\cite{GSW,misner,weinberg-grav,wald}, rather than the particle physics and field theory communities, as in \cite{Parker}.  Mainly, the metric tensor convention is $( -+++)$ throughout this paper. However, the Dirac gamma matrices are defined as in most field theory textbooks~\cite{peskin,schwartz}.)
The origin and meaning of curvature involving the $ \omega _{\mu }^{\alpha \beta } $
will be considered below. 
For a vector field, the usual covariant derivative provides invariance under a coordinate transformation. The vierbein introduced above has both coordinate and tangent-space indices, so~\cite{GSW,Parker}
\begin{equation}
\nabla _{\mu }e^{\alpha}_{\nu} = \partial_{\mu} e^{\alpha}_{\nu}  - \Gamma^{\rho}_{\mu \nu} e^{\alpha}_{\rho}  +\omega_{\mu \; \;  \beta} ^{\; \, \alpha} e^{\beta}_{\nu} \; .
\end{equation}

\textit{In the following we will interpret the generators in $\nabla _{\mu }$ as operators when acting on a specific member of a representation (i.e., each member of a multicomponent field).}

The above arguments also hold for bosons with (in the initial notation)
\begin{equation}
S_{b}=\int d^{D}x\,\left( -\frac{1}{2m_{0}}\Psi _{b}^{\prime \, \dagger }\partial
^{M}\partial _{M}\Psi' _{b} 
-\mu_{0} \Psi _{b}^{\prime \, \dagger }\Psi' _{b}+V_{0}\Psi
_{b}^{\prime \, \dagger }\Psi' _{b} \right)  \label{eq4.41}
\end{equation}
\begin{equation}
\Psi'_{b}\left( x^{\mu },x^{m}\right) =
\widetilde{\psi }_{b}^{r}\left(
x^{\mu }\right) \widetilde{\psi }_{int}^{r}\left( x^{m}\right)
\label{eq4.42}
\end{equation}
leading to the result
\begin{eqnarray}
S_{b} = \int d^{4}x \,  \overline{\mathcal {L}}_b \quad , \quad  \overline{\mathcal {L}}_b  = \psi _{b}^{\dagger }ie_{\alpha}^{\mu }\overline{\sigma} ^{\alpha }\nabla _{\mu }\psi _{b} 
\label{eq4.43} \; .
\end{eqnarray}

A central feature of the present theory is that the \textit{primitive} bosonic fields $\psi _{b}$ are 2-component spinors but the final \textit{physical} fields derived from these are the 1-component scalar-boson fields defined below.

Eqs. (\ref{invariant}) and (\ref{eq4.43}) are still restricted to the coordinate system of (\ref{eq4.29}), with a volume element equal to that of the original coordinate system. Each of these actions has the form
\begin{eqnarray}
S_{\chi} = \int d^{4}x \,  \overline{\mathcal {L}}_{\chi} \quad , \quad  \overline{\mathcal {L}}_{\chi}  = \chi^{\dagger } A_{\chi} \chi
\label{orig} 
\end{eqnarray}
and this form will still hold after mass and interaction terms are added. To permit transformations to a general coordinate system, let 
\begin{eqnarray}
\quad \overline{\chi} &=& e^{-1/2}\chi \label{redefine} \\
e &=& \left\vert \det \, e_{\mu }^{\alpha }\right\vert =\left( -\det \, g_{\mu \nu} \right) ^{1/2} 
\label{eq8.4b}
\end{eqnarray}
where the metric tensor $g_{\mu \nu}$ is defined as usual by 
\begin{eqnarray}
g^{\mu \nu }=\eta ^{\alpha \beta }e_{\alpha }^{\mu }e_{\beta }^{\nu } \quad , \quad \eta ^{\alpha \beta } = diag \left(-1, 1, 1, 1 \right)  \quad , \quad
g_{\mu \nu }g^{\nu \rho } = \delta_{\mu}^{\rho}   
\; .
\label{eq8.3z} 
\end{eqnarray}
(The (-1,1,1,1) signature of $\eta ^{\alpha \beta }$ is required for consistency with the form of the fermion action (\ref{eq4.39}), regardless of the convention chosen for the vierbein in (\ref{eq4.40x}).)

It is important to note that (for energies far below $m_0$) the observable geometry of external spacetime is entirely determined by the vierbein $e^{\mu}_{\alpha}$ or metric tensor $g_{\mu \nu}$ (with the geometry of the original coordinate system being unobservable and relevant only to the global topology).

Each operator $A_{\chi}$ considered here or in the following contains a term with one or two factors of $\nabla _{\mu }$. But even this term commutes with $e^{-1/2}$, because (\ref{torsion-free}) implies that the covariant derivative of a function of only the vierbein (or metric tensor) is zero:
\begin{align}
\nabla _{\mu } \left( e^{-1/2} \chi \right) = [ \nabla _{\mu } e^{-1/2} ] \chi +  e^{-1/2} \nabla _{\mu } \chi = e^{-1/2} \nabla _{\mu } \chi  \; .
\label{var44}
\end{align}
Then (\ref{orig}) can be replaced by
\begin{eqnarray}
S_{\overline{\chi}} = \int d^{4}x \,  e \, \overline{\mathcal {L}}_{\overline{\chi}} \quad , \quad  \overline{\mathcal {L}}_{\overline{\chi}}  = \overline{\chi}^{\dagger } A_{\chi} \overline{\chi} \; .
\label{changed} 
\end{eqnarray}
According to (\ref{orig}), the contribution of $\chi$ to the gravitational energy-momentum tensor in the present theory is 
\begin{align}
T^{\mu \nu }_{\chi} =  2e^{-1}  \frac{\delta \overline{\mathcal{L}}_{\chi}} {\delta g_{\mu \nu }} = 2 \, e^{-1} \chi^{\dag} \frac{\delta A_{\chi} } {\delta g_{\mu \nu} } \chi 
= 2 \, \overline{\chi}^{\dag} \, \frac{\delta A_{\chi} } {\delta g_{\mu \nu} }  \, \overline{\chi} 
\label{var11}
\end{align}
since each term in $\delta A_{\chi} / \delta g_{\mu \nu}$, involving functions and covariant derivatives, also commutes with $e^{-1/2}$.

In (\ref{var11}) and the remainder of this paper, a covariant derivative $\nabla _{\mu }$ is treated as an operator which acts before $A_{\chi} $ is varied.

\textbf{For simplicity of notation $\overline{\chi}$ is renamed $\chi $ in the following.} Then (\ref{orig}) is changed to
\begin{eqnarray}
S_{\chi} = \int d^{4}x \,  e \, \overline{\mathcal {L}}_{\chi} \quad , \quad  \overline{\mathcal {L}}_{\chi}  = \chi^{\dagger } A_{\chi} \chi 
\label{change} 
\end{eqnarray}
and in particular
\begin{eqnarray}
S_{f} &=& \int d^{4}x \,  e \, \overline{\mathcal {L}}_f \quad , \quad  \overline{\mathcal {L}}_f  = \psi _{f}^{\dagger }ie_{\alpha}^{\mu }\overline{\sigma} ^{\alpha }\nabla _{\mu }\psi _{f} \label{fermion} \\
S_{b} &=& \int d^{4}x \,  e \, \overline{\mathcal {L}}_b \quad , \quad  \overline{\mathcal {L}}_b  = \psi _{b}^{\dagger }ie_{\alpha}^{\mu }\overline{\sigma} ^{\alpha }\nabla _{\mu }\psi _{b}  
\label{boson}
\end{eqnarray}
\textbf{where $\psi _{b}$ and $\psi _{f}$ have each been redefined to include a factor of $e^{-1/2}$ as in} (\ref{redefine}).

The present theory thus yields the basic form of the standard Lagrangian (\ref{fermion}) for Weyl fermions. The path integral still has a Euclidean form and 
the action for bosons is also not yet in standard form, but we will return to these points below.

Notice that, for the fields of (\ref{fermion}) and (\ref{boson}), the stress-energy tensor must now be calculated from 
 \begin{align}
T^{\mu \nu }_{\chi} &=  2  \, \chi^{\dag} \, \frac{\delta A_{\chi} } {\delta g_{\mu \nu} }  \, \chi
\label{var-final1}
\end{align} 
rather than the conventional
\begin{align}
T^{\mu \nu }_{\chi} &= 2 \, e^{-1} \frac{\delta \left( e \overline{\mathcal{L}}_{\chi} \right)} {\delta g_{\mu \nu }} \; .
\end{align}
It will be seen below that the two formulas give the same result for particles satisfying their classical equations of motion, but not in general, and not for the zero-point (vacuum) action of fermions and scalar bosons (if used in the conventional way). For fermions or scalar bosons with a bilinear action (\ref{var-final1}) is a natural definition of the stress-energy, or energy-momentum, tensor in a quantum description.

In the present picture spacetime is consistently 4-dimensional with one time coordinate because there are 3 Pauli matrices and one $2 \times 2$ unit matrix, and because an instanton-like topological defect can be defined in 4 dimensions with a preferred radial coordinate. An instanton-like defect can also be naturally defined in the 10-dimensional internal space. (The term ``instanton-like'' is used here in a very broad sense, to indicate that the ``superfluid velocity'' of the condensate in 2n dimensions is somewhat similar to the gauge potential $A_{\mu}$ of an instanton, with both represented in terms of the generators for rotations.) In the 3-dimensional internal space the order parameter is isotropic, in an $\ell=0$ state.

\section{\label{sec:sec5}Origin of gauge fields}

\textit{In most of this section and the following two sections, we will temporarily ignore the spin connection and write $\partial _{\mu }$ instead of $\nabla _{\mu } $, to avoid irrelevant complications in notation. We will also consider only $\Psi_{int}$, the field in the 10-dimensional internal space, so that
$\sum_m$ means $\sum_{m=1}^{10}$ and $\widetilde{\psi }_{int}^{r}$ is interpreted as a basis function in this space.}

For $\Psi_{int}$ we now relax assumption (\ref{eq4.20}) and allow $U_{int}$ to vary 
with the external coordinates $x^{\mu }$.
The more general version of (\ref{eq4.11b}) is satisfied if (\ref{eq4.21}) is generalized to
\begin{equation}
\left( -\frac{1}{2m_{0}}\partial ^{\mu }\partial _{\mu }-\mu _{ext}\right) \Psi
_{ext}\left( x^{\mu }\right) \Psi ^0_{int}\left( x^{m},x^{\mu }\right) =0
\label{eq5.2}
\end{equation}
with $\Psi ^0_{int}$ required to satisfy the internal equation of motion 
(at each $x^{\mu }$)
\begin{equation}
\left[ \sum_{m} \frac{1}{2m_{0}}\left( -\frac{\partial ^{2}}{\partial 
\left( x^{m}\right) ^{2}}+
\frac{a_{0}^{2}}{16}\frac{\partial ^{4}}{\partial \left( x^{m}\right) ^{4}}
\right) 
+V_{0}\left( x^{m}\right) -\mu _{int}\right] \Psi ^0_{int}\left( x^{m},x^{\mu
}\right) =0 \; .  \label{eq5.3}
\end{equation}
The higher-derivative 
term of (\ref{eq3.14}) has been retained and two integrations by 
parts have been performed. (In order to simplify the notation, we do not
explicitly show the weak parametric dependence of $\mu _{int}$, $V_{0}$, and 
$n_{int}$ on $x^{\mu }$.) This is a nonlinear equation because (at each 
$x^{\mu }$) $V_{0}\left( x^{m}\right) $ 
is mainly determined by 
$n_{int} = \Psi _{int}^{0 \, \dag } \Psi ^0_{int} $.

The internal basis functions satisfy the more general version of 
(\ref{eq4.32}) with $\varepsilon _{r}=0$: 
\begin{equation}
\left[ \sum_{m} \frac{1}{2m_{0}} \left( -\frac{\partial ^{2}}{\partial 
\left( x^{m}\right) ^{2}}+
\frac{a_{0}^{2}}{16}\frac{\partial ^{4}}{\partial \left( x^{m}\right) ^{4}}
\right) 
+V_{0}\left( x^{m}\right) -\mu _{int}\right] \widetilde{\psi }_{int}^{r}\left( x^{m},x^{\mu }\right) =0 \; .  \label{eq5.4}
\end{equation}
This is a linear equation because $V_{0}\left( x^{m}\right) $ is now
regarded as a known function. 

The behavior of the condensate and basis functions in this internal
space is discussed in Appendices \ref{sec:appA} and \ref{sec:appB}.
In (\ref{eq12.9}), the parameters $\overline{\phi } _{i}$ specify a rotation  
of the internal order parameter $\Psi ^0_{int}\left( x^{m},x^{\mu }\right)$ at fixed $x^{m}$ as the external coordinates 
$x^{\mu}$ are varied, and according to (\ref{eq12.9a}) the 
$\overline{J}_{i}$ satisfy the Spin(10) algebra
\begin{eqnarray}
\overline{J}_{i}\overline{J}_{j}-\overline{J}_{j}\overline{J}_{i}
=ic_{ij}^{k}\overline{J}_{k} \; .
\label{eq5.9}
\end{eqnarray}
For simplicity of notation, let 
\begin{equation}
\left\langle r\,|Q|r^{\prime }\right\rangle =\int d^{\overline{d}}x\,\widetilde{\psi }
_{int}^{r\dagger }Q\,\widetilde{\psi }_{int}^{r^{\prime}}\quad \mbox{with}
\quad \left\langle r\,|r^{\prime }\right\rangle =\delta _{rr^{\prime }}
\label{eq5.16}
\end{equation}
for any operator $Q$ (where here $\overline{d}=10$), and in particular let 
\begin{equation}
t_{i}^{rr^{\prime }}=\left\langle r\,|\,\overline{J}_{i}|r^{\prime }\right\rangle
\label{eq5.17}
\end{equation}
with the matrices $t_{i}^{rr^{\prime }}$ (which are constant according
to (\ref{eq12.9b})) inheriting the Spin(10) algebra: 
\begin{eqnarray}
\left( t_{i}t_{j}-t_{j}t_{i} \right) ^{r r^{\prime}}
&=&\sum_{r^{\prime \prime }}\left\langle
r\,|\,\overline{J}_{i}|r^{\prime \prime }\right\rangle 
\left\langle r^{\prime \prime }|\,\overline{J}_{j}|r^{\prime }\right\rangle 
-\sum_{r^{\prime \prime }}\left\langle r\,|\,\overline{J}_{j}|
r^{\prime \prime }\right\rangle \left\langle r^{\prime \prime }|
\,\overline{J}_{i}|r^{\prime }\right\rangle  \label{eq5.18} \\
&=&\left\langle r\,|\,\overline{J}_{i}\,\overline{J}_{j}|\,
r^{\prime} \,\right\rangle -\left\langle r\,|\,
\overline{J}_{j}\,\overline{J}_{i}|\,r^{\prime} \,\right\rangle  
\label{eq5.19}  \\
&=&ic_{ij}^{k}t_{k} ^{r r^{\prime}} \; . \label{eq5.20}
\end{eqnarray}
The $t_{i}$ are the generators in the $N_{g}$-dimensional reducible
representation determined by the physically significant solutions to 
(\ref{eq5.4}), which spans all the irreducible (physical) gauge 
representations, as in (\ref{repx}).

When $x^{\mu }\rightarrow x^{\mu }+\delta x^{\mu }$, 
$\Psi ^0_{int} $ and $\widetilde{\psi} _{int}^{r} $
rotate together, and (\ref{eq12.9}) implies that
\begin{eqnarray}
\partial _{\mu }\widetilde{\psi} _{int}^{r}\left( x^{m},x^{\mu }\right) &=&
\frac{\partial \overline{\phi } _{i}}{\partial x^{\mu }} \frac{\partial }
{\partial \overline{\phi } _{i}} \widetilde{\psi} _{int}^{r} 
\left( x^{m},x^{\mu }\right) 
\label{eq5.21}  \\
&=&-i\,A_{\mu }^{i}\overline{J}_{i}\, \widetilde{\psi} _{int}^{r} 
\left( x^{m},x^{\mu }\right) 
\label{eq5.22}
\end{eqnarray}
where
\begin{eqnarray}
A_{\mu }^{i}= \frac{\partial \overline{\phi } _{i}}{\partial x^{\mu }} \; . 
\label{eq5.23}
\end{eqnarray}
The $A_{\mu }^{i}$ will be interpreted below as gauge potentials. In other
words, the gauge potentials are simply the rates at which the internal 
order parameter $\Psi ^0_{int}\left( x^{m},x^{\mu }\right) $ (at fixed $x^{m}$) is rotating as a 
function of the external coordinates $x^{\mu }$. 

Let us return to the fermionic action. 
If (\ref{eq4.31}) is written in the more general form
\begin{eqnarray}
\hspace{-0.5cm}
\Psi _{f}\left( x^{\mu },x^{m}\right) 
=\widetilde{\psi }_{f}^{r}\left(
x^{\mu }\right) \widetilde{\psi }_{int}^{r}\left( x^{m}, x^{\mu }\right)
=U_{ext}\left( x^{\mu }\right) \psi _{f}^{r}\left( x^{\mu }\right)
\widetilde{\psi }_{int}^{r} \left( x^{m}, x^{\mu } \right) 
\label{eq5.25}
\end{eqnarray}
we now have, from (\ref{eq4.17}) and (\ref{eq5.22}),
\begin{equation}
\partial _{\mu }\Psi _{f}
=U_{ext}\left( x^{\mu }\right) \left( \partial _{\mu }^{\, \prime}
+im_{0}v_{\alpha \mu }\sigma ^{\alpha }-i\,A_{\mu }^{i}\overline{J}_{i} \,\right) \psi _{f}^{r}\,\widetilde{\psi }_{int}^{r}   \label{eq5.26}
\end{equation}
where the prime indicates that $\partial _{\mu }^{\, \prime}$ 
does not operate on $\widetilde{\psi }_{int}^{r}$. 

Straightforward repetition of the treatment in the preceding section, with inclusion of the additional term involving
\begin{eqnarray}
\int d^{\overline{d}}x\,\widetilde{\psi }_{int}^{r\dagger } \, \left( -i\,A_{\mu }^{i} \overline{J}_{i} \right) \,\widetilde{\psi }_{int}^{r^{\prime}} =
-i\,A_{\mu }^{i} \left\langle r\,|\,\overline{J}_{i}|r^{\prime }\right\rangle = -i\,A_{\mu }^{i} t_{i}^{rr^{\prime }}
\end{eqnarray}
and with the $\psi _{f}$ and $\psi _{b}$ again redefined to include a factor of $e^{-1/2}$, gives
\begin{align}S_{f} =\int d^{4}x\, e \,  \overline{\mathcal{L}}_f \quad , \quad 
\overline{\mathcal{L}}_f  &= \psi _{f} ^{\dagger }ie_{\alpha }^{\mu }\overline{\sigma}^{\alpha }\widetilde{D}_{\mu }\psi _{f} \label{eq5.31} \\
S_{b}= \int d^{4}x\, e \, \overline{\mathcal{L}}_b \quad , \quad 
\overline{\mathcal{L}}_b &= \psi _{b} ^{\dagger }ie_{\alpha }^{\mu }\overline{\sigma}^{\alpha }\widetilde{D}_{\mu }\psi _{b}  \; . 
\label{eq5.32}
\end{align}
where 
\begin{equation}
\widetilde{D}_{\mu } = \nabla_{\mu} - iA_{\mu }^{i}t_{i} \; . \label{eq5.29}
\end{equation}
This is the generalization of (\ref{fermion}) and (\ref{boson}) 
when the internal order parameter 
is permitted to vary as a function of the external coordinates $x^{\mu }$.

For simplicity, we revert to a locally inertial coordinate system in the following two sections, where $e \rightarrow 1$ and the covariant derivative is just
\begin{equation}
D_{\mu } = \partial_{\mu} - iA_{\mu }^{i}t_{i} \; . \label{eq5.29x}
\end{equation}
\textit{In the following the generators $t_i$ will themselves be treated as operators.}

\section{\label{sec:sec6}Transformation to Lorentzian path integral: fermions}

All of the foregoing is within a Euclidean picture, but we will now show
that, in the case of fermions, there is a relatively trivial transformation
to the more familiar Lorentzian description. A key point is that the
low-energy \textit{operator} $ie_{\alpha }^{\mu }\overline{\sigma} ^{\alpha }D_{\mu }$
in $S_{f}$ is automatically in the correct Lorentzian form, even though the
initial \textit{path integral} is in Euclidean form. It is this fact which
permits the following transformation to a Lorentzian path integral. Within
the present theory, neither the fields nor the operators (nor the meaning of
the time coordinate) need to be modified in performing this transformation.

The operator within $S_{f} $ can be diagonalized to give 
\begin{eqnarray}
S_{f} =\sum\nolimits_{s}\,\,\overline{\psi }_{f}^{\, \ast }\left( s\right)
\,a\left( s\right) \,\overline{\psi }_{f}\left( s\right)  \label{eq6.2}
\end{eqnarray}
where 
\begin{equation}
\psi _{f}\left( x\right) =\sum\limits_{s}U\left( x,s\right) \,\overline{\psi 
}_{f}\left( s\right) \quad ,\quad \overline{\psi }_{f}\left( s\right) =\int
d^{4}x \,U^{\dag }\left( x,s \right) \,\psi _{f}\left( x\right)  
\label{eq6.3}
\end{equation}
with 
\begin{eqnarray}
ie_{\alpha }^{\mu }\overline{\sigma} ^{\alpha }D_{\mu }U\left( x,s\right) &=& 
a\left( s\right) U\left( x,s\right)   \label{eq6.4} \\
\int d^{4}x\, U^{\dag }\left( x,s \right) U\left( x,s^{\prime }\right)
&=&\delta _{ss^{\prime }}\;\quad ,\quad \;\sum\limits_{s}U\left( x,s\right)
U^{\dag }\left( x^{\prime },s \right) =\delta \left( x-x^{\prime }\right)
\; . \label{eq6.5}
\end{eqnarray}
\textit{Here, and in the following, $x$ represents a point in external 
spacetime, and $U\left( x,s\right) $ is a multicomponent 
eigenfunction.} There is an implicit inner product in 
\begin{eqnarray}
U^{\dag }\left( x,s \right) \,\psi _{f}\left( x\right)
&=&\sum\limits_{r}U_{r}^{\dag }\left( x,s \right) \,\psi _{f}^{r}\left( x\right)
 \label{eq6.7} 
\end{eqnarray}
with the $2 N_{g}$ components of $\psi _{f}\left( x\right) $ 
labeled by $r=1,...,N_{g}$ (spanning all components of all irreducible gauge 
representations) and $a=1,2$ (labeling the components of Weyl spinors), 
and with $s$ and $\left( x,r,a \right) $ each having the same number of values. 
Also, the delta function in (\ref{eq6.5}) implicitly multiplies the $2 N_{g} \times 2 N_{g}$ 
identity matrix.

Evaluation of the present Euclidean path integral (a Gaussian integral with
Grassmann variables) is then trivial for fermions; as usual, 
\begin{eqnarray}
Z_{f} &=&\int \mathcal{D}\,\psi _{f}^{\dag }\left( x\right) \,\mathcal{D}
\,\psi _{f}\left( x\right) \,\,\,e^{-S_{f}}   \label{eq6.9}  \\
&=&\prod_{x,ra}\int d\,\psi _{f}^{ra \, \ast }\left( x\right) \int d\,\psi
_{f}^{ra}\left( x\right) \,e^{-S_{f}}   \label{eq6.10}  \\
&=&\prod_{s}z_{f}\left( s\right)  \label{eq6.11}
\end{eqnarray}
with 
\begin{eqnarray}
z_{f}\left( s\right) &=&\int d\,\overline{\psi }_{f}^{\, \ast }\left( s\right)
\,\int d\,\overline{\psi }_{f}\left( s\right) \,e^{-\overline{\psi }
_{f}^{\, \ast }\,\left( s\right) \,a\left( s\right) \,\overline{\psi }
_{f}\left( s\right) }  \label{eq6.12}  \\
&=&a\left( s\right)  \label{eq6.13}
\end{eqnarray}
since the transformation is unitary~\cite{peskin}. Now let 
\begin{eqnarray}
Z_{f}^{L} &=& \int \mathcal{D}\,\overline{\psi }_{f}^{\, \dag }\left( s\right) 
\, \mathcal{D}\,\overline{\psi }_{f}\left( s\right) \,\,e^{iS_{f}} 
\label{eq6.16} \\
&=& \prod_{s}z_{f}^{L}\left( s\right)  \label{eq6.17}
\end{eqnarray}
where 
\begin{eqnarray}
z_{f}^{L}\left( s\right) &=&\int d\,\overline{\psi }_{f}^{\, \ast }\left(
s\right) \int \,d\,\overline{\psi }_{f}\left( s\right) \,e^{i\,\overline
{\psi }_{f}^{\, \ast }\,\left( s\right) \,a\left( s\right) \,\overline{\psi }
_{f}\left( s\right) }  \label{eq6.18}  \\
&=&-ia\left( s\right)  \label{eq6.19}
\end{eqnarray}
so that 
\begin{eqnarray}
Z_{f}^{L}=c_{f}Z_{f}\;\quad ,\quad c_{f}=\prod_{s}\left( -i\right) 
\; . \label{eq6.20}
\end{eqnarray}
This result holds for the path integral over an arbitrary time interval, 
with the fields, operator, and meaning of time left unchanged.

The transition amplitude from an initial state to a final state is equal 
to the path integral between these states, so transition 
probabilities are the same with the Lorentzian and Euclidean forms of the path integral. 
This result is consistent with the fact 
that the classical equations of motion are also the same, since they 
follow from extremalization of the same action. Furthermore, using 
the method on pp. 290-291 or 302-303 of Ref.~\cite{peskin}, it is easy to 
show that the magnitude $\left| G \left( x,x^{\prime} \right) \right| $ of 
the $2$-point function is again the same, so particles propagate the 
same way in both descriptions. This result is also obtained in Appendix 
\ref{sec:appC} with a different method.

It may seem strange that the Lorentzian and Euclidean forms of the path integral yield the same physical results, 
but perusal of the standard arguments in e.g. field theory textbooks shows that the physically significant features 
of the results derive from the Lorentzian form of the \textit{action} rather than the path integral. 

When the inverse transformation from $\overline{\psi }_{f}$ to $\psi _{f}$
is performed, we obtain 
\begin{equation}
Z_{f}^{L}=\int \mathcal{D}\,\psi _{f}^{\dag }\left( x\right) \,\mathcal{D}
\,\psi _{f}\left( x\right) \,e^{iS_{f}}  \label{eq6.21}
\end{equation}
with $S_{f}$ having its form (\ref{eq5.31}) in the coordinate representation.

One may perform calculations in either the path-integral formulation or the
equivalent canonical formulation, which can now be obtained in the 
standard way: Let us use the notation $\int_{a}^{b}$ to indicate that the
fields in a path integral are specified to begin in a state $\left\vert
a\right\rangle $ at time $t_{a}$ and end in state 
$\left\vert b\right\rangle $ at time $t_{b}$, and also to indicate that 
a path integral showing these limits has its conventional definition 
(so that it may differ by a normalization constant from $Z_{f}^{L}$ 
as defined above). Then the Hamiltonian $H_{f}$ is defined by 
\begin{eqnarray}
\left\langle b\right\vert \,U_{f}\left( t_{b},t_{a}\right) \left\vert
a\right\rangle  &=&\int_{a}^{b}\mathcal{D}\,\psi _{f}^{\dag }\left( x\right)
\,\mathcal{D}\,\psi _{f}\left( x\right) \,e^{iS_{f}} \label{eq6.24} \\
i\frac{d}{dt}U_{f}\left( t,t_{a}\right)  &=&H_{f}\left( t\right)
U_{f}\left( t,t_{a}\right) \quad ,\quad U_{f}\left( t_{a},t_{a}\right) =1 \; .
\label{eq6.25} 
\end{eqnarray}
I.e., the time evolution operator 
$U_{f}\left( t_{b},t_{a}\right) $ 
is defined to have the same effect as the path 
integral over intermediate states, and it is then straightforward to 
reverse the usual logic which leads from canonical quantization to 
path-integral quantization~\cite{peskin,weinberg}.
The Schr{\"o}dinger equation
\begin{eqnarray}
i \frac{d}{dt} \left\vert \psi_f \left( t \right) \right\rangle  = H_{f} \left( t\right)  \left \vert \psi_f \left( t \right) \right\rangle
\label{sch}  
\end{eqnarray}
follows from (\ref{eq6.25}) and
\begin{eqnarray}
\left\vert \psi_f \left( t \right) \right\rangle = U_{f}\left( t,t_a \right) \left\vert \psi_f \left( t_a \right) \right\rangle \; .
\end{eqnarray}
The Hamiltonian $H_{f}$ can be written in terms of field operators, and then creation and destruction operators, in a standard way (and essentially just as for the harmonic oscillator).

\section{\label{sec:sec7x}Scalar-boson fields}

The primitive (2-component) bosonic fields $\psi _{b}^r$ are clearly unphysical because they violate the spin-statistics connection implied by Lorentz invariance (and would have negative-energy excitations). In the following we will show that these primitive fields can be naturally transformed to physical (1-component) scalar-boson fields $\phi_b^r$ and $F_b^r$ consistent with Lorentz invariance and a stable vacuum, with the number of degrees of freedom preserved.

As will be seen below, a full $32=16+\overline{16}$ spinorial representation is required for an extension of each generation of standard model fermions and their bosonic partners (sfermions), with a $10=5+\overline{5}$ vectorial  representation for each set of Higgs-related particles and their fermionic partners (higgsinos). (With all fields initially left-handed, those of the 16 and $\overline{16}$, or $5$ and $\overline{5}$, are independent.) For simplicity of nomenclature, even the color triplets of a 5 and $\overline{5}$ are called ``Higgs-related''.

The complete transformation from primitive to physical fields requires two sets of steps: In the appendices, the fields are first rearranged and rescaled, while their quantum numbers are left unchanged. The fields are then combined below to achieve scalar-boson fields consistent with Lorentz invariance (still with the same gauge quantum numbers). 

Physically this set of transformations is viewed as automatically occurring when, in the extremely early universe, the bosonic fields in an initially unstable vacuum reform and reorganize themselves -- becoming the physical fields revealed by these transformations -- in order to achieve the stability of the current vacuum. 

The joining of primitive fields to satisfy a required symmetry has various well-known precedents. For example, left-and right-handed Weyl fields must be joined to form a massive Dirac field, in order to achieve Lorentz invariance, and two real fields must be joined to form the real and imaginary parts of an ordinary charged scalar field, to achieve gauge invariance.

The final fields $\phi$ and $\varphi$ are amplitude modes of penultimate 4-component fields $\Phi$ and $\overline{\Phi}$. These final fields describe physical excitations at accessible energies. corresponding to observable particles, and the 4-component fields are underlying fundamental fields in the vacuum. There are are many analogies in condensed matter physics, in which the low-energy excitations have a very different character than the original fundamental fields. The best-known example in high energy physics is the transformation of the fundamental $A^1_{\mu}$, $A^2_{\mu}$, $A^3_{\mu}$, and $B_{\mu}$ fields into the modified fields of a stable vacuum after Higgs condensation (the $W^+_{\mu}$, $W^-_{\mu}$, $Z^0_{\mu}$, and photon fields). In supersymmetry, higgsinos and electroweak gauginos must similarly be combined to from chargino and neutralino mass eigenstates.

The detailed steps begin in Appendix \ref{sec:newapp}, where half the fields of $\psi _{b}$ in each full representation are converted to right-handed fields, with the resulting action given by (\ref{eq702})-(\ref{eq704}). \textit{There, and in all of the following equations, we explicitly show summations over $r$, the label for 2-component or 1-component fields.}

Then a further transformation in Appendix \ref{sec:appD} gives
{\allowdisplaybreaks
\begin{align}
S'_b &= S_{\phi} + S_F \label{final0} \\
S_{\phi} &= \int d^{4}x \left( \phi_{\uparrow} ^{ \, \dag } \left( x \right)  B \, \phi_{\uparrow}  \left( x \right)  + \phi_{\downarrow} ^{\, \dag } \left( x \right)  B \, \phi_{\downarrow}  \left( x \right) \right) \label{final1} \\
&= \int d^{4}x \, \sum_{r}  \left( \phi_{\uparrow} ^{ \, r \, \dag } \left( x \right)  B \, \phi_{\uparrow}^r   \left( x \right)  + \phi_ {\downarrow} ^{r \, \dag } \left( x \right)  B \phi_{\downarrow}^r \left( x \right) \right) \label{final2} \\
S_{F} &= \int d^{4}x \left( F_{\Uparrow} ^{\, \dag } \left( x \right)  \, F_{\Uparrow}  \left( x \right)  + F_{\Downarrow} ^{\, \dag } \left( x \right)  \, F_{\Downarrow}  \left( x \right) \right) \label{final3} \\
&= \int d^{4}x \, \sum_{r}  \left( F_{\Uparrow} ^{\, r \, \dag } \left( x \right)  \, F_{\Uparrow}^{\, r }  \left( x \right)  + F_ {\Downarrow} ^{r \, \dag } \left( x \right)  F_{\Downarrow}^r \left( x \right) \right) \label{final4} \\
B  &= D^{\mu } D_{\mu } \quad , \quad r=1,2, ..., N
\label{final5}
\end{align}}
where $N=  5$ or $16$, $S'_b = S_b - S_{neg}$, $S_{neg}$ is a constant vacuum contribution, and the notation is further defined in Appendices \ref{sec:newapp} and \ref{sec:appD}. \textit{For example, the gauge generators $t_i$ in $D_{\mu}$ are treated as operators.} The fields of (\ref{final1}) and (\ref{final3}) have $2N$ components, and each is composed of the N 2-component fields of (\ref{final2}) or (\ref{final4}).

The path integral has also been transformed from Euclidean to Lorentzian form, and one can ultimately shift from the present path-integral formulation to canonical quantization as described in the preceding section for fermions.

After this set of steps, in the appendices, the fields have been rearranged and scaled. (More precisely, the Fourier modes are rearranged and scaled, in order to form appropriate new fields with quantum numbers left unchanged.) It is clear, however, that  a final step is required to achieve physically acceptable fields, since excitations of $\phi_{\uparrow} ^{\, r}$ etc. would be spin 1/2 bosons, still violating Lorentz invariance. 
We therefore now combine the 2-component $\uparrow$ and $\downarrow$ fields (which have the same gauge quantum numbers) to obtain proper scalar-boson fields: first with 4-component combined fields -- which will be called $\Phi ^{r}$, $\overline{\Phi }^{\, r}$, and $\cal{F}$$^r$ -- and then their 1-component (complex) amplitudes -- called $\phi ^{r}$, $\varphi ^{\, r}$, and $F^r$. Scalar bosons are interpreted as excitations of these amplitude modes, which are analogous to the Higgs/amplitude modes in superconductors~\cite{Varma1,Varma2,Shimano}.

In the following we will explicitly consider only the dynamical fields $\phi$ and $\varphi$, since the auxiliary fields $F$ can be treated in essentially the same way.

Proper dynamical scalar-boson fields can be achieved in either of two ways:

\textit{A standard scalar-boson field} $\phi^r$ can be obtained by first combining the 
fields $\phi_{\uparrow}^{\, r}$ and $\phi_{\downarrow}^{\, r}$ (having the same gauge quantum numbers but opposite spins):
\begin{align}
\Phi ^{r}\left( x \right) &=\left( 
\begin{array}{c}
\phi_{\uparrow}^{\, r} \left( x \right) \\ 
 \phi_{\downarrow}^{\, r} \left( x \right)
\end{array} \right) 
\quad , \quad 
\phi_{\downarrow}^{\, r \, \dag} \left( x \right)  \phi_{\downarrow}^{\, r} \left( x \right)  = \phi_{\uparrow}^{\, r \, \dag} \left( x \right)  \phi_{\uparrow}^{\, r} \left( x \right)  \quad , \quad r=1,2, ..., N \
\end{align}
with
\begin{align}
S_{\phi}  = \int d^{4}x \, \sum_{r} \, \Phi ^{r \, \dag} \left( x \right)  B \; \Phi ^{r}\left( x \right) = \int d^{4}x \, \Phi ^{\dag} \left( x \right)  B \; \Phi \left( x \right)  \; .
\end{align}

We can define amplitude modes $\phi _i^{r}$ by
\begin{align}
\phi _i^{r}\left( x\right) &= \xi _{i}^{r\,\dag } \, \Phi^{r}\left( x\right)     \qquad  
\mathrm{with } \qquad   \xi _{i}^{r\,\dag }\,\xi _{i'}^r = \delta_{i i'} 
\label{e8x}
\end{align}
where $\xi ^r_{i}$ has $4$ constant components.  
Of the 4 orthonormal basis vectors $\xi^r_{i}$, we can choose $\xi_{3}^r$ and $\xi _{4}^r$  to be orthogonal to $\Phi ^{r}$, so that only $\phi _1^{r}\left( x\right)$ and $\phi _2^{r}\left( x\right)$ are nonzero.
For example, with the basis for spin up and down chosen such that, with all 4 components shown explicitly, 
\begin{align}
\Phi ^{r}\left( x \right) &=\left( 
\begin{array}{c}
\left[ \,\phi_{\uparrow}^{\, r} \left( x \right) \right]_1  \\ 
0 \\
0 \\
\left[ \, \phi_{\downarrow}^{\, r} \left( x \right) \right] _2
\end{array} \right) 
\end{align}
we can take the basis vectors to be 
\begin{align}
\xi ^r_{1} = \frac{1}{\sqrt{2}} \left( 
\begin{array}{c}
1 \\ 
0 \\
0 \\
1
\end{array} \right) \quad , \quad
\xi^r_{2} = \frac{1}{\sqrt{2}} \left( 
\begin{array}{c}
1 \\ 
0 \\
0 \\
- 1
\end{array} \right) \quad , \quad
\xi ^r_{3} = \frac{1}{\sqrt{2}} \left( 
\begin{array}{c}
0\\ 
1 \\
1 \\
0
\end{array} \right) \quad , \quad
\xi^r_{ 4} = \frac{1}{\sqrt{2}} \left( 
\begin{array}{c}
0\\ 
1 \\
-1 \\
0
\end{array} \right) \; .
\label{basis}
\end{align}
There are then just the in-phase and out-of-phase amplitude modes:
\begin{align}
\Phi ^{r}\left( x\right) &=\phi _1^{r}\left( x\right) \,\xi ^r_{1} + \phi _2^{r}\left( x\right) \,\xi^r_{2} 
\end{align}
so that
\begin{align}
S_{\phi} &=\int d^{4}x \, \sum_{r} \, \left[ \phi _1 ^{r \, *} \left( x \right)  B \; \phi _1^{r} \left( x \right) 
+ \phi _2^{r \, *} \left( x \right)  B \; \phi _2^{r }\left( x \right)  \right] \\
&=\int d^{4}x \, \sum_{r} \, \left[ \phi _1^{r \, *} \left( x \right)  B \; \phi _1^{r} \left( x \right) 
+ \phi _2^{r \, c \, *} \left( x \right)  B \; \phi _2^{r \, c }\left( x \right)  \right] \\
&=\int d^{4}x \, \left[ \phi _1^{ \dag} \left( x \right)  B \; \phi _1 \left( x \right) 
+ \phi _2^{c\, \dag} \left( x \right)  B \; \phi _2^{ c}\left( x \right)  \right] \label{ampl}
\end{align}
where 
\begin{align}
\phi_i^{\, r \, c}  \left( x \right) = C \phi _i ^{\, r \, *} \left( x \right) \; . 
\end{align}
We have used the fact that
\begin{align}
\int d^{4}x \, \phi _i^{r \, *} \left( x \right)  B \; \phi _i^{r }\left( x \right)  =  \int d^{4}x \, \phi _i^{r \, c \, *} \left( x \right)  B \; \phi _i^{r \, c}\left( x \right) 
\end{align}
follows from a simpler version of the argument in (\ref{conj})-(\ref{eqA20}).
(When the conjugate fields are placed in an array, they are, of course, reordered so that each has its appropriate place in the gauge multiplet.)

For a $5 + \overline{5}$ representation, $\phi_1$ and $\phi _2^{c}$ each consist of 5 one-component scalar-boson fields; and for a $16 + \overline{16}$ representation,  $\phi_1$ and $\phi _2^{c}$  each consist of 16 one-component scalar-boson fields.

From a $5 + \overline{5}$ we obtain the usual two Higgs doublets of supersymmetry. 

For each of the three $16 + \overline{16}$ families we can obtain 16 sfermions to match the 16 fermions of the standard model, and 16 conjugate sfermions to match the additional 16 conjugate fermions of the $\overline{16}$. (Again, with all fermion fields initially left-handed, those of the 16 and $\overline{16}$ are independent.) But since only the families of the 16 are observed we assume that the $\overline{16}$ fermions have larger masses from larger Yukawa couplings.

Higgsinos belong to the same $5 + \overline{5}$ representation as the Higgs, and gauginos to the $45$. 

As mentioned above, we regain conventional (broken) SUSY by postulating a conventional superpotential in which the postulated interactions have the same symmetry as the kinetic terms derived above.

\textit{Nonstandard scalar-boson fields} $\varphi$ can be constructed by first combining 
$\phi_{\uparrow}^{\, r}$ and the charge-conjugate field 
\begin{align}
\phi_{\downarrow}^{\, r \, c}  \left( x \right) = C \phi_{\downarrow}^{\, r \, *} \left( x \right)
\end{align}
(having both opposite spin and opposite gauge quantum numbers):
\begin{align}
\overline{\Phi} ^{r} \left( x \right) &=\left( 
\begin{array}{c}
\phi_{\uparrow}^{\, r} \left( x \right) \\ 
 \phi_{\downarrow}^{\, r \, c} \left( x \right)
\end{array} \right) 
\quad , \quad 
\phi_{\downarrow}^{\, r \, c \, \dag} \left( x \right)  \phi_{\downarrow}^{\, r \, c} \left( x \right)  = \phi_{\uparrow}^{\, r \, \dag} \left( x \right)  \phi_{\uparrow}^{\, r} \left( x \right)  \label{same}\; .
\end{align}
This scenario can be formulated within a full Spin(10) representation, but below we will consider the simplest final (low-energy) version, in which the relevant representations (after symmetry breakings) are the fundamental representations of the SU(3), SU(2), and U(1) subgroups of the standard model, using
\begin{align}
\{ t^j,t^{j'} \}=\frac{1}{N} \delta^{jj'} + d_{jj'j''} t^{j''}
\label{anticomm}
\end{align}
which holds for a fundamental representation of any $SU(N)$, where the $d_{jj'j''}$ are structure constants. 

\textit{To avoid cumbersome notation, from the above paragraph through the end of this section the $t^{j}$ are generators for an arbitrary SU(N)}, with
(\ref{anticomm}) giving
\begin{align}
 A^{\mu }  A_{\mu} =  A^{\mu \, j} t^j A_{\mu}^{j' } t^{j'} = \frac{1}{2} g^{\mu \nu}  A_{\mu }^{j } A_{\nu}^{j' } \{ t^j , t^{j'} \}  =  A^{\mu \, j} A_{\mu}^{j' }\left( \frac{1}{2N} \delta^{jj'} +  \frac{1}{2} d_{jj'j''} t^{j''} \right) \; .
\end{align}
The results below also hold for a U(1) representation with $1/(2N) \rightarrow 1$.

We will also need, for the Fourier coefficients,
\begin{align}
 \phi_{\downarrow} ^{\, r \, c \, \dag } \left( p \right) \, t^j \; \phi_{\downarrow} ^{\, r \, c} \left( p \right) 
= - \; \phi_{\downarrow} ^{\, r \, \dag } \left( p \right) \, t^j \; \phi_{\downarrow} ^{\, r } \left( p \right) 
= - \; \phi_{\uparrow} ^{\, r \, \dag } \left( p \right) \, t^j \;  \phi_{\uparrow} ^{\, r} \left( p \right) 
\end{align}
which follows from (\ref{same}) because $\phi_{\downarrow} ^{\, r}  \left( p \right)$ (as defined here) and $\phi_{\uparrow}^{\, r} \left( p \right) $ have the same amplitude and gauge quantum numbers..

Since  
{\allowdisplaybreaks
\begin{align}
\int d^{4}x \, \phi_{\downarrow} ^{\, r \, \dag }  \left( x \right)  B \, \phi_{\downarrow} ^{\, r}  \left( x \right)  = 
\int d^{4}x \, \phi_{\downarrow} ^{\, r \, c \, \dag }  \left( x \right)  B \, \phi_{\downarrow} ^{\, r \, c}  \left( x \right)
\end{align}
again follows from a simpler version of the argument in (\ref{conj})-(\ref{eqA20}), we have
\begin{align}
S_{\varphi}  &= \int d^{4}x \, \sum_{r} \left( \phi_{\uparrow} ^{\, r \,\dag } \left( x \right) \, B \, \phi_{\uparrow} ^{\, r} \left( x \right) + \phi_{\downarrow} ^{\, r \, c \, \dag } \, B \phi_{\downarrow} ^{\, r \, c} \left( x \right)  \right)  \label{invar}\\
&= \int d^{4}x \, \sum_{r}  \bigg[ \sum_{p} \phi_{\uparrow} ^{\, r \,\dag } \left( p \right) e^{i p \cdot x } \, \left(  \partial^{\mu}  - i A^{\mu \, j} t^j \right)\left(  \partial_{\mu}  - i A_{\mu}^{j' }t^{j'} \right)   \, \sum_{p'} \phi_{\uparrow} ^{\, r} \left( p' \right) e^{i p' \cdot x } \nonumber \\
& \hspace{2.2cm} + \sum_{p}  \phi_{\downarrow} ^{\, r \, c \, \dag } \left( p \right) e^{i p \cdot x }  \, \left(  \partial^{\mu}  - i A^{\mu \, j} t^j \right)\left(  \partial_{\mu}  - i A_{\mu}^{j' }t^{j'} \right)   \sum_{p'} \phi_{\downarrow} ^{\, r \, c} \left( p' \right) e^{i p' \cdot x }    \bigg]   \\
&= {\cal V} \sum_{r} \sum_{p } \bigg[ \phi_{\uparrow} ^{\, r \,\dag } \left( p \right)  \, \left(  \left( i p^{\mu}  - i A^{\mu \, j} t^j \right)\left( i p_{\mu}  - i A_{\mu}^{j' }t^{j'} \right) - i \left( \partial^{\mu} A_{\mu}^{j' } \right) t^{j'}   \right) \, \phi_{\uparrow} ^{\, r}  \left( p \right)  \nonumber \\
&  \hspace{1.2cm} + \phi_{\downarrow} ^{\, r \, c \, \dag } \left( p \right)   \, \left(  \left( i p^{\mu}  - i A^{\mu \, j} t^j \right)\left( i p_{\mu}  - i A_{\mu}^{j' }t^{j'} \right)  - i \left( \partial^{\mu} A_{\mu}^{j' } \right) t^{j'}    \right) \phi_{\downarrow} ^{\, r \, c} \left( p \right)  \bigg]  \label{v1}\\
&={\cal V}  \sum_{r} \sum_{p} \bigg[ \phi_{\uparrow} ^{\, r \,\dag } \left( p \right)  \, \left( - p^{\mu} p_{\mu}  - A^{\mu j} A^j_{\mu}/\left( 2 \, N \right) \right)    \, \phi_{\uparrow} ^{\, r}  \left( p \right)  \nonumber \\
&   \hspace{2.2cm}+ \phi_{\downarrow} ^{\, r \, c \, \dag } \left( p \right)   \,  \left( - p^{\mu} p_{\mu}  - A^{\mu j} A^j_{\mu}/\left( 2 \, N \right)  \right)    \phi_{\downarrow} ^{\, r \, c} \left( p \right)    \bigg]  \label{v2}\\
&= \int d^{4}x \, \sum_{r} \bigg[ \phi_{\uparrow} ^{\, r \,\dag } \left( x \right) \, \left(   \partial^{\mu} \partial_{\mu}  - A^{\mu j} A_{\mu j}/\left( 2 \, N \right) \right)   \, \phi_{\uparrow} ^{\, r}  \left( x \right)   \nonumber \\
&   \hspace{2.2cm}+ \phi_{\downarrow} ^{\, r \, c \, \dag }  \left( x\right)   \, \left(  \partial^{\mu} \partial_{\mu}  - A^{\mu j} A_{\mu j}/\left( 2 \, N \right)   \right) \phi_{\downarrow} ^{\, r \, c} \left( x\right)    \bigg]  \\
&= \int d^{4}x \, \sum_{r} \overline{\Phi}  ^{\, r \,\dag } \left( x\right)  {\cal B }\, \overline{\Phi}  ^{\, r}  \left( x \right)  \quad , \quad  {\cal B } = \partial^{\mu}  \partial_{\mu}  - A^{\mu j} A^j_{\mu}/\left( 2 \, N \right)  \label{calB} \\
&= \int d^{4}x \, \overline{\Phi} ^{\,\dag } \left( x \right) \, {\cal B}  \, \overline{\Phi}  \left( x \right)  
\end{align}
}where ${\cal V}$ is a 4-dimensional normalization volume and \textit{we have renamed $S_{\phi} \rightarrow S_{\varphi} $ in the present context}.

We can again define amplitude modes $\varphi _1^{r}$ and $\varphi _2^{r}$, by
\begin{align}
\varphi _i^{r}\left( x\right) = \zeta _{i}^{r\,\dag } \, \overline{\Phi} ^{\, r}\left( x\right)     \quad \quad
\mathrm{with } \quad  \quad \zeta _{i}^{r\,\dag }\,\zeta _{i'}^r = \delta_{i i'} 
\label{e8y}
\end{align}
where $\zeta ^r_{i}$ has $4$ constant components.  
Of these 4 orthonormal basis vectors (like those of (\ref{basis})), we choose $\zeta _{3}^r$ and $\zeta _{4}^r$  to be orthogonal to $\overline{\Phi} ^{\, r}$, so that only $\varphi _1^{r}\left( x\right)$ and $\varphi _2^{r}\left( x\right)$ are nonzero:
\begin{align}
\overline{\Phi }^{\, r}\left( x\right) &=\varphi _1^{r}\left( x\right) \,\zeta ^r_{1} + \varphi _2^{r}\left( x\right) \,\zeta^r_{2} \; .
\label{e8yy}
\end{align}
Then (\ref{calB}) gives
\begin{align}
S_{\varphi}  &= \int d^{4}x \, \sum_{r} \left[ \varphi_1 ^{\, r \, * }\left( x\right)  {\cal B }^{r}\, \varphi_1 ^{r}\left( x\right)   + \varphi_2 ^{\, r \, * }\left( x\right)  {\cal B }^{r}\, \varphi_2 ^{r}\left( x\right)  \ \right] \\
&= \int d^{4}x \, \sum_{r} \left[ \varphi_1 ^{\, r \, * }\left( x\right)  {\cal B }^{r}\, \varphi_1 ^{r}\left( x\right)   + \varphi_2 ^{\, r \, c \, * }\left( x\right)  {\cal B }^{r}\, \varphi_2 ^{r \, c}\left( x\right)   \right] \\
&= \int d^{4}x \, \sum_{r} \left[ \varphi_1 ^{\, \dag }\left( x\right)  {\cal B }^{r}\, \varphi_1 \left( x\right)   + \varphi_2 ^{\, c \, \dag }\left( x\right)  {\cal B }^{r}\, \varphi_2 ^{c}\left( x\right)  \ \right] \label{var}
\end{align}
since $\int d^{4}x \, \varphi_2 ^{r \, *} \left( x \right)  {\cal B }^{r}  \, \varphi_2 ^{r }\left( x \right) $ = $ \int d^{4}x \, \varphi _2^{r \, c \, *} \left( x \right)  {\cal B }^{r}  \, \varphi _2^{r \, c}\left( x \right) $ as before, with $\varphi_2 ^{r \, c}\left( x \right)  = C \varphi _2^{r \, *} \left( x \right)$.  For a complete $5 + \overline{5}$ representation, $\varphi_1$ and $\varphi _2^{c}$ would each consist of 5 complex scalar-boson fields; and for a complete $16 + \overline{16}$ representation,  $\varphi_1$ and $\varphi _2^{c}$ would each consist of 16 complex scalar-boson fields.

The lowest-mass $\varphi$ particle from a $5+\overline{5}$ representation (assumed to be neutral) is the dark matter candidate $h^0$ of our previous papers~\cite{DM2021a,Tallman}, where we have called it a higgson because the gauge couplings are the same as for the Higgs $H^0$.
In the simplest picture, there are two most relevant $5+\overline{5}$ multiplets, with one containing the same two Higgs doublets as conventional SUSY, and the other containing two corresponding higgson doublets. The lowest-mass particle from the conventional $\phi$ pair of Higgs doublets is then the observed 125 GeV Higgs boson, and the lowest-mass particle from the corresponding $\varphi$ pair of higgson doublets is our predicted $\approx 70$ GeV dark matter WIMP.

Written more explicitly, (\ref{var}) is
\begin{align}
S_{\varphi} &=\int d^{4}x \, \sum_{r} \bigg[ \varphi _1^{r \,  *} \left( x \right) \left[ \left(  \partial^{\mu} \partial_{\mu}  - A^{\mu j} A^j_{\mu}/\left( 2 \, N \right) \right)  \right]\, \varphi _1^r \left( x \right) \nonumber \\
& \hspace{2.3cm}+ \varphi _2 ^{r \, c\, *} \left( x \right) \left[  \left(  \partial^{\mu} \partial_{\mu}  - A^{\mu j} A^j_{\mu}/\left( 2 \, N \right) \right)  \right]\, \varphi _2^{r \, c}\left( x \right) \bigg] 
  \label{var2}
\end{align}
with, again, $N =$ dimension of the representation. Notice that the free-particle \textit{propagator} for an unconventional scalar-boson field $ \varphi$, with $A^j_{\mu}=0$, is the same as that for for a conventional scalar-boson field $ \phi$.

The action for the fundamental fields is gauge invariant in the usual way, as can be seen from (\ref{invar}) in the form
\begin{align}
S_{\varphi} &= \int d^{4}x \, \sum_{r} \overline{\Phi}^{\, r \,\dag } \left( x \right)  D^{\mu } D_{\mu } \, \overline{\Phi}^{\, r} \left( x \right)   \label{invarx} \; .
\end{align}
Then it is immediately obvious that (\ref{var2}) is also gauge invariant: When $A^j_{\mu} \rightarrow A^{\prime \, j}_{\mu}$ and $\overline{\Phi}^{\, r}  \left( x\right) \rightarrow \overline{\Phi}^{\prime \, r}\left( x\right)$ in (\ref{invarx}),
\begin{align}
\varphi^{r}_{1}\left( x\right) \rightarrow \varphi^{\prime \, r}_{1}\left( x\right) = \zeta _{1}^{r \, \dag } \, \overline{\Phi}^{\prime \, r}\left( x\right)  \quad , \quad \varphi^{r}_{2}\left( x\right) \rightarrow \varphi^{\prime \, r}_{2}\left( x\right) = \zeta _{2}^{r \, \dag } \, \overline{\Phi} ^{\prime \, r}\left( x\right)  
\end{align}
so that (\ref{var2}) still holds with the transformed fields and potentials.

For the SU(3)$\times$SU(2)$\times$U(1) fields of the standard model, if coupling constants are displayed rather than absorbed into the gauge potentials, (\ref{var2}) implies that the Lagrangian for the interaction of nonstandard scalar-boson fields $\varphi_b^r$ with gauge fields is
\begin{align}
\overline{\mathcal{L}}^{int} &=  - \varphi_b^{r\, * } \left( \, \sum_n \overline{g}_n^2 \, A_n^{\mu  j} A_{n \mu}^j  \, \right) \varphi_b^r \quad , \quad \overline{g}_{3}^2 = g_{3}^2 /6 , \quad \overline{g}_{2}^2 = g_2^2/4, \quad \overline{g}_{1}^2 = g_1^2
\label{eq30}
\end{align}
where $g_{3}$, $g_{2}$, $g_{1}$ are the original SU(3), SU(2), U(1) coupling constants. 

When the original $SU(2) \times U(1)$ fields are rotated into those of the electroweak theory after symmetry breaking
\begin{align}
W^{\pm}_{\mu} = \frac{1}{\sqrt{2}} \left( A^1_{2 \mu } \mp i A^2_{2 \mu } \right) \;, \;
Z_{\mu} = \frac{1}{\sqrt{g_1^2 + g_2^2}} \left( g_2 A^3_{2 \mu } - g_1 A_{1 \mu } \right) \; , \;
\bar{A}_{\mu} = \frac{1}{\sqrt{g_1^2 + g_2^2}} \left( g_1 A^3_{2 \mu } + g_2 A_{1 \mu } \right) \label{complex}
\end{align}
with the covariant derivative~\cite{peskin}
\begin{align}
D_{\mu }=\partial _{\mu } -i\frac{g}{\sqrt{2}}\left( W_{\mu}^{+}\tau^{+}+W_{\mu }^{-}\tau^{-}\right)   - i \frac{g}{\cos \theta _{w}}Z_{\mu }\left( \tau^{3}-\sin ^{2}\theta_{w}\,Q\right) -ie\bar{A}_{\mu }\,Q  \label{e5}
\end{align}
where $\tau^{\pm} = \tau^1 \pm i \tau^2$, (\ref{e5}) and (\ref{eq30}) give 
\begin{align}
- \frac{g_s^2}{6} \,  {\cal A}^{\mu  i}  {\cal A} _{\mu}^i  \quad , \quad  - \frac{g^2}{2} \, W^{+ \mu } W^- _{\mu}  \quad , \quad - \frac{g_Z^2}{4}  \, Z^{\mu } Z_{\mu} \quad , \quad - \left( Qe \right)^2 \, \bar{A}^{\mu } \bar{A}_{\mu}   \label{eq43} 
\end{align}
for the strong, weak, and electromagnetic interactions respectively.
Here $g_s=g_3$ and $g=g_2$ are the usual strong and weak coupling constants, $g_Z=g/\cos \theta_W$, $\cos \theta_W = g_2/\sqrt{g_1^2 + g_2^2}$, $Q e$ is the electric charge, $\bar{A}_{\mu }$ is the electromagnetic vector potential, and ${\cal A}_{\mu } = {\cal A}_{\mu}^i T_i $ is the QCD gauge field containing gluon fields ${\cal A}_{\mu }^i$. Recall that the full Lagrangians have the gauge invariance demonstrated above. 

Here we will not explore possible further  implications of this unconventional scalar-boson sector, or assume any members beyond the dark matter candidate $h^0$ of \cite{DM2021a} and \cite{Tallman}.

\section{\label{sec:sec8}Fundamental action for fermions and scalar bosons}

Recall that $\widetilde{D}_{\mu } = \nabla_{\mu} - i A_{\mu} $ is the full covariant derivative, with the generators in 
$\nabla_{\mu}$ and $A_{\mu}$ now regarded as operators. 
When the various components in the preceding sections are assembled, the total action for fermion fields $\psi _{f}^r$, standard scalar-boson fields $\phi _b^r$,  auxiliary fields $F_b^r$, and nonstandard scalar-boson fields $\varphi_b ^{r}$ is
\begin{eqnarray}
 S_{matter}  =\int d^{4}x\, e \, \overline{\mathcal{L}}_{matter} 
 \label{eq8.3a}
 \end{eqnarray}
 \begin{align}
\overline{\mathcal{L}}_{matter} = \sum_r \psi _{f}^{r \, \dagger } \left( x \right) ie_{\alpha }^{\mu }\,\overline{\sigma} ^{\alpha } 
 \widetilde{D}_{\mu }\,\psi _{f}^r\left( x\right) +  \sum_r  \phi_b^{r \,* } \left( x\right)  g^{\mu \nu } \widetilde{D}_{\mu }  \widetilde{D}_{\nu } \,
 \phi_b^r \left( x\right) 
+ \sum_r F_b^{r \, * }\left( x\right) F_b^r\left( x\right)  
  +\overline{\mathcal{L}}_{\varphi}
\label{eq8.3}
\end{align}
or 
 \begin{align}
\overline{\mathcal{L}}_{matter} = 
\psi _{f}^{\dagger }  
ie_{\alpha }^{\mu }\,\overline{\sigma} ^{\alpha } 
 \widetilde{D}_{\mu }\,\psi _{f} +  \phi_b ^{\dagger } \,  g^{\mu \nu } \widetilde{D}_{\mu }  \widetilde{D}_{\nu } \, \phi_b 
 + F_b^{\dagger } F_b
  + \overline{\mathcal{L}}_{\varphi}
\label{eq8.3x}
\end{align}
after transformation to a general coordinate system, and before masses and further interactions result from symmetry breakings, radiative corrections, and other effects. Here $r$ labels all the fields of the 16, $\overline{16}$, 5, and $\overline{5}$ representations, and $\overline{\mathcal{L}}_{\varphi}$ is the Lagrangian for  the action of (\ref{var2}).

The spin 1/2 fermion fields in $\psi _{f}$ span the various physical
representations of the most fundamental gauge group, which is Spin(10) in
the present theory. The same is true for the combined spin 0 fields in $\phi_b$, $F_b$, and $\varphi_b$. All of (\ref{eq8.3}) corresponds to a Lorentz-invariant,  gauge-invariant, and supersymmetric action which is also invariant under general coordinate transformations. (Other aspects of supersymmetry are beyond the scope of the present paper, but are consistent with the present theory.)

As mentioned above, gravitational and gauge curvatures must ultimately originate from ``rapidly fluctuating'' $4$-dimensional topological defects (analogous to vortices and vortex rings, or extended and closed flux tubes) associated with the gauge potentials of (\ref{eq5.23}) and the vierbein of (\ref{eq4.40x}). Here ``rapidly fluctuating'' means that the above $A^i_{\mu}$ and 
$e_{\alpha }^{\mu }$ (or $g_{\mu \nu}$) span many topological configurations of the field $\Psi_0$. As mentioned below (\ref{eq4.11b}), and exhibited in the paragraph following this one, the path integral over all these configurations is replaced by a path integral over the $A_{\mu}$ and $e_{\alpha }^{\mu }$ (or $g_{\mu \nu}$). The gauge and gravitational fields then vary over all possibilities, and this is how these force fields are quantized in the present theory. A detailed discussion of the topological defects is beyond the scope of this paper, but all that is required here is the fact that topological defects permit the curvature associated with the the gauge potentials and vierbein to be nonzero. I.e., the gauge curvature (\textit{with the coupling constant no longer absorbed into} $A_{\mu} $)
\begin{align}
F^i_{\mu \nu} = \partial_{\mu}A^i_{\nu} - \partial_{\nu}A^i_{\mu} +g f^i_{j \, k} A^j_{\mu} A^k_{\mu}
\label{curv-1}
\end{align}
and the gravitational spin connection and curvature (see p.~274 of Ref.~\cite{GSW})
\begin{align}
& \omega_{\mu}^{\alpha \beta} = \frac{1}{2}e^{\nu \alpha} \left( \partial_{\mu} e^{\beta}_{\nu} - \partial_{\nu} e^{\beta}_{\mu} \right) - \frac{1}{2}e^{\nu \beta} \left( \partial_{\mu} e^{\alpha}_{\nu} - \partial_{\nu} e^{\alpha}_{\mu} \right)
- \frac{1}{2}e^{\rho \alpha} e^{\sigma \beta}  \left( \partial_{\rho} e_{\sigma \gamma} - \partial_{\sigma} e_{\rho \gamma}  \right) e^{\gamma}_{\mu} \\
& R_{\mu \nu \;\;  \beta}^{ \;\; \; \;  \alpha} = \partial_{\mu} \omega_{\nu \;\;  \beta}^{ \;\;   \alpha} - \partial_{\nu} \omega_{\mu \;\;  \beta}^{ \;\;   \alpha} + \left[\omega_{\mu }  , \omega_{\nu } \right]^{\alpha}_{\;\; \beta}
\quad , \quad
R_{\mu \nu \;\;  \beta}^{ \;\; \; \;  \alpha} = e^{\alpha}_{\sigma} e^{\tau}_{\beta} R_{\mu \nu \;\;  \tau}^{ \;\; \; \;  \sigma}
\label{curv-4}
\end{align}
originate from vortex-like configurations in the same way that the vorticity of a superfluid 
\begin{align}
\omega^s_{k \ell} =\partial_k v^s_{\ell} - \partial_{\ell} v^s_k \quad  \mathrm{or} \quad \omega^s_z =\partial_x v^s_y - \partial_y v^s_x
\label{curv-5}
\end{align}
originates from ordinary vortices, as pointed out by Feynman and Onsager. The simplest case is a magnetic field with
\begin{align}
B_z = \partial_x A_y - \partial_y A_x
\label{curv-6}
\end{align}
but all the force fields above have the same basic form. In each case, the curvature is nonzero in a region penetrated by flux lines that are interpreted as vortex lines.

From a more fundamental point of view, it is not necessary to consider specific topological defects, since, as mentioned below (\ref{eq4.11b}), the path integral automatically includes all possible configurations of the condensate order parameter $\Psi_0$ consistent with this equation of motion, and thus all physically independent (and physically acceptable) values of the gauge potentials and metric tensor.  These configurations of $\Psi_0$ span all such topologies, including those with vortex-like (and monopole-like) topological defects. The path integral over $\Psi_0$ can then be replaced by a path integral over the gauge potentials $A_{\mu}^i$ and vierbein $e_{\alpha }^{\mu }$, with a global transformation of integration variables
\begin{align}
d A_{\mu}^i \left( x' \right)  = \sum_{I , x}\frac{\partial A_{\mu}^i \left( x' \right) }{\partial \Psi_0^I \left( x \right)} d \Psi_0^I \left( x \right) \quad , \quad
d e_{\alpha }^{\mu }\left( x' \right) = \sum_{I , x} \frac{\partial e_{\alpha }^{\mu }\left( x' \right) } {\partial \Psi_0^I \left( x \right)} d \Psi_0^I \left( x \right)
\label{PI-trans}
\end{align}
where $I$ labels a component of the full order parameter. 
\textit{Here, in this one context,  $x$ represents all of the $D$ coordinates $x^M$, but $x' $ is limited to the 4 external spacetime coordinates $x^{\prime \, \mu}$.} 
(If the values of $x$ and $x'$ were continuous, $\sum_{x }$ would be replaced by $\int d^4 x $ and the partial derivatives by functional derivatives. But in the present description a path integral is over very closely spaced but discrete spacetime points.) The physically distinct $A_{\mu}^i $ of (\ref{eq5.23}) and $e^{\mu}_{\alpha}$ of (\ref{eq4.39z}) can be varied freely, because they correspond to variations in the \textit{phase} of $\Psi_0$, which are allowed even with a constraint like (\ref{eq4.11b}), since the other quantities (including the condensate \textit{density} in a specific path-integral trajectory, at each time and in each region of space) can vary in response to satisfy the constraint (just as in an ordinary superfluid with vortices). 

The restriction to physically distinct $A_{\mu}^i $ and $e^{\mu}_{\alpha}$ means eliminating copies which differ only by a gauge transformation in the usual way -- e.g. with Faddeev–Popov gauge fixing.
The path integral can be schematically represented by 
\begin{align}
\int D \psi \; D \phi \; D A \; D e \; e^{iS} = \int D  \psi \; D  \phi \; D \Psi_0 \; J \;  e^{iS} 
\label{PIx}
\end{align}
where $J$ represents the Jacobian for the transformation of (\ref{PI-trans}). (The justification for changing to $A_{\mu}^i $ and $e^{\mu}_{\alpha}$ as the primary variables in the path integral is that they give the force fields experienced by the matter fields. The change in the measure represented by $J$ is therefore physically justified, because they are the physically relevant variables.) (\ref{PIx}) is the appropriate path integral for transition amplitudes between states defined by the fermion, scalar-boson, gauge-boson, and gravitational fields. These fields are then to be varied over all values (subject to gauge-fixing for $A_{\mu}^i $ and $e^{\mu}_{\alpha}$), with $J$ specifying how each in the  vast number of relevant topologies of $\Psi_0$ is to be mixed in to satisfy this equality; small contributions correspond to small values of the derivatives in (\ref{PI-trans}).
A detailed treatment is beyond the scope of the present paper, but it is only required that such a transformation is possible. All physically distinct possibilities are then realized for the gauge and gravitational fields, just as in standard physics.

\section{\label{sec:sec8+}Cosmological constant, gravitational action, gauge-field action, black hole entropy, and dark matter}

\subsection{\label{sec:cos}Cosmological constant, gravitational action, and gauge-field action}

In conventional physics, the contribution of fermion and scalar-boson fields to the vacuum energy corresponds to  a Lagrangian $e \, \overline{\mathcal{L}}_{vac} $, with $\overline{\mathcal{L}}_{vac} $ constant. The resulting gravitational 
stress-energy tensor is
\begin{eqnarray}
T^{\mu \nu }_{vac}=2 e^{-1} \delta \left( e \overline{\mathcal{L}}_{vac} \right) / \delta g_{\mu \nu}  = g^{\mu \nu} \overline{\mathcal{L}}_{vac} \quad , \quad
\overline{\mathcal{L}}_{vac} = - \left( 8\pi \ell_{P}^2  \right)^{-1}   \Lambda 
 \label{lambda}
\end{eqnarray}
since
\begin{eqnarray}
\delta e = \frac{1}{2} \, e \,  g^{\mu \nu} \, \delta g_{\mu \nu}
 \label{lvariation}
\end{eqnarray}
where $\ell_{P}$ is the Planck length, and this produces a term $\Lambda g_{\mu \nu} $ in the Einstein field equations with a cosmological constant $\Lambda$. 

In the present description, on the other hand, the left-hand side of (\ref{var11}) implies that
\begin{eqnarray}
T^{\mu \nu }_{vac}=2 e^{-1} \delta \overline{\mathcal{L}}_{vac} / \delta g_{\mu \nu}=0
\label{vac}
\end{eqnarray}
so there is no direct contribution to a cosmological constant from these fields.

The predictions of the present theory are identical to those of standard general relativity for the motion of all particles and waves in gravitational fields, and for gauge bosons acting as a source of gravity. We will now show that they are also the same for matter acting as a classical gravitational source. 
The usual energy-momentum tensor is
\begin{eqnarray}
 T^{\mu \nu }_{\chi}=
2 \, e^{-1} \frac{\delta \left( e \overline{\mathcal{L}}_{\chi} \right) }{ \delta g_{\mu \nu } }= 2 \, \overline{\mathcal{L}}_{\chi} \, e^{-1} \frac{\delta e }{\delta g_{\mu \nu }} +
2 \frac{\delta \overline{\mathcal{L}}_{\chi} }{ \delta g_{\mu \nu }} = 2 \frac{\delta \overline{\mathcal{L}}_{\chi}}{ \delta g_{\mu \nu } } 
= 2 \, \chi^{\dag} \frac{\delta A_{\chi} } {\delta g_{\mu \nu} }  \chi  
\label{eq8.7x}
\end{eqnarray}
since the bilinear forms of (\ref{eq8.3}) -- or similar bilinear forms including mass and interaction terms, or for composite objects like protons or planets -- imply that $\overline{\mathcal{L}}_{\chi} = 0 $ if the classical equations of motion are satisfied. 
(Classical in the present context means that particles or composite bodies remain on the mass shell, and the energy-momentum tensor here corresponds to the energy and momentum of quantum fields, particles, or composite objects satisfying their quantum equations of motion, which yield classical trajectories according to Ehrenfest's theorem.) 
Since the result of (\ref{eq8.7x}) is identical to that of (\ref{var-final1}), the role of classical matter in acting as a source of gravity is unchanged.

The predictions of classical general relativity are thus unchanged in the present picture, except that the cosmological constant of (\ref{vac}) is zero if the vacuum Lagrangian density $\overline{\mathcal{L}}_{vac}$ is taken to be fixed.

On the other hand, $\overline{\mathcal{L}}_{vac}$ 
is not really fixed, since the fields in the vacuum will be modified when they are forced to accommodate the curvature associated with
the gauge potentials of (\ref{eq5.23}) and the vierbein of (\ref{eq4.40x}) (or metric tensor of (\ref{eq8.3z})). (A nontrivial gravitational or gauge field means  spacetime curvature of $e^{\alpha}_{\mu}$ or $A^i_{\mu}$, as in (\ref{curv-1})-(\ref{curv-4}); a change in $e^{\alpha}_{\mu}$ or $A^i_{\mu}$ with no curvature is just a gauge transformation, with the vacuum action left unchanged.)
Within the present description,  the Maxwell-Yang-Mills action 
\begin{eqnarray}
S_g= \int d^4 x \, \mathcal{L}_{g} \quad , \quad 
\mathcal{L}_{g}=-\frac{1}{4}g_{0}^{-2} \, e \, 
g^{\mu \rho }g^{\nu \sigma } \,F_{\mu \nu }^{i}F_{\rho \sigma}^{i} 
\label{eq8.8}
\end{eqnarray}
and the Einstein-Hilbert action 
\begin{eqnarray}
S_G= \int d^4 x \, \mathcal{L}_{G} \quad , \quad 
\mathcal{L}_{G}= \left( 16\pi \ell_{P}^2 \right)^{-1} e \, R 
\label{eq8.7}
\end{eqnarray}
must result from this modification of the vacuum states (and the terms in their action).
This form is analogous to the shift in the free energy
\begin{align}
\Omega \left( B \right) - \Omega \left( 0 \right) =  \frac{e^2 A}{24 \pi m c^2} B^2 \; ,
\end{align}
when electrons in a metal (with area $A$) respond to the gauge curvature of an applied magnetic field $\overrightarrow{B}$, exhibiting Landau diamagnetism. 
(In the present case, there is an additional contribution from the topological defects discussed in the paragraph containing (\ref{curv-1})-(\ref{PI-trans}).)

It appears that it is nontrivial to obtain (\ref{eq8.8}) and (\ref{eq8.7}) from a proper calculation, and that a detailed knowledge of the vacuum states is required. 
(A truly quantitative calculation, based on a complete understanding of the vacuum, after renormalization would in principle predict the gravitational constant $G$ and fundamental charge $e$.) However, (i)~these states will surely be even qualitatively (topologically) modified when they are perturbed by the curvature of the vierbein and gauge fields (as are the states of electrons in a metal), (ii)~(\ref{eq8.8}) and (\ref{eq8.7}) are the simplest forms consistent with the symmetries of the vacuum (including invariance under coordinate, Lorentz, and gauge transformations), and (iii)~within the present picture (\ref{eq8.8}) and (\ref{eq8.7}) can originate only from the response of the vacuum to external fields. 

With this interpretation, $\mathcal{L}_{g}$ must necessarily vanish when these fields vanish -- i.e., in the bare vacuum with no gauge fields:
\begin{eqnarray}
\langle \mathcal{L}_{g} \rangle_{vac} = 0 \; .
\label{eq8.9x}
\end{eqnarray}
(This means that when (\ref{eq8.8}) is treated in canonical quantization, the field operators must be normal-ordered.) It follows that there is no cosmological constant resulting from the gauge fields. On the other hand, virtual processes will still be affected by a change in their boundary conditions; a detailed treatment of this aspect, and of the observed Casimir effect~\cite{Jaffe,casimir1,casimir2},  would be inappropriately long here, but see the discussion of this point in Section \ref{sec:sec1a}.

Finally, there is still another related nuance: The physical vacuum will contain a very weak bath of virtual gauge bosons being exchanged between vacuum fields that might account for the dark energy, although a quantitative calculation would again be difficult.

\subsection{\label{sec:BH}Thermodynamics of black holes}

The Hawking temperature $T_H$ and Bekenstein-Hawking entropy $S_{BH}$ of a general black hole have the well-known forms
\begin{align}
T_H=\frac{\kappa }{2 \pi } \quad , \quad S_{BH} = \frac{A}{4 \ell_P^2}  \label{H}
\end{align}
(in units with $ \hbar = c = k = 1$) where $\kappa$ is the surface gravity and $A$ is the area of the event horizon. No convincing explanation of this entropy has been provided for physical black holes, despite a large number of attempts in various theoretical frameworks~\cite{Strominger,Rovelli,Wald}.

In 1977 Gibbons and Hawking~\cite{Gibbons-Hawking} showed that these results would (rather amazingly) follow if the Euclidean path integral \textemdash \, which they evaluated by calculating the Euclidean action for the most important physical black holes and de Sitter space \textemdash \, can be interpreted as a thermodynamic partition function $Z$, which yields a free energy $W = - T \log Z$ with
\begin{align}
W=M-TS_{BH}-\Phi Q -\Omega J 
\label{W}
\end{align}
where $M$, $T$, $S_{BH}$, $\Phi$, $Q$, $\Omega$, and $J$ are respectively the mass (or energy), temperature, entropy, electrostatic potential, electric charge, angular velocity, and angular momentum. 
A recent paper by Witten~\cite{Witten-BH} contains an extensive critical review of the field (with some new results).

What is missing is a statistical origin of the inferred thermodynamic entropy $S_{BH}$ from microscopic degrees of freedom. 

In the present theory, any black hole is a macrostate of the gravitational field with the Boltzmann entropy 
\begin{align}
S_{BH} = \log \, W_{BH}
\end{align}
where $W_{BH}$ is the number of microstates of the dits that are consistent with the specification of this macrostate, just as in the second paragraph of Section \ref{sec:sec2x}, with the entropy of any field configuration initially given by (\ref{eq2.5}), (\ref{eq2.2}), and (\ref{eq2.3}).

The present theory thus inevitably predicts a black hole entropy, consistent with the results of Bekenstein, Hawking, and others~\cite{Wald}, although the remarkable simplicity of (\ref{H}) is revealed only by this earlier work \textemdash \, including the demonstration of \cite{Gibbons-Hawking} that, if a black hole has entropy, it must be given by (\ref{H}) in the context of their highly sophisticated analysis (which is further considered and amplified in \cite{Witten-BH}).

The present theory also suggests new ways of viewing other issues like the ``information paradox''~\cite{Preskill}, but this and related topics~\cite{Fulling-Unruh} are beyond the scope of the present paper.

\subsection{\label{sec:families}Families, Yukawa couplings, scalar-boson masses, and quartic couplings}

As mentioned above, in the present theory there are 3 families of standard-model fermions, with Yukawa couplings to only one Higgs family,  because the SO(3) rotational group for the 3-dimensional internal space has $\ell=1$ and $\ell=0$ representations. It is now appropriate to detail the field/particle content according to the 10-dimensional and 3-dimensional representations respectively: (i) 16 and $\ell=1$ for standard model fermions and their sfermions, $\overline{16}$ and $\ell=1$ for new predicted higher-mass fermions and and their sfermions; (ii) $10=5+\overline{5}$ and $\ell=0$ for Higgs fields (so that conservation of internal angular momentum, or family number $m_{\ell}$, implies that there are Yukawa couplings only between fermion fields in the same family before radiative corrections) and for higgsinos; (iii) 45 and $\ell =0$ for gauge bosons and gauginos (so that fermion gauge couplings are also within families); (iv) $10=5+\overline{5}$ and $\ell=1$ for higgsons; (v) 16 and $\ell=0$ for the primordial condensate, which then belongs to no family, has no standard model (or even SU(5)) couplings, and has a special status different from all other fields.

As mentioned above, scalar-boson masses (as in e.g. Eq. (1.4) of \cite{Drees}, limited by supersymmetry, which is nonzero even if the bare masses are zero for the scalar boson and the fermions to which it is coupled), scalar-boson quartic couplings (as for the Higgs, renormalized from a small negative value at high energy to one that is sizable and positive), Yukawa couplings, and gaugino and higgsino masses must all result from radiative corrections (or additional fields as in conventional SUSY). 

\subsection{\label{sec:DM2}Dark matter}

There are a vast number of hypothetical dark matter
candidates, most of which do not have well-defined masses or couplings, and
many of which have already been ruled out by experiment -- or at least found to be subdominant species in a multicomponent scenario. For example, the simplest supersymmetric models which have ``natural'' values for the
parameters, and which are also compatible with limits from the LHC, are
found to be in disagreement with both the abundance of dark matter and the
limits from direct-detection 
experiments~\cite{Baer-Barger-2016,Roszkowski-2018,Baer-Barger-2020,Tata-2020} -- if the lightest supersymmetric particle (LSP) is assumed to be the dominant constituent. But, as mentioned in Section \ref{sec:sec1a}, the present picture requires (broken) supersymmetry, and the LSP (as a subdominant component, with R-parity conservation) can stably coexist with the present dark matter candidate -- the lightest higgson, $h^0$~\cite{DM2021a,Tallman}.

More generally, this dark matter candidate is consistent with all current experiments, and observable in the near or foreseeable future through indirect and collider detection experiments. To review the conclusions of Ref.~\cite{Tallman}, which presents the results of detailed calculations of cross-sections and observables for present, near-term, and longer-term experiments: This particle is unique in that it has precisely defined couplings and a well-defined mass, providing specific experimental signatures as targets for clean experimental tests. It has not yet been detected because it has no low-energy interactions other than second-order gauge couplings, to $W$ and $Z$ bosons. However, these weak couplings are still sufficient to enable observation by experiments which are currently taking data and planned for the foreseeable future. 

This lightest higgson $h^0$ is stable because of the form of the interaction in (\ref{var2}) or Eq. (47) of \cite{DM2021a}: It can radiate gauge bosons, annihilate into gauge bosons, scatter via exchange of gauge bosons, and be created in pairs, but not decay, since a single initial $h^0$ implies a final state containing $h^0$ and two gauge bosons. 

Detection should be achievable at the high-luminosity LHC, as well as the even more powerful hadron and lepton colliders now being planned~\cite{Tallman}. There is already a strong case that the present dark matter candidate has been observed via indirect detection: Several analyses of gamma rays from the Galactic center, observed by Fermi-LAT~\cite{Leane-2,Goodenough,Fermi,Fermi-GCE,Leane-1,Leane}, and of antiprotons, observed by AMS-02~\cite{Cuoco2,Cuoco,Cui,AMS-1,AMS-2}, have shown consistency with the interpretation that these result from annihilation of dark matter particles having approximately the same mass and annihilation cross-section as the present candidate. 
Finally, there is consistency with the observations of Planck, which have ruled out various possible candidates with larger masses and cross-sections~\cite{Planck,Hooper-positron}, but are quite consistent with the present candidate.

\section{\label{sec:sec9}Conclusion}

Starting with the simplest imaginable picture, and interpreting our universe as the product of three spaces (with two having central topological singularities), we obtain the following results, which are discussed more extensively in Section~\ref{sec:sec1a} and the sections following it:
\begin{itemize}
\item elimination of the usual enormous cosmological constant (Section~\ref{sec:sec1a}, topic (9), and Section~\ref{sec:sec8+})
\item a dark matter candidate with well-defined interactions and mass (Section~\ref{sec:sec8+} and Ref.~\cite{Tallman})
\item other new scalar bosons and spin 1/2 fermions 
\item SO(10) grand unification 
\item 4-dimensional spacetime 
\item one time coordinate 
\item spin 1/2 fermion fields with the correct relativistic action 
\item spin 0 boson fields with the correct relativistic action 
\item spin 1 gauge fields 
\item Lorentzian path-integral quantization of these fields 
\item correct couplings of matter fields to the gauge fields 
\item a gravitational vierbein and metric tensor 
\item correct coupling of matter to gravity 
\item Lorentz invariance 
\item 3 families of standard model fermions 
\item Yukawa couplings to only one Higgs family
\item Yukawa couplings only within each fermion family (before radiative corrections)
\item gauge couplings only within each fermion family
\item a supersymmetric action including auxiliary fields (with broken symmetry for masses)
\item Bekenstein-Hawking entropy of black holes 
\item a natural explanation of why the internal spaces responsible for gauge fields and family replication are extremely small 
\end{itemize}
Other issues are just touched upon: the origin of the Einstein-Hilbert gravitational action (with a connection to Starobinsky inflation and dark energy); the origin of the Maxwell-Yang-Mills action for the gauge fields; and the probability interpretation of quantum mechanics.

The theory is ultimately founded on counting arguments with positive integers, and the continuum is only an approximation to a discrete set of (closely spaced) points. The mathematical foundations are therefore consistent and well-defined, including the meaning of path integrals. Singularities are not a problem in the present theory, because the density of the primordial condensate underlying standard physics goes to zero at each point where there is a singularity. 

However, the above are largely theoretical issues, and the ultimate objective of any theory should be testable experimental predictions. Quantitative predictions already obtained for the present theory can be found in the calculations of Tallman et al.~\cite{Tallman} for a dark matter WIMP with well-defined mass and couplings, which is currently consistent with all experiments and observations. It should be within reach of collider-detection experiments in the near future \textemdash \, nominally detectable at the $5 \sigma$ level with 500 fb$^{-1}$ integrated luminosity at the high-luminosity LHC \textemdash \, and may have already been observed via indirect detection, by Fermi-LAT and AMS-02. (This estimate of statistical significance omits systematic experimental uncertainties and additional background from events with undetected ``soft'' leptons, but the ultimate goal of the high-luminosity LHC is 3000 fb$^{-1}$, and we are currently discussing the optimization of cuts with experimentalists.) A qualitative prediction is other new particles, including new scalar bosons and new fermions.

In conclusion, and in anticipating future work, it should be emphasized that the present paper is primarily meant to propose a new perspective on some problems that have resisted solution through conventional approaches for half a century or longer. Many other well-known problems are not considered here~\cite{pdg,LCDM}, but the present picture does not appear to be inconsistent with potential solutions, including those that have already been proposed.

\appendix

\section{\label{sec:sec2}Primordial fields and action}

We begin with the initial picture of Section \ref{sec:sec2x}, in which coordinates and fields are defined by (\ref{eq2.1}), (\ref{eq2.2}), and (\ref{eq2.3}).

Let $S\left( x\right) $ be the entropy at a fixed point $x$, as defined by 
$S\left( x\right) =\log \,W\left( x\right) $. Here $W\left( x\right) $ is the total
number of microstates for fixed occupation numbers $n_{i}$: $W\left(
x\right) =N\left( x\right) !/\Pi _{i}\,n_{i}\left( x\right) !$, with 
\begin{equation}
N\left( x\right) =\sum_{i}n_{i}\left( x\right) \quad ,\quad i=1,2,..., d
\; .
\label{eq2.4}
\end{equation}
The total number of available microstates for all points $x$ is $W=\Pi
_{x}\,W\left( x\right) $, so the total entropy is 
\begin{eqnarray}
S = \sum_{x}\,S\left( x\right) \quad , \quad
S\left( x\right) = \log \Gamma \left( N\left( x\right) +1\right)
-\sum_{i}\log \Gamma \left( n_{i}\left( x\right) +1\right) \; .
\label{eq2.5}
\end{eqnarray}

We will see below that $n_{k}\left( x\right) $ can be approximately treated
as a continuous variable when it is extremely large, with 
\begin{eqnarray}
\frac{\partial S}{\partial n_{k}\left( x\right) } &=&\psi \,\left( N\left(
x\right) +1\right) -\psi \left( n_{k}\left( x\right) +1\right) 
\label{eq2.6}\\
\frac{\partial ^{2}S}{\partial n_{k^{\prime }}\left( x\right) \partial
n_{k}\left( x\right) } &=&\psi \,^{\left( 1\right) }\left( N\left( x\right)
+1\right) -\psi ^{\left( 1\right) }\left( n_{k}\left( x\right) +1\right)
\delta _{k^{\prime }k}\; .  \label{eq2.7}
\end{eqnarray}
The functions $\psi \,\left( z\right) =d\log \Gamma \left( z\right) /dz$ and 
$\psi ^{\left( 1\right) }\,\left( z\right) =d^{2}\log \Gamma \left( z\right)
/dz^{2}$ have the asymptotic expansions
\begin{eqnarray}
\psi \,\left( z\right) = \log z-\frac{1}{2z}-\sum_{l=1}^{\infty }
\frac{B_{2l}}{2l\,z^{2l}} \quad , \quad 
\psi ^{\left( 1\right) }\,\left( z\right) = \frac{1}{z}+\frac{1}{2z^{2}}
+\sum_{l=1}^{\infty }\frac{B_{2l}}{\,z^{2l+1}}  \label{eq2.8}
\end{eqnarray}
as $z\rightarrow \infty $. It will be assumed that 
each $n_{k}\left( x\right) $ has some characteristic value $\overline{n}
_{k}\left( x\right) $ which is vastly larger than nearby values:
\begin{equation}
n_{k}\left( x\right) =\overline{n}_{k}\left( x\right) +\Delta n_{k}
\left( x \right) \quad ,\quad \overline{n}_{k}\left( x\right) \ggg 
\left | \Delta n_{k}\left( x\right) \right | \label{eq2.9}
\end{equation}
where ``$\ggg$'' means ``is vastly greater than'', as in $10^{100} \ggg 1$. 
(This assumption is consistent with the fact that the initial action below
has no lower bound as $n_{k}\left( x\right) \rightarrow \infty $ before an extra stochastic term is added.) 
Then it is an extremely good approximation to use the asymptotic formulas 
above and write 
\begin{eqnarray}
\hspace{-1cm}
S=S_{0}+\sum_{x,k}a_{k}\left( x\right) \Delta n_{k}\left( x\right)
-\sum_{x,k}a_{k}^{\prime }\left( x\right) \left[ \Delta n_{k}\left( x\right) 
\right] ^{2} + \sum_{x,k,k^{\prime }\ne k}
a_{kk^{\prime }}^{\prime }\left( x\right) 
\Delta n_{k}\left( x\right) \Delta n_{k^{\prime}}\left( x\right)
\label{eq2.10}
\end{eqnarray}
\begin{eqnarray}
a_{k}\left( x\right) &=& \log \overline{N}\left( x\right) -\log \overline{n}
_{k}\left( x\right) \label{eq2.11a} \\
a_{k}^{\prime }\left( x\right) &=& \left( 2
\overline{n}_{k}\left( x\right) \right) ^{-1}
- \left( 2\overline{N}\left( x \right) \right) ^{-1}  \quad , \quad
a_{kk^{\prime }}^{\prime }\left( x\right) 
= \left( 2\overline{N}\left( x \right) \right) ^{-1}  
\label{eq2.11b}
\end{eqnarray}
where $\overline{N}\left( x\right) $ is the value of $N\left( x\right) $
when $n_{k}\left( x\right) =$ $\overline{n}_{k}\left( x\right) $ for all $k$, 
and the higher-order terms have been separately neglected in $a_{k}\left(
x\right) $ and $a_{k}^{\prime }\left( x\right) $. (The above results then also follow immediately 
from Stirling's approximation for factorials.) For simplicity, we will 
neglect the terms involving 
$\left( 2\overline{N}\left( x \right) \right) ^{-1}$, assuming $\overline{N}(x) \ggg \overline{n}_k(x)$ (as is consistent with the limit $\overline{n}_M  \rightarrow \infty$ below (\ref{eq2.1})). Since there is initially
no distinction between the fields labeled by $k$, it is consistent to assume
that they all have the same $\overline{n}_{k}\left( x\right) =\overline{n}
\left( x\right) $, and that $\overline{n}\left(
x\right) $ is independent of $x$: 
$\overline{n}\left( x\right) =\overline{n}$ and 
$\overline{N}\left( x\right) =\overline{N}$, so that 
\begin{eqnarray}
a_{k}\left( x \right) &=& a = \log \left( \overline{N}/\overline{n} \right)
\label{eq2.11c} \\
a_{k}^{\prime }\left( x\right) &=& a^{\prime } = 
\left( 2\overline{n}\right) ^{-1} \; .  \label{eq2.11d}
\end{eqnarray}

It is not conventional or convenient to deal with $\Delta n_{k}\left(
x\right) $ and $\left[ \Delta n_{k}\left( x\right) \right] ^{2}$, so let us
instead write $S$ in terms of the fields $\phi _{k}$ and their derivatives 
$\partial \phi _{k}/\partial x^{M}$ via the following procedure: First, we
can switch to a new set of points $\overline{x}$, defined to be the corners
of the $D$-dimensional hypercubes centered on the original points $x$. It is
easy to see that 
\begin{equation}
S=S_{0}+\sum_{\overline{x},k}a\left\langle \Delta n_{k}\left( x\right)
\right\rangle -\sum_{\overline{x},k}a^{\prime }\left\langle \left[ \Delta
n_{k}\left( x\right) \right] ^{2}\right\rangle  \label{eq2.12}
\end{equation}
where $\left\langle \cdots \right\rangle $ in the present context indicates
an average over the $2^{D}$ boxes labeled by $x$ which have the common
corner $\overline{x}$. Second, at each point $x$ we can write $\Delta
n_{k}=\Delta \rho _{k}a_{0}^{D}=\left( \left\langle \Delta \rho
_{k}\right\rangle +\delta \rho _{k}\right) a_{0}^{D}$, with $\left\langle
\delta \rho _{k}\right\rangle =0$: 
\begin{eqnarray}
S &=&S_{0}+\sum_{\overline{x},k}a\left\langle \left\langle
\Delta \rho _{k}\right\rangle +\delta \rho _{k}\right\rangle a_{0}^{D}-\sum_
{\overline{x},k}a^{\prime }\left\langle \left( \left\langle \Delta \rho
_{k}\right\rangle +\delta \rho _{k}\right) ^{2}\right\rangle \left(
a_{0}^{D}\right) ^{2} \label{eq2.13} \\
&=&S_{0}+\sum_{\overline{x},k}a\left\langle \Delta \rho
_{k}\right\rangle a_{0}^{D}-\sum_{\overline{x},k}a^{\prime }\left[
\left\langle \Delta \rho _{k}\right\rangle ^{2}+\left\langle \left( \delta
\rho _{k}\right) ^{2}\right\rangle \right] \left( a_{0}^{D}\right) ^{2}
\; .
\label{eq2.14}
\end{eqnarray}
Each of the points $x$ surrounding $\overline{x}$ is displaced by 
$\delta x^{M}=$ $\pm a_{0}/2$ along each of the $x^{M}$ axes, so
\begin{eqnarray}
\hspace{-1cm}
\left\langle \left( \delta \rho _{k}\right) ^{2}\right\rangle
&=&\left\langle \left( \delta \phi _{k}^{2}\right) ^{2}\right\rangle 
\label{eq2.15} \\
&=&\left\langle \sum_{M}\left( \frac{\partial \phi _{k}^{2}}{\partial x^{M}}
\delta x^{M}+\frac{1}{2}\frac{\partial ^{2}\phi _{k}^{2}}{\partial \left(
x^{M}\right) ^{2}}\left( \delta x^{M}\right) ^{2}\right) ^{2}\right\rangle 
\label{eq2.16} \\
&=&\left\langle \sum_{M}\left( 2\phi _{k}\frac{\partial \phi _{k}}{\partial
x^{M}}\delta x^{M}+\left( \frac{\partial \phi _{k}}{\partial x^{M}}\right)
^{2}\left( \delta x^{M}\right) ^{2}+\phi _{k}\frac{\partial ^{2}\phi _{k}}
{\partial \left( x^{M}\right) ^{2}}\left( \delta x^{M}\right) ^{2}\right)
^{2}\right\rangle  \label{eq2.17}
\end{eqnarray}
to lowest order, where it is now assumed that 
at normal energies the fields are slowly varying
over the extremely small distance $a_{0}$. This assumption is justified by
the prior assumption that $\overline{n}$ is extremely large: $\phi
_{k}^{2}\left( x\right) =\rho _{k}\left( x\right) =n_{k}\left( x\right)
/a_{0}^{D}$ implies that $2\delta \phi _{k}/\phi _{k}\approx \delta
n_{k}/n_{k}$ and $\phi _{k}=n_{k}^{1/2}a_{0}^{-D/2}$, so that $\delta \phi
_{k}\sim \delta n_{k}\,n_{k}^{-1/2}a_{0}^{-D/2}$. The minimum change in 
$\phi _{k}$ is given by $\delta n_{k}=1$:
\begin{eqnarray}
\delta \phi _{k}^{\min } \sim n_{k}^{-1/2} a_{0}^{-D/2}
\label{eq2.17xx}
\end{eqnarray}
which means that $\delta \phi _{k}^{\min }$ is
extremely small if $n_{k} $ is extremely large. 

In other words, the fields $\phi _{k}$ have effectively continuous values as 
$\overline{n}\rightarrow \infty $.

For extremely large $\overline{n}$ it is an extremely good approximation to
neglect the middle term in (\ref{eq2.17}), and to replace $\phi _{k}^{2}$ by 
\begin{equation}
\overline{\phi }^{2}=\overline{\rho }=\overline{n}/a_{0}^{D}
\label{eq2.17a}
\end{equation}
giving 
\begin{equation}
a^{\prime }\left\langle \left( \delta \rho _{k}\right) ^{2}\right\rangle =
\frac{1}{2a_{0}^{D}}\sum_{M}\left[ \left( \frac{\partial \phi _{k}}{\partial
x^{M}}\right) ^{2}a_{0}^{2}+\left( \frac{\partial ^{2}\phi _{k}}{\partial
\left( x^{M}\right) ^{2}}\right) ^{2}\frac{a_{0}^{4}}{16}\right] \; .
\label{eq2.18}
\end{equation}
It is similarly an extremely good approximation to neglect the term in
(\ref{eq2.14}) 
involving $a^{\prime }\left( a_{0}^{D}\right) ^{2}
\left\langle \Delta \rho _{k}\right\rangle ^{2}
= \left\langle \Delta n_{k}\right\rangle ^{2}/2\overline{n}$ in
comparison to that involving $\left\langle \Delta \rho _{k}\right\rangle
a_{0}^{D}=\left\langle \Delta n_{k}\right\rangle $, so that 
\begin{equation}
S=S_{0}+\sum_{\overline{x},k}a_{0}^{D}\frac{\mu_{0}}{m_{0}} 
\left( \phi _{k}^{2} - \overline{\phi }^{2}\right) 
-\sum_ {\overline{x},k}\sum_{M}a_{0}^{D}
\frac{1}{2m_{0}^{2}}\left[ \left( \frac{\partial
\phi _{k}}{\partial x^{M}}\right) ^{2}+\frac{a_{0}^{2}}{16}\left( \frac
{\partial ^{2}\phi _{k}}{\partial \left( x^{M}\right) ^{2}}\right) ^{2}\right]
\label{eq2.19}
\end{equation}
where 
\begin{equation}
m_{0} = a_{0}^{-1} \quad ,\quad \mu_{0} = 
m_{0} \log \left( \overline{N}/\overline{n} \right) \; .  \label{eq2.21}
\end{equation}

The philosophy behind the above treatment is simple: We essentially wish to
replace $\left\langle f^{2}\right\rangle $ by $\left( \partial f/\partial
x\right) ^{2}$, and this can be accomplished because 
\begin{equation}
\left\langle f^{2}\right\rangle -\left\langle f\right\rangle
^{2}=\left\langle \left( \delta f\right) ^{2}\right\rangle \approx
\left\langle \left( \partial f/\partial x\right) ^{2}\left( \delta x\right)
^{2}\right\rangle =\left( \partial f/\partial x\right) ^{2}
\left( a_{0}/2\right)^{2} \; . \label{eq2.22}
\end{equation}
The form of (\ref{eq2.19}) also has a simple interpretation: The entropy 
$S$ increases with the number of dits, but decreases when the dits are 
not uniformly distributed.

In the continuum limit, 
\begin{equation}
\sum_{\overline{x}}a_{0}^{D}\rightarrow \int d^{D}x   \label{eq2.23}
\end{equation}
(\ref{eq2.19}) becomes 
\begin{equation}
S=S_{0}+\int d^{D}x\,\,\sum_{k}
\left\{ \frac{\mu_{0}}{m_{0}}
\left( \phi _{k}^{2} - \overline{\phi }^{2}\right) 
-\frac{1}{2m_{0}^{2}}\sum_{M}\left[ \left( \frac{\partial \phi _{k}}{\partial
x^{M}}\right) ^{2}+\frac{a_{0}^{2}}{16}\left( \frac{\partial ^{2}\phi _{k}}
{\partial \left( x^{M}\right) ^{2}}\right) ^{2}\right] \right\} \; .
 \label{eq2.24}
\end{equation}

A physical configuration of all the fields $\phi _{k}\left( x\right) $
corresponds to a specification of all the density variations $\Delta \rho
_{k}\left( x\right) $. In the present picture, the probability of such a
configuration is proportional to $W=e^{S}$. In a path integral with Euclidean form,
the probability is proportional to $e^{-\overline{S}_{b}}$, where 
$\overline{S}_{b}$ is the action for these bosonic fields. 
We conclude that 
\begin{equation}
\overline{S}_{b}=-S+\mathrm{constant} 
\label{eq2.25}
\end{equation}
and we will choose the constant to equal $S_{0}$.

In the following it will
be convenient to write the action in terms of $\widetilde{\phi }_{k}
= m_{0}^{-1/2}\phi _{k}$. For simplicity, we assume that the number of
relevant $\widetilde{\phi }_{k}$ is even, so that we can group these real
fields in pairs to form $N_{f}$ complex fields $\widetilde{\Psi }_{b,k}$. 
It is also convenient to subtract out the enormous contribution of 
$\overline{\phi }$ by defining
\begin{eqnarray}
\Psi _{b}=\widetilde{\Psi }_{b}-\overline{\Psi }_{b}
\label{eq2.25a}
\end{eqnarray}
where $\widetilde{\Psi }_{b}$ is the vector with components $\widetilde{\Psi 
}_{b,k}$ and $\overline{\Psi }_{b}$ is similarly defined with 
$\phi_{k}\rightarrow \overline{\phi }$. Then the action can be written
\begin{eqnarray}
\hspace{-0.8cm}\overline{S}_{b}=\int d^{D}x\,\Bigg\{\frac{1}{2m_{0}}\left[ 
\frac{\partial \Psi _{b}^{\dagger }}{\partial x^{M}}\frac{\partial \Psi _{b}
}{\partial x^{M}}+\frac{a_{0}^{2}}{16}\frac{\partial ^{2}\Psi _{b}^{\dagger }
}{\partial \left( x^{M}\right) ^{2}}\frac{\partial ^{2}\Psi _{b}}{\partial
\left( x^{M}\right) ^{2}}\right] -\mu _{0}\,\left( \widetilde{\Psi }
_{b}^{\dagger }\widetilde{\Psi }_{b}-\overline{\Psi }_{b}^{\dagger }
\overline{\Psi }_{b}\right) \Bigg\}
\label{eq2.25b}
\end{eqnarray}
since $\overline{\Psi }_{b}$ is constant, with summation now implied over
repeated indices like $M$.

As described above, in Section \ref{sec:sec1a}, we now add an extra imaginary term 
$i\widetilde{V}\,\Psi _{b}^{\dagger }\Psi _{b}$ in the integral giving the action. 
Here $\widetilde{V}$ is a potential which has a Gaussian distribution, with 
\begin{equation}
\left\langle \widetilde{V}\,\right\rangle =0\quad ,\quad \left\langle 
\widetilde{V}\left( x\right) \widetilde{V}\left( x^{\prime }\right)
\right\rangle =b\,\delta \left( x-x^{\prime }\right)   
\label{eq2.26}
\end{equation}
where $b$ is a constant, with
\begin{eqnarray}
b \rightarrow 0+
\label{eq2.26xx}
\end{eqnarray}
at the end of the calculations.

Then the complete action has the form 
\begin{align}
\widetilde{S}_{B}\left[ \Psi _{b}^{\dagger },\Psi _{b}\right]
=\int d^{D}x\,\Bigg\{\frac{1}{2m_{0}}\bigg[ \frac{\partial \Psi
_{b}^{\dagger }}{\partial x^{M}}\frac{\partial \Psi _{b}}{\partial x^{M}}  +
\frac{a_{0}^{2}}{16} & \frac{\partial ^{2}\Psi _{b}^{\dagger }}{\partial \left(
x^{M}\right) ^{2}}  \frac{\partial ^{2}\Psi _{b}}{\partial \left( x^{M}\right)
^{2}}\bigg] \nonumber \\ 
& -\mu _{0}\,\left( \widetilde{\Psi }_{b}^{\dagger }\widetilde{\Psi }_{b}-
\overline{\Psi }_{b}^{\dagger }\overline{\Psi }_{b}\right)  +i\widetilde{V}
\,\Psi _{b}^{\dagger }\Psi _{b}\Bigg\} \; .
\label{eq2.25c}
\end{align}
In the following we will assume that the only fields which make an
appreciable contribution in (\ref{eq2.25c}) are those for which 
$\int d^{D}x\, \overline{\Psi } _{b}^{\dagger }\Psi _{b} = 
\overline{\Psi } _{b}^{\dagger } \int d^{D}x\, \Psi _{b} = 0$. 
This assumption is justified by the fact that $\overline{\Psi } _{b}$ 
is constant with respect to all the coordinates and, in the present
picture, fields $\Psi _{b}$ corresponding to  physical gauge representations 
have nonzero angular momenta in the internal space of Appendices \ref{sec:appA} and \ref{sec:appB}. Then (\ref{eq2.25c})
simplifies to 
\begin{eqnarray}
\hspace{-0.7cm} \widetilde{S}_{B}\left[ \Psi _{b}^{\dagger },\Psi _{b}\right]
=\int d^{D}x\,\Bigg\{\frac{1}{2m_{0}}\left[ \frac{\partial \Psi
_{b}^{\dagger }}{\partial x^{M}}\frac{\partial \Psi _{b}}{\partial x^{M}}+
\frac{a_{0}^{2}}{16}\frac{\partial ^{2}\Psi _{b}^{\dagger }}{\partial \left(
x^{M}\right) ^{2}}\frac{\partial ^{2}\Psi _{b}}{\partial \left( x^{M}\right)
^{2}}\right]  
-\mu _{0}\,\Psi_{b}^{\dagger }\Psi_{b}
 +i\widetilde{V}\,\Psi _{b}^{\dagger }\Psi _{b}\Bigg\} \; .
\label{eq2.27}
\end{eqnarray}

\section{\label{sec:sec3}Primordial supersymmetry}

If $F$ is any functional of the fundamental fields $\Psi _{b}$, 
its average value is given by 
\begin{equation}
\left\langle F\right\rangle =\left\langle \frac{\int \mathcal{D}\,\Psi
_{b}^{\dagger }\,\mathcal{D}\,\Psi _{b}\,F\left[ \Psi _{b}^{\dagger },\Psi _{b}
\right] \,e^{-\widetilde{S}_{B}\left[ \Psi _{b}^{\dagger },\Psi _{b}\right] }}
{\int \mathcal{D}\,\underline{\Psi } _{\,b}^{\dagger }\,\mathcal{D}
\,\underline{\Psi } _{\,b}
\,e^{-\widetilde{S}_{B}\left[ \underline{\Psi } _{\,b}^{\dagger },
\underline{\Psi } _{\,b}\right] }}\right\rangle   \label{eq3.1}
\end{equation}
where $\left\langle \cdots \right\rangle $ now represents an average over
the perturbing potential $i\widetilde{V} \,$ and $\int \mathcal{D}\,\Psi
_{b}^{\dagger }\,\mathcal{D}\,\Psi _{b}$ is to be interpreted as 
$\prod\nolimits_{x,k}\int_{-\infty }^{\infty }d \, 
\mathrm{Re} \, \Psi _{b,k}\left(
x\right) \,\int_{-\infty }^{\infty }d  \, 
\mathrm{Im} \, \Psi _{b,k}\left( x\right) $.
The transition from the original summation over $n_{k}\left( x\right) $ to
this Euclidean path integral has the form (with $\Delta n=1$ here) 
\begin{eqnarray}
\hspace{-0.5cm}\sum_{n=0}^{\infty }f\left( n\right) \Delta n\rightarrow
\int_{0}^{\infty }fdn\rightarrow \int_{0}^{\infty }f\;d\left( a_{0}^{D}\phi
^{2}\right) \rightarrow 2\overline{\phi }a_{0}^{D}\int_{0}^{\infty }f\;d\phi 
\rightarrow 2\overline{\phi }\,a_{0}^{D}m_{0}^{1/2}\int_{-\infty }^{\infty
}f\;d\phi ^{\prime }
\label{eq3.2}
\end{eqnarray}
where $\phi ^{\prime }=\widetilde{\phi }-m_{0}^{-1/2}\overline{\phi }$,
since $d\left( \phi ^{2}\right) \approx 2\overline{\phi }\,d\phi $ 
is an extremely good approximation for physically relevant fields, 
and since $\phi ^{\prime } $ effectively ranges from $-\infty $ 
to $+\infty $. Each $\phi ^{\prime }$ then
becomes a $\mathrm{Re} \,  \Psi _{b,k}\left( x\right) $ or 
$\mathrm{Im} \,  \Psi _{b,k}\left( x\right) $, and the constant factors 
cancel in the numerator and denominator of (\ref{eq3.1}).

The presence of the denominator makes it difficult to perform the average of
(\ref{eq3.1}), but the 
bosonic degrees of freedom $\underline{\Psi } _{\,b}$ 
in the denominator can be replaced by fermionic degrees
of freedom $\Psi _{f}$ in the numerator, using a method that is similar to those used in previous treatments of other systems~\cite{parisi,efetov,huang}:
After integration by parts (with boundary terms usually assumed 
either to vanish or to be irrelevant in this paper), (\ref{eq2.27}) can 
be written in the form 
$\widetilde{S}_{B}=\int d^{D} x \,\Psi _{b}^{\dagger } A \Psi _{b} $. 
Then, since 
\begin{equation}
\int \mathcal{D}\,\underline{\Psi } _{\,b}^{\dagger }\,\mathcal{D}\,
\underline{\Psi } _{\,b}
\,e^{-\widetilde{S}_{B}\left[ \underline{\Psi } _{\,b}^{\dagger },
\underline{\Psi } _{\,b} \right] }= C \left( \det \, \cal A \right) ^{-1}  
\label{eq3.3}
\end{equation}
\begin{equation}
\int \mathcal{D}\,\Psi _{f}^{\dagger }\,\mathcal{D}\,\Psi _{f}\,
e^{-\widetilde{S}_{B}\left[ \Psi _{f}^{\dagger },\Psi _{f}\right] }=
\det \, \cal A
\label{eq3.4}
\end{equation}
where the matrix $\cal A$ corresponds to the operator $A$ and $C$ is a
constant, it follows that 
\begin{eqnarray}
\left\langle F\right\rangle &=& \frac{1}{C} \left\langle \int \mathcal{D}\,\Psi
_{b}^{\dagger }\,\mathcal{D}\,\Psi _{b}\,\mathcal{D}\,\Psi _{f}^{\dagger }\,
\mathcal{D}\,\Psi _{f}\,F\,e^{-\widetilde{S}_{B}\left[ \Psi _{b}^{\dagger
},\Psi _{b}\right] }e^{-\widetilde{S}_{B}\left[ \Psi _{f}^{\dagger },\Psi _{f}
\right] }\right\rangle \label{eq3.5} \\
&=&  \frac{1}{C} \left\langle \int \mathcal{D}\,\Psi ^{\dagger }\,
\mathcal{D}\,\Psi
\,F\,e^{-\widetilde{S}_{bf}\left[ \Psi ^{\dagger },\Psi \right]
}\right\rangle  \label{eq3.6}
\end{eqnarray}
where $\Psi _{b}$ and $\Psi _{f}$ have been combined into 
\begin{equation}
\Psi =\left( 
\begin{array}{c}
\Psi _{b} \\ 
\Psi _{f}
\end{array}
\right)  \label{eq3.7}
\end{equation}
and
\begin{equation}
\widetilde{S}_{bf}\left[ \Psi ^{\dagger },\Psi \right] =\int d^{D}x\,\left\{ 
\frac{1}{2m_{0}}\left[ \frac{\partial \Psi ^{\dagger }}{\partial x^{M}}
\frac{\partial \Psi }{\partial x^{M}}+\frac{a_{0}^{2}}{16}\frac{\partial ^{2}
\Psi^{\dagger }}{\partial \left( x^{M}\right) ^{2}}\frac{\partial ^{2}\Psi }
{\partial \left( x^{M}\right) ^{2}}\right] -\mu_{0} \,\Psi ^{\dagger }\Psi 
+i\widetilde{V}\,\Psi ^{\dagger }\Psi \right\} \; .  \label{eq3.8}
\end{equation}
In (\ref{eq3.7}), $\Psi _{f}$ consists of Grassmann variables $\Psi _{f,k}$, 
just as $\Psi _{b}$ consists of ordinary variables $\Psi _{b,k}$, and $\int 
\mathcal{D}\,\Psi ^{\dagger }\,\mathcal{D}\,\Psi $ is to be interpreted as
\begin{eqnarray}
\prod\nolimits_{x,k}\int_{-\infty }^{\infty }d  \, \mathrm{Re} \,  
\Psi _{b,k}\left( x\right) \,\int_{-\infty }^{\infty }d  \, \mathrm{Im} \,  
\Psi _{b,k}\,\int d\,\Psi
_{f,k}^{\, \ast }\left( x\right) \,\int d\,\Psi _{f,k}\left( x\right) 
\; . 
\label{eq3.9}
\end{eqnarray}
Recall that $\Psi _{b}$ and $\Psi _{f}$ each have $N_{f}$ components. 

In this early stage of the theory, the bosonic fields in the denominator -- which are 
of critical importance for proper normalization -- have effectively been transformed into fermionic fields in the numerator, where they perform the same function. This is the origin of supersymmetry in the present picture. 

For a Gaussian random variable $v$ whose mean value is zero, the result 
\begin{equation}
\left\langle e^{-iv}\right\rangle =e^{-\frac{1}{2}\left\langle
v^{2}\right\rangle }  \label{eq3.10}
\end{equation}
implies that 
\begin{eqnarray}
\left\langle e^{-\int d^{D}x \, i  \widetilde{V}\,\Psi ^{\dagger
}\Psi }\right\rangle &=&e^{-\frac{1}{2}\int d^{D}x\,\,\,d^{D}x\,^{\prime
}\,\Psi ^{\dagger }\left( x\right) \Psi \left( x\right) \left\langle 
\widetilde{V}\left( x\right) \widetilde{V}\left( x^{\prime }\right)
\right\rangle \Psi ^{\dagger }\left( x^{\prime }\right) \Psi \left(
x^{\prime }\right) } \label{eq3.11}  \\
&=&e^{-\frac{1}{2}b\int d^{D}x\,\,\,\left[ \Psi ^{\dagger }\left( x\right)
\Psi \left( x\right) \right] ^{2}} \; . \label{eq3.12}
\end{eqnarray}
It follows that 
\begin{equation}
\left\langle F\right\rangle = \frac{1}{C} \int \mathcal{D}\,
\Psi ^{\dagger }\,\mathcal{D} \,\Psi \,F\,e^{-S_{E}}  \label{eq3.13}
\end{equation}
with
\begin{equation}
S_{E}=\int d^{D}x\,\left\{ \frac{1}{2m_{0}}\left[ \frac{\partial \Psi ^{\dagger }
}{\partial x^{M}}\frac{\partial \Psi }{\partial x^{M}}+\frac{a_{0}^{2}}{16}
\frac{\partial ^{2}\Psi ^{\dagger }}{\partial \left( x^{M}\right) ^{2}}
\frac{\partial ^{2}\Psi }{\partial \left( x^{M}\right) ^{2}}\right] -\mu_{0} \,
\Psi^{\dagger }\Psi +\frac{1}{2}b\left( \Psi ^{\dagger }\Psi \right) ^{2}
\right\} \; . \label{eq3.14}
\end{equation}
A special case (with $F=1$) is 
\begin{equation}
Z= \frac{1}{C} \int \mathcal{D}\,\Psi ^{\dagger }\,\mathcal{D}\,
\Psi e^{-S_{E}} \label{eq3.15}
\end{equation}
but according to (\ref{eq3.1}) $Z=1$, so 
$C=\int \mathcal{D}\,\Psi ^{\dagger }\,\mathcal{D}\,\Psi e^{-S_{E}} $
and
\begin{equation}
\left\langle F\right\rangle =\frac{\int \mathcal{D}\,\Psi ^{\dagger }\,
\mathcal{D}\,\Psi \,F\,e^{-S_{E}} }
{\int \mathcal{D}\,\Psi ^{\dagger }\,\mathcal{D}\,\Psi \; e^{-S_{E}}}
\; . \label{eq3.17}
\end{equation}

Again, notice that the fermionic fields $\Psi _{f}$ are effectively a transformed version of the bosonic fields $\underline{\Psi }_{\,b}$.
The coupling between the fields $\Psi _{b}$ and $\Psi _{f}$
(or $\underline{\Psi } _{\,b}$) is due to the random perturbing potential 
$i\widetilde{V}$. In the replacement of (\ref{eq3.1}) by
(\ref{eq3.17}), $F$ essentially serves as a test functional. The
meaning of this replacement is that the action (\ref{eq3.14}), with
both bosonic and fermionic fields, must be used instead of the
original action (\ref{eq2.27}), with only bosonic fields, in treating
all physical quantities and processes, if the average over random
fluctuations in (\ref{eq3.1}) is to disappear from the theory. 

Notice that the two steps above serve two independent purposes: The transformation of $\underline{\Psi }_{\,b}$ in the denominator 
to $\Psi _{f}$ in the numerator provides a more convenient formulation because all fields now have equal status in the numerator, and can be treated in the same way. The introduction of 
an infinitesimal perturbing potential is then preparation for the formation of a condensate at finite energy (in the very early universe), as discussed in Section \ref{sec:sec1a}. Of course, it is the 
conjunction of these two steps that makes the following developments possible.

The basic idea is to search for a stable path (through the space of coordinates and fields) along which $S_E$ maintains a coherent local minimum, using the representation of fields that works best for characterizing events along that path. The underlying reality consists of the states of dits \textemdash \, the ``atoms'' that constitute all of Nature \textemdash \, but we can now conveniently describe this reality with fermion and boson fields.

\section{\label{sec:appA}The internal space}

The full internal space of Section \ref{sec:sec5} is $(10+3)$-dimensional, with 
Spin(10) and SO(3) rotation groups  
and their various representations. For the 
10-dimensional space these are the 
vector $10=5+\overline{5}$, spinor $32=16+\overline{16}$, adjoint $45$, etc. representations. For the 3-dimensional space we need only the scalar and vector representations. 

It may be
helpful to begin with an analogy, in which external spacetime is
replaced by the $z$-axis. The 
internal space is replaced by an $xy$-plane, with internal states described
by $2$-dimensional vector fields (rather than the higher-dimension vector
and spinor fields considered below). One of these states is occupied by the
condensate, and is represented by a vector $\boldsymbol{v}_{1}$ which
points radially outward from the origin at all points in the
$xy$-plane when $z=0$. (This is the internal order parameter for the condensate.) 
The other state is an additional basis function, represented
by a vector $\boldsymbol{v}_{2}$ which is everywhere perpendicular to 
$\boldsymbol{v}_{1}$. But $\boldsymbol{v}_{1}$ is allowed to rotate
as a function of $z$, so it has both radial and tangential components after
a displacement along the $z$-axis. Then $\boldsymbol{v}_{2}$ is forced
to rotate with $\boldsymbol{v}_{1}$ -- i.e., the condensate -- in order
to preserve orthogonality.

Now let us turn to a $\bar{d}$-dimensional internal space (with a single rotation group), first considering a
set of $\bar{d}$-dimensional vector fields $\widetilde{\psi }_{vec}^{r}$, \textit{where, in the present context, $r$ labels a field (or basis function) rather than a component}. 

Let 
$\widetilde{\psi }_{vec}^{0}$ represent the state occupied by a hypothetical bosonic
condensate. (Since the actual primordial condensate of Section \ref{sec:sec4} is in a spinorial 16 representation, as discussed below, this is still only an illustrative analogy.) In the simplest picture, and at some fixed $x_{0}^{\mu }$, only
the $r\,$th component of the field $\widetilde{\psi }_{vec}^{r}$ is nonzero
along some radial direction in the internal space, making the fields
trivially orthogonal in that direction. Then, with $x^{\mu }$ still fixed, 
$\widetilde{\psi }_{vec}^{r}\left( x^{m}\right) $ in all other radial
directions is obtained from the original 
$\widetilde{\psi }_{vec}^{r}\left( x_{0}^{m}\right) $ by rotating it 
to $ x^{m}$. 
In other words, the field at each point in the internal space is identical
to the field that would be obtained at that point 
if the original field $\widetilde{\psi } _{vec}^{r}\left( x_{0}^{m}\right) $ 
were subjected to a rotation about the origin. This produces
an isotropic configuration for the condensate density and the magnitude-squared of each basis function. 
As in (\ref{eq4.14}) we can write
\begin{eqnarray}
\widetilde{\psi }_{vec}^{r}\left( x^{m}\right) =U_{vec}\left(
x^{m},x_{0}^{m}\right) \widetilde{\psi }_{vec}^{r}\left( x_{0}^{m}\right) \;.
\label{eq12.1}
\end{eqnarray}
Just as in the simpler analogy, a field that is radial at 
$x_{0}^{m}$ will also be radial at all other points $x^{m}$. However, 
a general $\widetilde{\psi }_{vec}^{r}\left( x_{0}^{m}\right)$ 
permits a general vortex-like configuration of the condensate (for e.g. $\bar{d}=10$).

Also as in the simpler analogy, the state $\widetilde{\psi }_{vec}^{0}$ of the
condensate is allowed to rotate as a function of $x^{\mu }$ (because such 
a rotation does not alter the internal action). Since the other
basis functions $\widetilde{\psi }_{vec}^{r}$ in the same (irreducible) representation are required to remain
orthogonal to $\widetilde{\psi }_{vec}^{0}$ and each other, they are
required to rotate with the condensate. Then (\ref{eq12.1}) becomes more 
generally
\begin{eqnarray}
\widetilde{\psi }_{vec}^{r}\left( x^{m},x^{\mu }\right) =U_{vec}
\left( x^{m},x_{0}^{m};x^{\mu },x_{0}^{\mu }\right) \widetilde{\psi }
_{vec}^{r}\left( x_{0}^{m},x_{0}^{\mu }\right) 
\label{eq12.2}
\end{eqnarray}
with
\begin{eqnarray}
\widetilde{\psi }_{vec}^{r\,\dag }\left( x^{m},x^{\mu }\right) 
\,\widetilde{\psi }_{vec}^{r^{\prime }}\left( x^{m},x^{\mu }\right) 
=\widetilde{\psi }
_{vec}^{r\,\dag }\left( x_{0}^{m},x_{0}^{\mu }\right) \,\widetilde{\psi }
_{vec}^{r^{\prime }}\left( x_{0}^{m},x_{0}^{\mu }\right) 
= \delta_{r r^{\prime }}
\label{eq12.2a}
\end{eqnarray}
since
\begin{eqnarray}
U_{vec}^{\dag }\left( x^{m},x_{0}^{m};x^{\mu },x_{0}^{\mu }\right)
U_{vec}\left( x^{m},x_{0}^{m};x^{\mu },x_{0}^{\mu }\right) =1 \; .
\label{eq12.2b}
\end{eqnarray}

In general (but first with $x^{\mu }$ fixed), let 
$\widetilde{\psi }\left( \boldsymbol{x}\right)$ 
represent a multicomponent basis function with
angular momentum $j$ at a point $\boldsymbol{x}$ in the $\bar{d}$-dimensional
internal space. As before, we define the relevant basis functions by
\begin{eqnarray}
\widetilde{\psi }\left( \boldsymbol{x}\right) =\mathcal{R}
 \,\widetilde{\psi }\left( \boldsymbol{x}
_{0}\right) \quad ,\quad \boldsymbol{x=R\,x}_{0}
\label{eq12.1f}
\end{eqnarray}
where $\mathcal{R} $ is the rotation matrix for a basis function with angular momentum $j$.
With this definition, each component $\widetilde{\psi }_{p}\left( \boldsymbol{x} \right) $
is a single-valued function of the coordinates only if $j$
is an integer. If $j=1/2$, e.g., $\widetilde{\psi }_{p}\left( \boldsymbol{x} \right) $
acquires a minus sign after a rotation of $2\pi $,
but it is single-valued on the group manifold (which is a parallelizable spin manifold -- the kind of compact manifold sought in previous Kaluza-Klein-like attempts to obtain a gauge theory from an internal space). 

Multivalued functions are 
well-known in other similar contexts, such as the behavior of the
phase of an ordinary superfluid order parameter 
$\psi _{s}=e^{i\theta _{s}}n_{s}^{1/2}$ around a
vortex, which becomes discontinuous if it is required to be a 
single-valued function of the coordinates~\cite{kleinert}. In the same
way, $z^{1/2}$ exhibits a discontinuity across a branch cut if it
is required to be a single-valued function and $z$ is restricted to a 
single complex plane. I.e., 
$z^{1/2} = \left| z \right|^{1/2} e^{i\phi /2}$ gives 
$+\left| z \right|^{1/2}$ for $\phi=0$ and 
$-\left| z \right|^{1/2}$ for $\phi=2 \pi$. But when defined on a pair
of Riemann sheets, $z^{1/2}$ is a continuous function, and 
the same is true of 
$\widetilde{\psi }\left( \boldsymbol{x}\right) $ or each component $\widetilde{\psi }_{p}\left( \boldsymbol{x} \right) $ as we
have defined it above, on the group manifold. The key idea in either 
case is to extend the manifold over which the function is defined, 
so that there are no artificial discontinuities.
A similar principle holds in condensed matter physics, where a spinor 
can be a multivalued function of position.

The above discussion, however, is only relevant to mathematical consistency. 
Although the mathematical quantity $\widetilde{\psi }_{p}\left( \boldsymbol{x} \right) $ can be multivalued, the more fundamental and physically significant 
quantity $\widetilde{\psi }_{p}^{\, \ast}\left( \boldsymbol{x} \right) \widetilde{\psi }_{p}\left( \boldsymbol{x} \right) $ 
is a single-valued function of position. The same is true regarding the SU(2)$\times$U(1) order parameter and fields in external spacetime.

In the 10-dimensional internal space, spinorial fields $\widetilde{\psi }_{sp}^{r}$ are appropriate basis functions for standard model fermions and their sfermions, and for the primordial bosonic condensate of Section \ref{sec:sec4}.
Again, let $\widetilde{\psi }_{sp}^{r}\left(
x_{0}^{m}\right) $ represent a field along some radial direction in the
internal space at some fixed  $x_{0}^{\mu }$. Then the field configuration 
for every point $x^{m}$ is obtained by taking 
$\widetilde{\psi }_{sp}^{r}\left( x^{m}\right) $ 
to be identical to the field that would be obtained at that point if 
$\widetilde{\psi }_{sp}^{r}\left( x_{0}^{m}\right) $ were subjected 
to a rotation, with 
\begin{eqnarray}
\widetilde{\psi }_{sp}^{r}\left( x^{m}\right) =U_{sp}\left(
x^{m},x_{0}^{m}\right) \widetilde{\psi }_{sp}^{r}\left( x_{0}^{m}\right) 
\label{eq12.3}
\end{eqnarray}
as in (\ref{eq12.1f}).

Again, the state $\widetilde{\psi }_{sp}^{0}$ of the condensate is allowed
to rotate as a function of $x^{\mu }$, and since the other basis functions 
$\widetilde{\psi }_{sp}^{r}$ in the same 16 representation must remain orthogonal to 
$\widetilde{\psi }_{sp}^{0}$ they are required to rotate 
with the condensate. The general version of (\ref{eq12.3}) is then
\begin{eqnarray}
\widetilde{\psi }_{sp}^{r}\left( x^{m},x^{\mu }\right) =U_{sp}\left(
x^{m},x_{0}^{m};x^{\mu },x_{0}^{\mu }\right) \widetilde{\psi }
_{sp}^{r}\left( x_{0}^{m},x_{0}^{\mu }\right) \;. 
\label{eq12.4}
\end{eqnarray}

The same reasoning applies to each representation, and thus to
the combined set of all physically relevant basis functions $\widetilde{\psi }_{int}^{r}\left( x^{m},x^{\mu
}\right) $, which are defined to rotate with the primordial condensate in the internal space (just as the $\widetilde{\psi} ^r_{f}\left( x^{\mu }\right)$ of (\ref{eq4.36}) are defined to rotate with the condensate in external spacetime):
\begin{eqnarray}
\widetilde{\psi }_{int}^{r}\left( x^{\prime \, m},x^{\prime \, \mu }\right)
=U_{int}\left( x^{\prime \, m},x^{m};x^{\prime \, \mu },x^{\mu }\right) 
\widetilde{\psi }_{int}^{r}\left( x^{m},x^{\mu }\right) 
\label{eq12.5}
\end{eqnarray}
with
\begin{eqnarray}
\widetilde{\psi }_{int}^{r\,\dag }\left( x^{\prime \,m},x^{\prime \,\mu
}\right) \,\widetilde{\psi }_{int}^{r^{\prime }}
\left( x^{\prime \,m},x^{\prime \,\mu
}\right) =\widetilde{\psi }_{int}^{r\,\dag }\left( x^{m},x^{\mu }\right) \,
\widetilde{\psi }_{int}^{r^{\prime }}\left( x^{m},x^{\mu }\right) 
= \delta_{r r^{\prime }} \; .
\label{eq12.6}
\end{eqnarray}

So that the internal action will be unaffected as $x^{\mu}
\rightarrow x^{\prime \,\mu }$, we require that the order parameter
experience a uniform rotation, described by a matrix $\overline{\mathcal{R}}
_{int}$ which is independent of $x^{m}$. Then $U_{int}$ has the form
\begin{eqnarray}
U_{int}\left( x^{\prime \, m},x^{m};x^{\prime \, \mu },x^{\mu }\right)  =
\overline{\mathcal{R}}_{int}\left( x^{\prime \,\mu },x^{\mu }\right) \,
\mathcal{R}_{int}\left( x^{\prime \,m},x^{m}\right) 
\label{eq12.7}
\end{eqnarray}
with the special case
\begin{eqnarray}
\widetilde{\psi }_{int}^{r}\left( x^{m},x^{\mu }\right) =\overline{
\mathcal{R}}_{int}\left( x^{\mu },x_{0}^{\mu }\right) \,\widetilde{\psi 
}_{int}^{r}\left( x^{m},x_{0}^{\mu }\right) \;.
\label{eq12.8}
\end{eqnarray}

In (\ref{eq12.8}) the order parameter (at each fixed $x^m$) has been rotated as $x_0^{\mu} \rightarrow x^{\mu}$.
We define the parameters $\delta \overline{\phi }_{i}$ by 
\begin{eqnarray}
\overline{\mathcal{R}}_{int}\left( x^{\mu}+\delta x^{\mu },x_{0}^{\mu}
\right) = \overline{\mathcal{R}}_{int}\left( x^{\mu}, x_{0}^{\mu } \right) 
\left( 1-i\,\delta \overline{\phi }_{i}\,J_{i}  \right)
\label{eq12.8b}
\end{eqnarray}
or
\begin{eqnarray}
\delta \widetilde{\psi }_{int}^{r}\left( x^{m}, x^{\mu}\right) &=& -i\,\delta 
\overline{\phi }_{i}\,\overline{J}_{i}\,\widetilde{\psi }_{int}^{r}
\left( x^{m} , x^{\mu}\right) \quad \text{as}\quad 
x^{\,\mu }\rightarrow x^{\,\mu }+\delta x^{\,\mu }
\label{eq12.9} \\
\overline{J}_{i} &=& 
\overline{\mathcal{R}}_{int}\left( x^{\mu}, x_{0}^{\mu } \right) J_{i} 
\,\overline{\mathcal{R}}_{int}^{-1}\left( x^{\mu}, x_{0}^{\mu } \right)
\label{eq12.9a} 
\end{eqnarray}
where the matrices $J_{i}$ are the generators in the reducible representation 
corresponding to all of the $\widetilde{\psi }_{int}^{r}$ (in all physical representations).  
The matrix elements of $\overline{J}_{i}$ are independent of
$x^{\mu}$:
\begin{eqnarray}
\hspace{-1.6cm} \int d^{\bar{d}}x\,\widetilde{\psi }_{int}^{r\dagger } 
\left( x^{m},x^{\mu }\right) \overline{J}_{i} \,
\widetilde{\psi }_{int}^{r^{\prime}} \left( x^{m},x^{\mu }\right)
= \int d^{\bar{d}}x\,\widetilde{\psi }_{int}^{r\dagger } 
\left( x^{m},x_{0}^{\mu }\right) J_{i} \,
\widetilde{\psi }_{int}^{r^{\prime}} \left( x^{m},x_{0}^{\mu }\right) 
\; .
\label{eq12.9b}
\end{eqnarray}
The primordial condensate is in a specific representation, but the
basis functions in other representations are chosen to rotate with it 
according to (\ref{eq12.8}) and (\ref{eq12.9}). 

It may be helpful to illustrate the above ideas by returning to the
$2$-dimensional analogy. Equation (\ref{eq12.1f}) becomes  
\begin{eqnarray}
\hspace{-1.5cm}
\boldsymbol{v}\left( \boldsymbol{x}\right) =\mathcal{R}_{vec}\,
\boldsymbol{v}\left( \boldsymbol{x}_{0}\right) \;,\;\mathcal{R}_{vec}
=\left( 
\begin{array}{cc}
\cos \phi  & -\sin \phi  \\ 
\sin \phi  & \cos \phi 
\end{array}
\;\right) \;,\;\boldsymbol{v}\left( \boldsymbol{x}_{0}\right) =\left( 
\begin{array}{c}
R\left( r\right)  \\ 
0
\end{array}
\right) \;\mathrm{or}\;\left( 
\begin{array}{c}
0 \\ 
R\left( r\right) 
\end{array}
\right) 
\label{eq12.10}
\end{eqnarray}
for the vector representation and 
\begin{eqnarray}
s\left( \boldsymbol{x}\right) =\mathcal{R}_{sp}\,s\left( \boldsymbol{x}
_{0}\right) \;,\;\mathcal{R}_{sp}=e^{-i\sigma _{3}\phi /2}\;,\;s\left( 
\boldsymbol{x}_{0}\right) =\left( 
\begin{array}{c}
R\left( r\right)  \\ 
0
\end{array}
\right) \;\mathrm{or}\;\left( 
\begin{array}{c}
0 \\ 
R\left( r\right) 
\end{array}
\right) 
\label{eq12.11}
\end{eqnarray}
for the spinor representation. The matrices corresponding to the 
$J_{i}$ are 
\begin{eqnarray}
J_{vec}=\left( 
\begin{array}{cc}
0 & -i \\ 
i & 0
\end{array}
\;\right) \quad \text{and }\quad J_{sp}=\frac{\sigma _{3}}{2}=\frac{1}{2}
\left( 
\begin{array}{cc}
1 & 0 \\ 
0 & -1
\end{array}
\;\right) \;.
\label{eq12.13}
\end{eqnarray}

Notice that $\phi_{i}$ is an angular coordinate in the internal space,
whereas $\overline{\phi }_{i}$ is a parameter specifying the rotation of 
$\widetilde{\psi }_{int}^{r}$ at each fixed $x^{m}$ as $x^{\mu }$ is varied.

\section{\label{sec:appB}Solutions in the internal space}

Our goal in this appendix is merely to show that there are solutions with
the form required in Appendix \ref{sec:appA}, so we will look first for 
solutions with the higher-derivative 
terms in (\ref{eq5.3}) and (\ref{eq5.4}) neglected, and with $\Psi ^0_{int}$
sufficiently small that $V_{0} \left( x^{m}\right) $ 
can also be neglected. Then (\ref{eq5.3}) and (\ref{eq5.4}) become  
\begin{eqnarray}
\left( -\frac{1}{2m_{0}}\partial _{m}\partial _{m}-\mu _{int}
\right)\Psi ^0_{int}\left( x^{m},x^{\mu }\right) 
= 0 \\
\left( -\frac{1}{2m_{0}}
\partial _{m}\partial _{m}-\mu _{int}\right) \widetilde{\psi }
_{int}^{r}\left( x^{m},x^{\mu }\right) =0 \; .
\label{eq11.1}
\end{eqnarray}

For simplicity of notation, let $\widetilde{\psi }_{int}^{r}\left(
x^{m},x^{\mu }\right) $ again be represented by 
$\widetilde{\psi }\left( \boldsymbol{x} \right) $, with components 
$\widetilde{\psi }_{p}\left( \boldsymbol{x} \right) $. Each component 
varies with position in the way specified by (\ref{eq12.1f}) (together
with the radial dependence of 
$\widetilde{\psi }\left( \boldsymbol{x} _{0}\right) $). 
It therefore has an orbital
angular momentum given by the usual orbital angular momentum 
operators $\widehat{J}_{i}$ in $\bar{d}$ 
dimensions~\cite{narumi,gallup,louck,herrick,dong,lee}, which essentially
measure how rapidly 
$\widetilde{\psi }_{p}\left( \boldsymbol{x} \right) $ 
varies as a function of the angles $\phi _{i}$.
(We could here redefine the generators $J_i$ of rotations to be operators acting on components, rather than matrices acting on multicomponent fields, as is done elsewhere in the text for the generators $t_i$ of gauge transformations and the generators in $\nabla_{\mu}$ for local Lorentz transformations, but explicitly writing $\widehat{J}_{i}$ may be clearer in the present context.)

The Laplacian $\partial _{m}\partial _{m}$ can be rewritten in terms of 
radial derivatives and the usual $\widehat{J}^{2}$, 
giving~\cite{narumi,gallup,louck}  
\begin{eqnarray}
\left( -\frac{1}{r^{2K}}\frac{\partial}{\partial r}\left( r^{2K}
\frac{\partial}{\partial r}\right) +\frac{
\widehat{J}^{2}}{r^{2}}-1\right) \widetilde{\psi }_{p}\left( \boldsymbol{x}
\right) =0\quad ,\quad K=\frac{\bar{d}-1}{2}
\label{eq11.3}
\end{eqnarray}
after rescaling of the radial coordinate $r$, where $\bar{d}$ is again the number of dimensions in this internal space.

 In addition, it is shown in
Narumi and Nakau~\cite{narumi}, Gallup~\cite{gallup}, and Louck~\cite{louck} that
\begin{eqnarray}
\widehat{J}^{2}\widetilde{\psi }_{p}\left( \boldsymbol{x}\right) 
=j\left( j+\bar{d}-2\right) \widetilde{\psi }_{p}\left( \boldsymbol{x}\right)
\label{eq11.4}
\end{eqnarray}
where $j$ is the orbital angular momentum quantum number, as defined on 
p. 677 of Gallup, but with this definition extended to half-odd-integer
values of $m_{\alpha}$ and $j$. Normally, of course, only 
integer values of these orbital quantum numbers are permitted. 
However, the functions $\widetilde{\psi }_{p}\left( \boldsymbol{x}\right) $
as defined in \ref{sec:appA} can have $j=1/2$ etc. (in which case 
they are multivalued functions of the coordinates but single-valued
functions on the group manifold, as discussed below (\ref{eq12.1f})). 
Also, the demonstration of (\ref{eq11.4}) of Gallup~\cite{gallup} 
can be extended in the present context to half-odd-integer 
$j$, because it employs raising and lowering operators. 
(At each $\boldsymbol{x}$, $\widetilde{\psi }_{p}$ is a linear combination 
of states with different values of $m_{\alpha}$, but (\ref{eq11.4})
still holds.) For each 
$\widetilde{\psi }_{p}\left( \boldsymbol{x}\right) $
the radial wavefunction then satisfies 
\begin{eqnarray}
\left[ -\frac{1}{r^{2K}}\frac{d}{dr}\left( r^{2K}\frac{d}{dr}\right) 
+\frac{j\left( j+\bar{d}-2\right) }{r^{2}}-1\right] R\left( r\right) =0\;.
\label{eq11.5}
\end{eqnarray}

It may be helpful once again to consider the $2$-dimensional analogy 
of \ref{sec:appA}, where the orbital angular momentum operator 
is
\begin{eqnarray}
\widehat{J}=-i \partial / \partial \phi \; .
\label{eq12.12}
\end{eqnarray}
For the vector representation, (\ref{eq12.10}) implies
\begin{eqnarray}
\partial _{m}\partial _{m}\boldsymbol{v}\left( \boldsymbol{x}\right)  &=&
\left[ \frac{1}{r}\frac{\partial}{\partial r}
\left( r\frac{\partial}{\partial r}\right) + 
\frac{1}{r^{2}}\frac{\partial ^{2}}{\partial \phi ^{2}}\right] \left( 
\begin{array}{cc}
\cos \phi  & -\sin \phi  \\ 
\sin \phi  & \cos \phi 
\end{array}
\;\right) \boldsymbol{v}\left( \boldsymbol{x}_{0}\right)  \\
&=&\left[ \frac{1}{r}\frac{\partial}{\partial r}
\left( r\frac{\partial}{\partial r}\right) - 
\frac{1}{r^{2}}\right] \boldsymbol{v}\left( \boldsymbol{x}\right) 
\label{eq11.5b}
\end{eqnarray}
in agreement with (\ref{eq11.5}) for $j=1$. For the spinor representation, 
(\ref{eq12.11}) gives
\begin{eqnarray}
\partial _{m}\partial _{m}s\left( \boldsymbol{x}\right)  &=&
\left[ \frac{1}{r}\frac{\partial}{\partial r}
\left( r\frac{\partial}{\partial r}\right) + 
\frac{1}{r^{2}}\frac{\partial ^{2}}{\partial \phi ^{2}}\right] e^{-i\sigma
_{3}\phi /2} \, s \left( \boldsymbol{x}_{0}\right)  \\
&=&\left[ \frac{1}{r}\frac{\partial}{\partial r}
\left( r\frac{\partial}{\partial r}\right) - 
\frac{1/4}{r^{2}}\right] s\left( \boldsymbol{x}\right) 
\label{eq11.5d}
\end{eqnarray}
in agreement with (\ref{eq11.5}) for $j=1/2$. 

Equation (\ref{eq11.5}) can be further reduced to~\cite{herrick,lee} 
\begin{eqnarray}
\left[ -\frac{d^{2}}{dr^{2}}+\frac{k\left( k-1\right) }{r^{2}}-1\right] \chi
\left( r\right) =0\quad ,\quad k=j+K=j+\frac{\bar{d}-1}{2}\;
\label{eq11.6}
\end{eqnarray}
where $\chi \left( r\right) \equiv r^{K}R\left( r\right) \;.$
It is then easy to show that 
\begin{eqnarray}
\chi \left( r\right) \propto r^{k}\quad \mathrm{as}\;r\rightarrow 0\quad
,\quad \chi \left( r\right) \propto \sin \left( r + \delta \right)
\quad \mathrm{as}\;r\rightarrow
\infty 
\label{eq11.7}
\end{eqnarray}
where $\delta$ is a phase. 

The higher derivatives in the full internal wave equation (\ref{eq5.4})
permit exponentially decaying solutions which are then normalizable  
and have finite action. (For completeness, it may be worth noting that the resulting 
differential equation would be even higher than fourth-order if higher than second-order terms 
were kept in the expansion of (\ref{eq2.16}). Also, an exact treatment would require a difference rather than differential equation, but this approximate treatment establishes the principle.) Suppose that the above equation at large $r$ is
modified  to
\begin{eqnarray}
\left( a\frac{d^{4}}{dr^{4}}-b\frac{d^{2}}{dr^{2}}-1\right) \chi
\left( r\right) =0 \; .
\label{equa11.1}
\end{eqnarray}
The solutions are 
\begin{eqnarray}
\chi \left( r\right) \propto e^{iq\,r}\quad ,\quad q^{2}=-\frac{b}{2a}\pm \frac{\sqrt{b^2+4a}}{2a} \; .
\label{equa11.2}
\end{eqnarray}
There is then an exponentially decaying solution with the form
$q=i/\alpha $ and
\begin{eqnarray}
\chi \left( r\right) \propto e^{-\,r/\alpha }
\label{equa11.4}
\end{eqnarray}
so both the order parameter and the basis functions fall to zero as $r \rightarrow \infty$.

In general, differential equations with higher derivatives will asymptotically have exponentially decaying, exponentially growing, and oscillating solutions. This provides a natural explanation for the picture that our universe has a very large external spacetime and a very small internal space: The solution for the internal condensate decays exponentially, while the external condensate is born in a nondecaying solution. Here the condensate and basis functions in the internal space are confined on an extremely small length scale of roughly $\text{\textit{few}} \times a_0$, so that the action is finite at every point in external spacetime.

\section{\label{sec:appC}Euclidean and Lorentzian Propagators}

For Weyl fermions, the Euclidean 2-point function is
\begin{eqnarray}
&&G_{f}\left( x_{1},x_{2}\right)  
= \left\langle \psi _{f}\left( x_{1}\right)
\psi _{f}^{\dag }\left( x_{2}\right) \right\rangle  
=\frac{\int \mathcal{D}\,\psi _{f}^{\dagger }\,\mathcal{D}\,\psi
_{f}\,\psi _{f}\left( x_{1}\right) \psi _{f}^{\dag }\left( x_{2}\right)
\,e^{-S_{f}}}{\int \mathcal{D}\,\psi _{f}^{\dagger }\,\mathcal{D}\,\psi
_{f}\;e^{-S_{f}}} \label{eq13.1}\\
\hspace{-1.6cm} &=&\frac{\prod\nolimits_{s}
\int d \overline{\psi }_{f}^{\, \ast } \left(
s\right) \int d\overline{\psi }_{f}\left( s\right) \,e^{-\overline{\psi }
_{f}^{\, \ast }\left( s\right) a\left( s\right) \overline{\psi }_{f}\left(
s\right) }\sum\nolimits_{s_{1},s_{2}}\overline{\psi }_{f}\,\left(
s_{1}\right)  \overline{\psi }_{f}^{\, \ast }\left( s_{2}\right) U\left(
x_{1},s_{1}\right) \,U^{\dag }\left( x_{2},s_{2}\right) }{
\prod\nolimits_{s}\int d \overline{\psi }_{f}^{\, \ast }\left( s\right) \int 
d \overline{\psi }_{f}\left( s\right) 
\,e^{-\overline{\psi }_{f}^{\, \ast
}\left( s\right) a\left( s\right) \overline{\psi }_{f}\left( s\right) }} \nonumber
\end{eqnarray}
where (\ref{eq6.2}) and (\ref{eq6.3}) have been used. In a term with  
$s_{2}\neq s_{1}$, the numerator contains the factor
\begin{eqnarray}
\int d \overline{\psi }_{f}^{\, \ast }\left( s_{1}\right) \int d\overline{\psi 
}_{f}\left( s_{1}\right) \,e^{- \overline{\psi }_{f}^{\, \ast }\left(
s_{1}\right) a\left( s_{1}\right) \overline{\psi }_{f}\left( s_{1}\right) }
\overline{\psi }_{f}\left( s_{1}\right) =0
\label{eq13.3}
\end{eqnarray}
according to the rules for Berezin integration. But a term with $s_{2}=s_{1}$
contributes
\begin{eqnarray}
\hspace{-1.0cm} \frac{\int d \overline{\psi }_{f}^{\, \ast }\left( s_{1}\right) \int 
d \overline{\psi }_{f}\left( s_{1}\right) 
\,e^{-\overline{\psi }_{f}^{\, \ast
}\left( s_{1}\right) a\left( s_{1}\right) \overline{\psi }_{f}\left(
s_{1}\right) }\overline{\psi }_{f}\left( s_{1}\right) \overline{\psi }
_{f}^{\, \ast }\left( s_{1}\right) }{\int d\overline{\psi}_{f}^{\, \ast
}\left( s_{1}\right) \int d\overline{\psi }_{f}\left( s_{1}\right) \,e^{-
\overline{\psi }_{f}^{\, \ast }\left( s_{1}\right) a\left( s_{1}\right) 
\overline{\psi }_{f}\left( s_{1}\right) }} \; U\left( x_{1},s_{1}\right)
\,U^{\dag }\left( x_{2},s_{1}\right)  \nonumber \\
=a\left( s_{1}\right) ^{-1}U\left( x_{1},s_{1}\right) \,U^{\dag }\left(
x_{2},s_{1}\right) 
\label{eq13.4}
\end{eqnarray}
so
\begin{eqnarray}
G_{f}\left( x_{1},x_{2}\right) =\sum\nolimits_{s}\overline{G}_{f}\left(
s\right) U\left( x_{1},s\right) \,U^{\dag }\left( x_{2},s\right) 
\quad , \quad 
\overline{G}_{f}\left( s\right) =a\left( s\right) ^{-1} \; .
\label{eq13.5}
\end{eqnarray}
If the $U\left( x,s\right)$ used to represent 
$\psi _{f}\left( x \right)$ are a complete set, the
propagator $G_{f}\left( x,x' \right)$ is a true Green's function: 
\begin{eqnarray}
L_{f} \left( x \right) U\left( x,s\right) = a \left( s\right)
U\left( x,s\right) \quad , \quad 
\psi _{f}\left( x\right) = \sum_{s} U\left( x,s\right) \,
\overline{\psi }_{f}\left( s\right)  
\label{eq13.6a}
\end{eqnarray}
and $\sum\nolimits_{s}
U\left( x,s\right) \,U^{\dag }\left( x',s \right) =
\delta \left( x-x' \right) $ imply that
\begin{eqnarray}
L_{f} \left( x \right) G_{f}\left( x,x' \right) = 
\delta \left( x-x' \right) 
\label{eq13.7}
\end{eqnarray}
as usual.

The treatment for scalar bosons is similar: For $S_b > 0$,
\begin{eqnarray}
\hspace{-1.6cm}&&G_{b}\left( x_{1},x_{2}\right)  =\left\langle \phi _{b}\left( x_{1}\right)
\phi _{b}^{\dag }\left( x_{2}\right) \right\rangle  
=\frac{\int \mathcal{D}\,\phi _{b}^{\dagger }\,\,\mathcal{D}\,\phi
_{b}\,\phi _{b}\left( x_{1}\right) \phi _{b}^{\dagger }\,\left( x_{2}\right)
\,e^{-S_{b}}}{\int \mathcal{D}\,\phi _{b}^{\dagger }\,\mathcal{D}\,\phi
_{b}\;e^{-S_{b}}} 
\label{eq13.8} \\
&=&\frac{\prod\nolimits_{s}\int\nolimits_{-\infty }^{\infty }
d \, \mathrm{Re} \, 
\overline{\phi }_{b}\left( s\right) \int\nolimits_{-\infty }^{\infty }
d \, \mathrm{Im} \, 
\overline{\phi }_{b}\left( s\right) \,e^{-b\left(
s\right) \left[ \left( \mathrm{Re} \,  \overline{\phi }_{b}
\left( s\right) \right)^{2}+\left(\mathrm{Im} \, \overline{\phi }_{b}
\left( s\right) \right) ^{2}\right] }
\sum\nolimits_{s_{1},s_{2}}\overline{\phi }_{b}\left( s_{1}\right) 
\overline{\phi }_{b}^{\, \ast }\,\left( s_{2}\right) }
{\prod\nolimits_{s}\int\nolimits_{- \infty }^{\infty }
d \, \mathrm{Re} \, \overline{\phi }_{b}\left( s\right)
\int\nolimits_{-\infty }^{\infty }
d \, \mathrm{Im} \, \overline{\phi }_{b}\left(
s\right) \,e^{-b\left( s\right) \left[ 
\left( \mathrm{Re} \, \overline{
\phi }_{b}\left( s\right) \right) ^{2}+
\left( \mathrm{Im} \, \overline{\phi }
_{b}\left( s\right) \right) ^{2}\right] }} \nonumber \\
&& \hspace{8cm}
\times U_{b}\left( x_{1},s_{1}\right)\,  
U_{b}^{\dag }\left( x_{2},s_{2}\right) 
\label{eq13.9}
\end{eqnarray}
where
\begin{eqnarray}
L_{b} \left( x \right) U_{b}\left( x,s\right) = b\left( s\right)
U_{b}\left( x,s\right) \quad , \quad 
\phi _{b}\left( x\right) = \sum_{s} U_{b}\left( x,s\right) \,
\overline{\phi }_{b}\left( s\right)    \; .
\label{eq13.9a}
\end{eqnarray}

In a term with $s_{2}\neq s_{1}$, the numerator contains the factor
\begin{eqnarray}
\int\nolimits_{-\infty }^{\infty }
d \, \mathrm{Re} \, \overline{\phi }_{b}\left(
s_{1}\right) \int\nolimits_{-\infty }^{\infty }
d \, \mathrm{Im} \, \overline{\phi }
_{b}\left( s_{1}\right) \,e^{-b\left( s_{1}\right) \left[ \left( 
\mathrm{Re} \, \overline{\phi }_{b}\left( s_{1}\right) \right) ^{2}
+\left( \mathrm{Im} \, 
\overline{\phi }_{b}\left( s_{1}\right) \right) ^{2}\right] }
\left[\mathrm{Re} \, \overline{\phi }_{b}\left( s_{1}\right) 
+i \, \mathrm{Im} \, \overline{\phi }_{b}\left( s_{1}\right) \right] 
\nonumber \\
 =0 \hspace{1cm}
 \label{eq13.10}
\end{eqnarray}
since the integrand is odd. But a term with $s_{2}=s_{1}$ contains
the factor
\begin{eqnarray}
\hspace{-1cm}
\frac{\int\nolimits_{-\infty }^{\infty }
d \, \mathrm{Re} \, \overline{\phi }
_{b}\left( s_{1}\right) e^{-b\left( s_{1}\right) 
\left( \mathrm{Re} \, 
\overline{\phi }_{b}\left( s_{1}\right) \right) ^{2}}
\left( \mathrm{Re} \, \overline{
\phi }_{b}\left( s_{1}\right) \right) ^{2}}{\int\nolimits_{-\infty }^{\infty}
d \, \mathrm{Re} \, \overline{\phi }_{b}\left( s_{1}\right) 
\,e^{-b\left( s_{1}\right) 
\left( \mathrm{Re} \, \overline{\phi }_{b}\left( s_{1}\right) \right) ^{2}}}
+\frac{\int\nolimits_{-\infty }^{\infty }
d \, \mathrm{Im} \, \overline{\phi }
_{b}\left( s_{1}\right) e^{-b\left( s_{1}\right) 
\left( \mathrm{Im} \, 
\overline{\phi }_{b}\left( s_{1}\right) \right) ^{2}}
\left( \mathrm{Im} \, \overline{
\phi }_{b}\left( s_{1}\right) \right) ^{2}}{\int\nolimits_{-\infty }^{\infty}
d \, \mathrm{Im} \, \overline{\phi }_{b}\left( s_{1}\right) 
\,e^{-b\left( s_{1}\right) 
\left( \mathrm{Im} \, \overline{\phi }_{b}\left( s_{1}\right) \right) ^{2}
}} \nonumber \\
= b\left( s_{1}\right) ^{-1} \hspace{1cm}
\label{eq13.11}
\end{eqnarray}
so
\begin{eqnarray}
G_{b}\left( x_{1},x_{2}\right) =\sum\nolimits_{s}\overline{G}_{b}
\left( s\right) U_{b}\left( x_{1},s\right) \, U^{\dag }_{b}
\left( x_{2},s\right) 
\quad , \quad 
\overline{G}_{b}\left( s\right) =b\left( s\right) ^{-1}  \; .
\label{eq13.12} 
\end{eqnarray}
($S_b < 0$ has to be treated with the ideas discussed in Appendix \ref{sec:appD}. As usual, $a\left( s\right) $ and $b\left( s\right) $ contain a 
$ +i\epsilon $ which is associated with a convergence factor in the path
integral and which gives a well-defined inverse corresponding to a causal propagator.)

The above are the propagators in the Euclidean formulation. The Lorentzian
propagators are obtained through the same procedure with 
$a\left( s\right) \rightarrow -ia\left(
s\right) $ and $b\left( s\right) \rightarrow -ib
\left( s\right) $:
\begin{eqnarray}
\overline{G}_{f}^{L}\left( s\right)  = ia\left( s\right) ^{-1} 
\quad , \quad 
\overline{G}_{b}^{L}\left( s\right)  = ib\left( s\right)
^{-1} \; .
\label{eq13.15}
\end{eqnarray}
The propagators in the Euclidean and Lorentzian formulations thus differ by
only a factor of $i$. More generally, in the present picture, the action,
fields, operators, classical equations of motion, quantum transition
probabilities, propagation of particles, and meaning of time are the same in
both formulations.

For a single noninteracting bosonic field with a mass $m_{b}$, the basis
functions are
\begin{eqnarray}
U_{b}\left( x,p\right) =\mathcal{V}^{-1/2}e^{ip\cdot x}=\mathcal{V}
^{-1/2}e^{-i\omega t}e^{i\overrightarrow{p}\cdot \overrightarrow{x}}
\label{eq13.16}
\end{eqnarray}
so with $s\rightarrow p$ we have
$b\left( p\right) =\omega ^{2}-\left\vert \overrightarrow{p}
\right\vert ^{2}-m_{b}^{2}+i\epsilon $
and
\begin{eqnarray}
\overline{G}_{b}\left( p\right)  = \frac{1}{\omega ^{2}-\left\vert 
\overrightarrow{p}\right\vert ^{2}-m_{b}^{2}+i\epsilon } \quad , \quad
\overline{G}_{b}^{L}\left( p\right)  = \frac{i}{\omega ^{2}-\left\vert 
\overrightarrow{p}\right\vert ^{2}-m_{b}^{2}+i\epsilon } \; .
\label{eq13.19}
\end{eqnarray}

\section{\label{sec:newapp}First steps in transformation from primitive to physical boson fields}

A reminder on notation: The initial gauge group is Spin(10), with spin 1/2 fermions belonging to spinorial $32=16+ \overline{16}$ representations (standard model fermions plus new predicted ones), vectorial $10 = 5 + \overline{5}$ representations (higgsinos), and the adjoint $45$ representation  (gauginos). The fundamental covariant derivative (in an inertial frame)  is $D_{\mu } = \partial_{\mu} - i A_{\mu}$,  where the coupling constant is absorbed into the gauge potentials $A_{\mu}=A_{\mu}^i t_i$ and the generators $t_i$ are treated as operators. The field strength tensor is $F_{\mu \nu} $, with $\mu, \nu = 0,1,2,3$. Summations are implied over coordinate and gauge-field indices like $\mu$, $k$, and $i$, but in the following not  labels of other fields like $r$. The metric tensor has the form $diag \left(-1, 1, 1, 1 \right)$. In $\sigma^{\mu}$ and  $\overline{\sigma }^{\mu}$, the $\sigma^k$ are Pauli matrices, $\sigma^0$ is the $2 \times 2$ identity matrix, $\overline{\sigma }^{0}=\sigma ^{0}$, and $\overline{\sigma }^{k}=-\sigma ^{k}$. 

All spin 1/2 fields are initially left-handed. But we will now show that a given left-handed field $\psi _L$ can be transformed into a right-handed charge-conjugate field $\psi_R$, with a result that is standard for fermions but with a minus sign acquired for bosons:
\begin{align}
 i \psi_L^{\dag } \, \overline{\sigma} ^{\mu} D_{\mu } \, \psi _L
\;  \longrightarrow \;  \pm \,  i \psi_R^{c \, \dag} \, \sigma ^{\mu} D_{\mu } \, \psi_R^c  \quad , 
\quad  \psi_R^c = - \sigma^2 C \psi_L^{*} \quad , 
\quad  \psi_L = \sigma^2 C \psi_R^{c \, *}
\label{conj}
\end{align}
where $C=C^{\dag}=C^{-1}$ represents charge conjugation and is here treated as an operator. We will use
\begin{align}
 \quad \sigma^2 \sigma ^k \sigma^2 = - \sigma ^{k \, *} \quad  \text{or} \quad \sigma^2 \overline{\sigma} ^{\mu} \sigma^2 = \sigma ^{\mu \, *}  \label{A50}
\end{align}
and 
\begin{align}
\quad  C A_{\mu} C = - A_{\mu}^* 
\end{align}
(as in e.g. the treatment on pp. 56-56 of \cite{Nanopoulos-1979} or pp. 36-37 of \cite{Ozer} in the case of a spinorial representation)
Here the potentials $A_{\mu}^i$ are real and the generators $ t_i $ Hermitian, with 
\begin{align}
C A_{\mu}^i  C = A_{\mu}^i \quad , \quad C  t_i C = - t_i^* \; .
\end{align}
(In the present formulation for path integral quantization using classical fields, $A_{\mu}^i $ is a real number; in canonical quantization it becomes a Hermitian operator, containing the destruction and creation operators for gauge bosons of species $i$. These real fundamental gauge fields are sometimes combined to form complex fields like $W^{\pm}_{\mu}$ of (\ref{complex}), which become nonhermitian operators with nonhermitian generators.)

The action in (\ref{eq5.31}) or (\ref{eq5.32}) for a single $r$, in a locally inertial coordinate system, is equivalent to (with an implied summation below over $\alpha$)
\begin{align}
2 {\cal L}^L  &=  i \psi _{L}^{\dagger } \, \overline{\sigma} ^{\mu} D_{\mu } \,  \psi _L   + h.c. \label{eqA10} \\
              &= i \left( \sigma^2 C \psi_R^{c \, *} \right)^{\dag} \, \overline{\sigma} ^{\mu} D_{\mu } \,  \left( \sigma^2 C \psi_R^{c \, *} \right)  + h.c. \\
             &=  i \, \psi_R^{c \, T} \, C \sigma^2 \overline{\sigma} ^{\mu} \sigma^2 D_{\mu } \, C \psi_R^{c \, *}  + h.c. \\
               &= i \, \psi_{R \, \alpha}^c \, \left( \sigma ^{\mu \, *} C D_{\mu } C \,  \psi_R^{c \, *} \right)_{\alpha} + h.c. \\
             &= \mp \, i \left( \sigma ^{\mu \, *} C D_{\mu } C \,  \psi_R^{c \, *} \right)_{\alpha} \psi_{R \, \alpha}^c  + h.c. 
                                                      \quad \text{upper\ sign\ for\ anticommuting\ fermion\ fields}  \\
              &= \mp  \, i \left( \sigma ^{\mu \, *} \left(  \partial_{\mu} - i C A_{\mu} C \right) \,  \psi_R^{c *} \right)_{\alpha} \psi_{R\, \alpha}^c  + h.c. \\
               &= \mp  \, i \left( \sigma ^{\mu \, *} \left(  \partial_{\mu} + i A_{\mu}^* \right) \,  \psi_R^{c \, *} \right)_{\alpha} \psi_{R\, \alpha}^c + h.c. \\
               &= \mp \, i \left( \sigma ^{\mu } D_{\mu} \,  \psi_R \right)^{c \, *}_{\alpha} \psi_{R\, \alpha}^c  + h.c. \\
               &= \mp  \, i  \left( \sigma ^{\mu } D_{\mu } \,  \psi_R^c \right)^{\dag}  \psi_{R}^c  + h.c. \\
               &= \pm \, i \psi_R^{c \, \dag} \sigma ^{\mu} D_{\mu }  \psi_R^c   + h.c.       \label{eqA20}                
\end{align}
Within the action $S^L = \int d^{4}x \, {\cal L}^L $, the second term of (\ref{eqA20} ) (represented by $h.c.$) gives the same contribution as the first after an integration by parts (with boundary contributions neglected),  so, in a locally inertial coordinate system, we can write (\ref{eq5.31}) and (\ref{eq5.32}) as
\begin{align}
S_{fb} = \sum_r S^r_{f} + \sum_r S^r_{b}
\end{align}
with either 
\begin{align}
S^r_{f} = \int d^{4}x \, i \, \psi _{f \, L}^{r \, \dagger } \, \overline{\sigma}  ^{\mu} D_{\mu } \, \psi _{f \, L}^r  \quad , \quad S^r_{b} = \int d^{4}x \, i \, \psi _{b \, L}^{r \, \dagger } \, \overline{\sigma}  ^{\mu} D_{\mu } \, \psi _{b \, L}^r 
\end{align}
or
\begin{align}
S^r_{f} = \int d^{4}x \, i \, \psi _{f \, R}^{r \, c \, \dagger } \, \sigma  ^{\mu} D_{\mu } \, \psi _{f \, R}^{r  \, c } \quad , \quad S^r_{b} = - \int d^{4}x \, i \, \psi _{b \, R}^{r  \, c \, \dagger } \, \sigma  ^{\mu} D_{\mu } \, \psi _{b \, R}^{r \, c}   \; .\label{minus}
\end{align}
Since fermions are treated just as in standard physics, we now focus exclusively on bosons in an $N + \overline{N}$ representation.
Let us leave the boson fields of the $N$ unchanged and left-handed, but change all those of the  $\overline{N}$ to right-handed via the above procedure. There are then N pairs of these new 2-component fields $\overline{\psi}_{b}^{r}$ with the same gauge quantum numbers, where one is left-handed and unchanged and the other is right-handed with the minus sign of (\ref{minus}):
\begin{align}
S_{b} &= \int d^{4}x \, \overline{\mathcal{L}}_{b} \quad , \quad 
\overline{\mathcal{L}}_{b} = \overline{\psi}_{b}^{\dagger } \left( x \right) \,A \, \overline{\psi}_{b} \left( x \right)
= \sum_{r} \overline{\psi}_{b}^{r \, \dagger } \left( x \right) \, A_{r} \, \overline{\psi}_{b}^{r} \left( x \right) \quad , \quad  r=1,2, ..., 2N  \label{eq702} \\
A_{r} &=   i \, \overline{\sigma} ^{\mu} D_{\mu } \quad \text{for\ field\ from\ N\ representation}  \label{eq703} \\
A_{r} &= -  i \, \sigma ^{\mu} D_{\mu } \quad \text{for\ field\ from\ $\overline{N}$ representation}. \label{eq704} 
\end{align}
$A$, with components $A_{r r'} = \delta_{r r'} A_r $, is Hermitian, so it has a complete orthonormal set of eigenfunctions:
\begin{align}
A \, U_i \left( x \right)  = a_i \, U_i \left( x \right) \; . \label{B2} 
\end{align}
Each 2N-component eigenfunction $U_i$ can be taken to be given by a 2-component spinor $u^{\sigma}_s \left( x \right) $ which has well-defined gauge quantum numbers $\underline{a}_{\, \mu}^s $, 4-momentum $p_{\mu}$ -- or frequency $\omega$ and 3-momentum $\vec{p}$ -- and 
helicity $\sigma=+$ or $-$, with the other components equal to 0:
\begin{align}
u^{\sigma}_s \left( x \right) &=  u^{\sigma}_s \left( 0 \right)  e ^{ i p_{\mu} x^{\mu} } = u^{\sigma}_s \left( 0 \right)  e ^{ - i \omega x^{0} + i \vec{p} \cdot \vec{x} } \\
A_{\mu} u^{\sigma}_s \left( x \right)  &= \underline{a}_{\, \mu}^s u^{\sigma}_s \left( x \right) \quad , \quad s, \sigma   \leftrightarrow i \label{B2b} \; .
\end{align}
(Again, the generators $t_i$ in $A_{\mu}$ are treated as operators. Helicity here is generalized to distinguish spin states with the full operator of (\ref{eq703}) or (\ref{eq704}), including the gauge fields.)

If $s$ corresponds to the N representation, this gives
\begin{align}
i \overline{\sigma} ^{\mu} D_{\mu } u^{\sigma}_s \left( x \right)  &=  \left[  \left( \omega +  \underline{a}_{\, 0}^s \right) + \left( \vec{p} -  \underline{\vec{a}}^s \right) \cdot \vec{\sigma} \right] u^{\sigma}_s \left( x \right) \label{B2c}  \\
&=  \left(  \tilde{\omega} \, \pm \, \tilde{p}  \right) u^{\sigma}_s \left( x \right) \label{B2d}  
\end{align}
where the upper (lower) sign holds for the spinor which has positive (negative) helicity, and
\begin{align}
\tilde{\omega} = \omega +  \underline{a}_{\, 0}^s \quad , \quad \tilde{p} = |\vec{p} -  \underline{\vec{a}}^s | \label{B2e} 
\end{align}
so that 
\begin{align}
a_i = a^{\sigma}_s =  \tilde{\omega} \, \pm \, \tilde{p}
\end{align}
where the upper sign holds for $\sigma=+$ (and the lower sign for $\sigma=-$).

If $r$ corresponds to the $\overline{N}$ representation, with $i \, \overline{\sigma} ^{\mu} \rightarrow -  i \, \sigma ^{\mu} $, the result is instead
\begin{align}
- i \sigma ^{\mu} D_{\mu } u^{\sigma}_s \left( x \right)  &=  \left[  - \left( \omega +  \underline{a}_{\, 0}^s \right) + \left( \vec{p} -  \underline{\vec{a}}^s \right) \cdot \vec{\sigma} \right] u^{\sigma}_s \left( x \right) \label{B2f}  \\
 &=  \left(  - \tilde{\omega} \, \pm \, \tilde{p}  \right) u^{\sigma}_s \left( x \right) \label{B2g}  
\end{align}
so that 
\begin{align}
a_i = a^{\sigma}_s =  - \tilde{\omega} \, \pm \, \tilde{p}
\end{align}
in this case.

For a given pair of 2-component spinors specified above with the same gauge quantum numbers (of the N such pairs), there are then 4 eigenfunctions, as listed in Table 1: the 2-component spinor from the N (16 or 5) can have helicity $+$ or $-$, and the same is true of the spinor from the $\overline{N}$ ($\overline{16}$ or $ \overline{5}$). 

As the first major step in the transformation to physical fields, we wish to transform $S_{b}$ to the form
\begin{align}
S_{b} = \int d^{4}x \, \left( \overline{\mathcal{L}}_{\phi} + \overline{\mathcal{L}}_{F} \right) \quad & , \quad 
\overline{ \mathcal{L}}_{\phi} = \overline{\phi} ^{\dag } \left( x \right) B \, \overline{\phi} \left( x \right)  
 \label{phi-F}\\
B = D^{\mu } D_{\mu }   \label{B33}
\end{align}
where  $\overline{\mathcal{L}}_{F}$ involves nondynamical auxiliary fields, as specified below.

$B$ also has a complete orthonormal set of eigenfunctions:
\begin{align}
B \, V_i \left( x \right)  = b_i \, V_i \left( x \right)   \; .   \label{B4}
\end{align}
(Each of the $2N$ components of $U_i$ or $V_i$  is itself a spinor with 2 complex components. For a fixed 4-momentum, $A$ and $B$ then each have 4 eigenfunctions for a given set of gauge quantum numbers, as indicated in Table 1.)
We wish to choose the eigenstates of $B$ to be essentially the same as those of $A$, in the sense that each $V_i$ is given by a 2-component spinor with the same gauge quantum numbers and the same 4-momentum (frequency and 3-momentum) as its progenitor in $U_i$. This means that each of the 4 modes in Table 1 for $A$ with a fixed $\tilde{\omega}$ and $\tilde{p}$ has to be matched to a corresponding mode for $B$ with the same $\tilde{\omega}$ and $\tilde{p}$, although the eigenvalues will, of course, be different for the different operators:
\begin{align}
v^{\sigma}_s\left( x \right) &=  v^{\sigma}_s\left( 0 \right)  e ^{ i p_{\mu} x^{\mu} } = v^{\sigma}_s\left( 0 \right)  e ^{ - i \omega x^{0} + i \vec{p} \cdot \vec{x} } \label{B4a}  \\
B v^{\sigma}_s\left( x \right)  &= b_{s}^{\sigma} v^{\sigma}_s \left( x \right) \quad , \quad s, \sigma   \leftrightarrow i  \label{B4b}  
\end{align}
with the other components of $V_i = V^{\sigma}_s$ equal to 0.
Here $\sigma$ is taken below to label the spin orientation of $v^{\sigma}_s$.
The special case
\begin{align}
b_i = b_{s}^{\sigma}  = \tilde{\omega}^2 - \tilde{p}^2  \quad \mathrm{if} \; A_{\mu} \; \mathrm{is\ constant} \label{B4c}
\end{align}
is emphasized in Table 1, but, as pointed out there, the results demonstrate that the required matching can always be accomplished, so that each of the fields $\phi_{\sigma}  \left( x \right) , F_{\sigma}  \left( x \right)$ defined below can be represented by a complete set of states.

With
\begin{align}
\overline{\psi}_{b} \left( x \right) = \sum_i \overline{\psi}_i \, U_i \left( x \right)
\label{expan}
\end{align}
(\ref{eq702}) can be rewritten as
 \begin{align}
S_b =  \sum_i \overline{\psi}_i ^{*} a_i \overline{\psi}_i \; .
\end{align}
Now define
 \begin{align}
\overline{\phi}_i &= \left( a_i / b_i \right) ^{1/2} \overline{\psi}_i \quad \; \text{if}  \; \; b_i / a_i > 0 \label{B19}\\
\overline{F}_i &=  \left| a_i \right|  ^{1/2} \overline{\psi}_i \quad \; \text{if}  \; \; b_i / a_i < 0 
\label{B20}
\end{align}
where each $b_i$ is matched to a corresponding $a_i$ in the way specified below, so that
 \begin{align}
S_b =  \sum_i \overline{\phi}_i ^{*} \, b_i \, \overline{\phi}_i  - \sum_i \sgn \left( b_i \right) \overline{F}_i^{*} \overline{F}_i   \label{B70} 
\end{align}
where the limitations on these summations are defined by (\ref{B19}) and (\ref{B20}). (The constant Jacobian for this transformation in the path integral cancels between numerator and denominator when it is used to calculate physical quantities or propagators, as in (\ref{eq13.8})).

For each case in Table 1, one of the two states labeled $\phi$ or $F$ can be rotated to have spin up, and the other to have spin down, with no change in the action (which involves just $B=D^{\mu}D_{\mu}$). These 2-component spinors can then be taken to be the $v^{\sigma}_s\left( x \right)$ of (\ref{B4a})-(\ref{B4b}), giving the $2N$-component eigenfunctions $V_i =  V^{\sigma}_s \left( x \right)$. To avoid confusion, let us write $\sigma = \uparrow, \downarrow$ respectively for the spin up and down states labeled $\phi$ in Table 1, and $\Uparrow, \Downarrow$ for the spin up and down states labeled $F$. Then the general 2N-component fields can be represented as
 \begin{align}
\phi'_{\uparrow} \left( x \right) &= \sum_s \overline{\phi} _s ^{\uparrow} V^{\uparrow}_s \left( x \right) \quad , \quad
\phi'_{\downarrow} \left( x \right) = \sum_s \overline{\phi} _s ^{\downarrow} V^{\downarrow}_s \left( x \right) 
\quad , \quad s=1,2,...,N \label{primed1} \\
F'_{\Uparrow} \left( x \right) &= \sum_s \overline{F} _s ^{\Uparrow} V^{\Uparrow}_s \left( x \right) \quad , \quad
F'_{\Downarrow} \left( x \right) = \sum_s \overline{F} _s ^{\Downarrow} V^{\Downarrow}_s \left( x \right) \label{primed2}
\end{align}
with 2-dimensional components $\phi ^{\, \prime \, r}_{\uparrow} \left( x \right)$, $\phi ^{\, \prime \, r}_{\downarrow} \left( x \right)$, 
$F ^{\, \prime \, r}_{\Uparrow} \left( x \right)$, $F ^{\, \prime \, r}_{\Downarrow} \left( x \right)$, $r=1, 2, ..., N$.
The expansion coefficients $\overline{\phi} _s ^{\uparrow}$, $\overline{\phi} _s ^{\downarrow}$, $\overline{F} _s ^{\Uparrow}$, and $\overline{F} _s ^{\Downarrow}$ are the more specifically renamed $\overline{\phi} _i$ and $\overline{F} _i$ of (\ref{B19}) and (\ref{B20}). These expansions (\ref{primed1}) and (\ref{primed2}) in terms of eigenstates of B correspond to the expansion (\ref{expan}) for A.

$S_b$ can now be written as 
\begin{align}
S_b &= S'_{\phi} + S'_F \label{nearfinal0} \\
S'_{\phi} &= \int d^{4}x \left( \phi_{\uparrow} ^{\, \prime\, \dag } \left( x \right)  B \, \phi'_{\uparrow}  \left( x \right)  + \phi_{\downarrow} ^{\, \prime\, \dag } \left( x \right)  B \, \phi'_{\downarrow}  \left( x \right) \right) \label{nearfinal1} \\
&= \int d^{4}x \, \sum_{r}  \left( \phi_{\uparrow} ^{\, \prime \, r \, \dag } \left( x \right)  B \, \phi_{\uparrow}^{\, \prime \,r }  \left( x \right)  + \phi_ {\downarrow} ^{\, \prime \,r \, \dag } \left( x \right)  B \phi_{\downarrow}^{\, \prime \, r} \left( x \right) \right) \label{nearfinal2} 
 \end{align}
 where $S'_F$ is still given by the second term of (\ref{B70}).

\noindent\rule{\linewidth}{0.4pt}
\begin{center}
\begin{tabular}{ |p{1.9 cm}| |p{1.6cm}| p{0.9cm}| p{0.8cm}| |p{2.3cm}|  p{1.2cm}|  p{1.2cm}|  p{1.5cm}|  }
 \hline
  \multicolumn{8}{|l|}{\hspace{2.2cm}modes of A reassigned to modes of B } \\
 \hline
 $\; \tilde{\omega}$ and   $\tilde{p}$ & \; \, rep & L, R &\hspace{0.015cm}  hel & $A$ eigenvalue & $A$ sign & $B$ sign & $B$ mode \\
 \hline
 $\tilde{\omega} > 0$ & 16 or 5 & L & \;  \,-- & $\tilde{\omega}$ -  $\tilde{p}$  & +  & + & $\phi $ \\
$|\tilde{\omega} | >  \tilde{p} $ & 16 or 5 & L & \,   +   & $\tilde{\omega}$ +  $\tilde{p}$    & + & +  & $\phi $ \\
& $\overline{16}$ or $\overline{5}$ & R & \,  + &  - $\tilde{\omega}$ +  $\tilde{p}$   & \,--  & + & \footnotesize{F} \\
& $\overline{16}$ or $\overline{5}$ & R & \;   \,--  & - $\tilde{\omega}$ -  $\tilde{p}$  & \,-- & + & \footnotesize{F} \\
\hline
$\tilde{\omega} > 0$ & 16 or 5  & L & \;  \,--  & $\tilde{\omega}$  -   $\tilde{p}$  &  \,--  &  \,--  & $\phi $ \\
 $|\tilde{\omega} | <  \tilde{p} $ &16 or 5 & L & \, +  & $\tilde{\omega}$ +  $\tilde{p}$   & + &  \,--  & \footnotesize{F} \\
& $\overline{16}$ or $\overline{5 }$  & R & \,  +  & - $\tilde{\omega}$+  $\tilde{p}$  & + &  \,--  & \footnotesize{F} \\
& $\overline{16}$ or $\overline{5 }$ & R & \;   \,--   &  - $\tilde{\omega}$ -  $\tilde{p}$   &  \,--  &  \,--  & $\phi $ \\
\hline
$\tilde{\omega} < 0$ & 16 or 5   & L & \;  \,--   & $\tilde{\omega}$ -   $\tilde{p}$  &  \,--   & + & \footnotesize{F}\\
$|\tilde{\omega} | >  \tilde{p} $ & 16 or 5 & L &  \,  +  & $\tilde{\omega}$ +  $\tilde{p}$    &  \,--  & + & \footnotesize{F}  \\
& $\overline{16}$ or $\overline{5 }$ & R & \,  +  & - $\tilde{\omega}$ +  $\tilde{p}$  & +  & + & $\phi $ \\
& $\overline{16}$ or $\overline{5 }$ & R& \;   \,-- &  - $\tilde{\omega}$ -  $\tilde{p}$   & + & + & $\phi $ \\
\hline
$\tilde{\omega} < 0$ & 16 or 5  & L & \;  \,--   & $\tilde{\omega}$ -  $\tilde{p}$  &  \,-- &  \,-- & $\phi $ \\
$|\tilde{\omega} | <  \tilde{p} $ & 16 or 5 & L & \,  +  & $\tilde{\omega}$ +  $\tilde{p}$   & + &  \,--  & \footnotesize{F} \\
& $\overline{16}$ or $\overline{5 }$ & R & \,  +  & - $\tilde{\omega}$ +  $\tilde{p}$ & + &  \,--  & \footnotesize{F} \\
& $\overline{16}$ or $\overline{5 }$ & R & \;   \,--   &  - $\tilde{\omega}$ -  $\tilde{p}$   &  \,--  &  \,-- & $\phi $ \\
 \hline
\end{tabular} \\
\end{center}
\noindent
Table 1. Here we consider how the eigenstates $u_i$ and $v_i$ of $A$ and $B$ -- the operators defined in (\ref{eq703})-(\ref{eq704}) and (\ref{B33}) -- can be matched. In the case of $A$, any one of the 16 or 5 pairs in the full $32 \times 32$ or $10 \times 10$ vector $U_i$ of (\ref{B2}) consists of two fields with the same gauge quantum numbers: a left-handed field from the 16 or 5 representation and a right-handed field from the $\overline{16}$ or $\overline{5 }$. For a specific $\omega$ 
and $\vec{p}$ in (\ref{B2e}), there are two eigenstates for the left-handed field and two for the right-handed field, corresponding to positive and negative helicities, labeled $+$ and $-$ respectively. These eigenvalues and their signs are given in the columns labeled ``$A$ eigenvalue'' and ``$A$ sign'', for the four possible combinations of $\tilde{\omega}$ and $\tilde{p} $ which yield different signs. As noted in (\ref{B4c}), in the special case that $A_{\mu}$ is constant, the corresponding eigenvalues of $B$ are $ \tilde{\omega}^2 - \tilde{p}^2 =\left( \left| \tilde{\omega}\right| +\tilde{p} \right) \left( \left| \tilde{\omega} \right| - \tilde{p} \right)$, with the sign of $b_i $ always determined by $\left( \left| \tilde{\omega} \right| - \tilde{p} \right)$, as indicated in the column ``$B$ sign''. If the signs for $A$ and $B$ agree, the matched eigenstate of $B$ is labeled $\phi$. If they do not agree, the matched state is labeled $F$. It is remarkable that in every case two matches are obtained for $\phi$ and two for $F$. Moreover, this matching holds regardless of the sign for $B$, since if this sign is reversed, for any of the 4 cases shown in the table, there are still two matching eigenvalues for $\phi$ and two for $F$  (since $A$ always has two $+$ and two $-$ eigenvalues). This implies that each of the fields $\phi'_{\uparrow} \left( x \right)$, $\phi'_{\downarrow} \left( x \right)$, $F'_{\Uparrow} \left( x \right)$, $F'_{\Downarrow} \left( x \right)$ of (\ref{primed1}) and (\ref{primed2}) is always represented by a complete set of states (with the specified spin).
\noindent\rule{\linewidth}{0.4pt}

In the transformations above (and in most of this paper) the last term in (\ref{eq4.8}) has been neglected, but it must be retained when using the terms in the action (\ref{nearfinal0}) which are negative in the original path integral, since it has Euclidean form. This is done below in Appendix \ref{sec:appD}, where it is shown that the final scalar-boson action $S_b$, appropriate for the Lorentzian path integral, has the same form as in (\ref{nearfinal0}), but with the negative-action modes redefined. 

\section{\label{sec:appD}Transformation to Lorentzian path integral: scalar bosons}

Here we consider only $\phi'_{\uparrow} (x)$ explicitly, since $\phi'_{\downarrow} (x)$ $F'_{\Uparrow} (x)$, and $F'_{\Downarrow} (x)$ can be treated the same way. Since the eigenvectors of (\ref{primed1}) are orthonormal, its contribution to the action of (\ref{nearfinal1}) or (\ref{B70}) is (in terms of the expansion coefficients $\overline{\phi} _s^{\uparrow}$ of $\phi'_{\uparrow}  \left( x \right)$)
\begin{align}
S_{\uparrow}=\sum_s \overline{\phi} _s^{\uparrow \, *} \, b_s\, \overline{\phi} _s^{\uparrow} \quad , \quad s=1,2,...,N \; .
\label{actup}
\end{align}
We will write the contribution of a single $s$ as 
\begin{align}
S_s = \phi^* \, b \, \phi = b \, x^2 + b \, y^2 \; ,
\end{align}
where $\phi = x + iy$, before the quartic term of (\ref{eq4.8}) is included. Symmetry under $x \rightarrow -x$ implies that, in integrating over a single variable $x$, this quartic interaction can be replaced by an effective action $c+ \left( a x^2 + \frac{1}{2} \lambda x^4 \right)$ (where $c$, $a$, and $\lambda$ depend on $s$ and include the effect of all the other variables), so the contribution from the single variable $x=\mathrm{Re} \, \phi$ to the path integral $Z_{\uparrow}=\prod_s z_s$ is
\begin{align}
z_x &= \int_{-\infty }^{\,\infty }dx \, e^{- S_x} \quad , \quad S_x= c+ b' x^2 + \frac{1}{2} \lambda x^4  \quad , \quad b' = b +a \; .
\end{align}

If $b' $ is non-negative, the minimum in the action is at $x^2=0$. But if $b'$ is negative, minimization gives 
\begin{align}
x_0^2 = |b'|/\lambda \; .
\label{min}
\end{align}
(In a full description, a minimum is first found for all the variables taken together, and then $x$ is varied in the path integral, with the minimum of 
(\ref{min}) 
automatically consistent with the full minimum.) 

A shift from the minimum is given by $x=x_0 + \delta x$,
and the lowest-order additional action is
\begin{eqnarray}
\delta S_x &=&\left[ \frac{\partial S_x}{\partial x}\right] _0 \delta x
+\frac{1}{2}\left[ \frac{\partial ^{2}S_x}{\partial x^2}\right] _0\left( \delta x \right) ^{2} \\
&=&0+\left[ - \left\vert b' \right\vert +3 \lambda
x_0^2 \right] \left( \delta x\right) ^{2} \\
&=&2\left\vert b' \right\vert \left( \delta
x\right) ^{2}\;.
\end{eqnarray}
Notice that the quartic coefficient $\lambda$ is not present in this result (so we could even formally take $\lambda \rightarrow 0$, consistent with (\ref{eq2.26xx}) in the present high-energy regime where the fields are reforming and the renormalization of $\lambda$ is very small).

We are then back to the original form of (\ref{nearfinal2}) or (\ref{B70}) with $b \rightarrow 2 |b'|$ (to a very good approximation, since the higher-order terms are proportional to $\lambda$) but with a positive action, making the modified path integral convergent. (The vacuum is left with a ``condensate'', having a constant action $c-|b'|^2 / 2 \lambda$, which is added to the other vacuum fields.)

Continuing with the case that $b' $ is negative, and letting 
\begin{align}
x' = c_b^{-1/2} \delta x \quad , \quad c_b = \frac{|b|}{2 |b'|} 
\label{rescale}
\end{align}
we have, for the contribution to the path integral from $x$, omitting a constant factor,
\begin{align}
z_x^E = c_b^{1/2} \int_{-\infty }^{\,\infty }dx' \, e^{- |b| x^{\prime \, 2}} = c_b^{1/2}  \sqrt{\frac{\pi}{|b|}} 
\label{zx}
\end{align}
and 
\begin{align}
z^E = z_x^E z_y^E = c_b \frac{\pi}{|b|} \; .
\end{align}
With the proper choice of a convergence factor (which also yields causal propagators \textemdash \, see e.g. pp. 286-288 of \cite{peskin}), the corresponding Lorentzian path integral is
\begin{align}
z^{L} = \int_{-\infty }^{\,\infty }dx \, e^{i\, \left( b+ i \epsilon \right) \, x^2}  
\int_{-\infty }^{\,\infty }dy \, e^{i\, \left( b+ i \epsilon \right) \,  y^2}
= i  \, \frac{\pi}{b + i \epsilon} 
\end{align}
so 
\begin{align}
z^{E} = i \, c_b \, z^L \quad \mathrm{as} \; \epsilon \rightarrow 0+
\end{align}
when $b$ is negative. Again, the extra constant factors including $c_b$ in the path integral cancel (between numerator and denominator) when it is used to calculate physical quantities or propagators, as in (\ref{eq13.8}).

When the eigenvalues for $\phi'_{\uparrow}$ are non-negative, the transformation to a Lorentzian path integral is more straightforward (with similar steps to those  above), and equivalent arguments transform the path integrals for $\phi'_{\downarrow}$, $F'_{\Uparrow}$, and $F'_{\Downarrow}$ to Lorentzian form. 

Now let $\overline{\phi} _s ^{\prime \, \uparrow}$, $\overline{\phi} _s ^{\prime \, \downarrow}$, $\overline{F} _s ^{\prime \, \Uparrow}$, 
and $\overline{F} _s ^{\prime \, \Downarrow}$ be the new versions of the expansion coefficients in (\ref{primed1}) and (\ref{primed2}) when they are shifted and rescaled according to the prescription (\ref{rescale}) above, and let 
 \begin{align}
\phi_{\uparrow} \left( x \right) &= \sum_s \overline{\phi} _s ^{\prime \, \uparrow} V^{\uparrow}_s \left( x \right) \quad , \quad
\phi_{\downarrow} \left( x \right) = \sum_s \overline{\phi} _s ^{\prime \, \downarrow} V^{\downarrow}_s \left( x \right) \\
F_{\Uparrow} \left( x \right) &= \sum_s \overline{F} _s ^{\prime \, \Uparrow}V^{\Uparrow}_s \left( x \right) \quad , \quad
F_{\Downarrow} \left( x \right) = \sum_s \overline{F} _s ^{\prime \, \Downarrow} V^{\Downarrow}_s \left( x \right) 
\label{unprimed}
\end{align}
so that (\ref{nearfinal0}) can be rewritten as
\begin{align}
S_b &= S'_b +S_{neg} \quad, \quad S'_b = S_{\phi} + S_F \label{final0x} \\
S_{\phi} &= \int d^{4}x \left( \phi_{\uparrow} ^{ \, \dag } \left( x \right)  B \, \phi_{\uparrow}  \left( x \right)  + \phi_{\downarrow} ^{\, \dag } \left( x \right)  B \, \phi_{\downarrow}  \left( x \right) \right) \label{final1x} \\
&= \int d^{4}x \, \sum_{r}  \left( \phi_{\uparrow} ^{ \, r \, \dag } \left( x \right)  B \, \phi_{\uparrow}^r   \left( x \right)  + \phi_ {\downarrow} ^{r \, \dag } \left( x \right)  B \phi_{\downarrow}^r \left( x \right) \right) \label{final2x} \\
S_{F} &= \int d^{4}x \left( F_{\Uparrow} ^{\, \dag } \left( x \right)  \, F_{\Uparrow}  \left( x \right)  + F_{\Downarrow} ^{\, \dag } \left( x \right)  \, F_{\Downarrow}  \left( x \right) \right) \label{final3x} \\
&= \int d^{4}x \, \sum_{r}  \left( F_{\Uparrow} ^{\, r \, \dag } \left( x \right)  \, F_{\Uparrow}^{\, r }  \left( x \right)  + F_ {\Downarrow} ^{r \, \dag } \left( x \right)  F_{\Downarrow}^r \left( x \right) \right) \; .
\label{final4x}
\end{align}
with a Lorentzian path integral, 
where $S_{neg}$ is the constant vacuum contribution of the negative-action modes. 



\end{document}